\documentclass[amsmath,amssymb,aps,pra,superscriptaddress,showpacs,10pt,reprint]{revtex4-1}

\usepackage{graphicx}
\usepackage{grffile} 
\usepackage{array}
\usepackage{bbm}
\usepackage{color}
\usepackage[utf8]{inputenc}
\usepackage[OT1]{fontenc}

\usepackage{units} 

\usepackage{dsfont}
\usepackage{hyperref}

\newcommand{\eq}[1]{Eq.~\eqref{#1}}
 \newcommand{\bra}[1]{\left\langle{#1}\right|}
 \newcommand{\ket}[1]{\left|{#1}\right\rangle}

\newcommand{\partdifffrac}[2]{\frac{\partial #1}{\partial #2}}

\newcommand{\e}[1]{\operatorname{e}^{#1}}
\newcommand{\ehoch}[1]{\operatorname{exp}{\left[ #1 \right]}}
\newcommand{\I}{i}

\newcommand{\D}{\text{d}}
\newcommand{\summe}[3]{\sum\limits^{#3}_{#1 = #2}}

\def\vec#1{\ensuremath{\mathchoice{\mbox{\boldmath$\displaystyle#1$}}
                            {\mbox{\boldmath$\textstyle#1$}}
                            {\mbox{\boldmath$\scriptstyle#1$}}
                            {\mbox{\boldmath$\scriptscriptstyle#1$}}}}

\def\ba#1\ea{\begin{align}#1\end{align}}

\usepackage{tikz}                       
\usepackage{pgfplots}                   
\usepgfplotslibrary{external}

\pgfplotsset{%
  every axis legend/.append style={%
    cells={anchor=west},
    at={(0.96,0.04)},
    anchor=south east,
    font=\scriptsize
  },
  every axis/.append style={%
    yticklabel style={/pgf/number format/fixed zerofill, /pgf/number format/precision=2}
  },
  width= \textwidth, height=8cm, xmajorgrids=true, xminorgrids=false, minor x tick num=1
}

\begin{document}
\title{Double Bragg diffraction: A tool for atom optics} 
    \author{E.~Giese}
    \author{A.~Roura}

\affiliation{Institut f\"ur Quantenphysik and Center for Integrated Quantum Science and Technology $\left(\text{IQ}^{\text{ST}}\right)$, Universit\"at Ulm, Albert-Einstein-Allee 11, D-89081, Germany.} 
 \author{G.~Tackmann}    
 \author{E.~M.~Rasel}
\affiliation{Institut f\"ur Quantenoptik, Leibniz Universit\"at Hannover, Welfengarten 1, D-30167 Hannover, Germany.} 
      \author{W.~P.~Schleich}
\affiliation{Institut f\"ur Quantenphysik and Center for Integrated Quantum Science and Technology $\left(\text{IQ}^{\text{ST}}\right)$, Universit\"at Ulm, Albert-Einstein-Allee 11, D-89081, Germany.} 

\begin{abstract}
The use of retro-reflection in light-pulse atom interferometry under microgravity conditions naturally leads to a double-diffraction scheme. The two pairs of counterpropagating beams induce simultaneously transitions with opposite momentum transfer that, when acting on atoms initially at rest, give rise to symmetric interferometer configurations where the total momentum transfer is automatically doubled and where a number of noise sources and systematic effects cancel out. Here we extend earlier implementations for Raman transitions to the case of Bragg diffraction. In contrast with the single-diffraction case, the existence of additional off-resonant transitions between resonantly connected states precludes the use of the adiabatic elimination technique. Nevertheless, we have been able to obtain analytic results even beyond the deep Bragg regime by employing the so-called ``method of averaging,'' which can be applied to more general situations of this kind. Our results have been validated by comparison to numerical solutions of the basic equations describing the double-diffraction process.
\end{abstract}

\maketitle

\section{Introduction}

There has recently been a growing interest in the possibilities that (ultra)cold atoms in microgravity offer for long-time interferometry and its application to high-precision measurements \cite{van-zoest10,geiger11,muentinga,STE-quest}.
This has contributed to stimulating the development of compact devices \cite{Bodart10,Hauth13,muquans,herr}, which are typically necessary for such environments and are also important for their use in navigation, geodesy, and geophysics, to name a few \cite{kasevich-patent,durfee06, Schmidt11}.
In this article we put forth a generalization of conventional Bragg diffraction in light-pulse atom interferometry which can be a valuable technique for compact set-ups and microgravity applications.

\subsection{Overview of atom interferometry and Bragg diffraction}

The diffraction of X rays by crystals studied a century ago by Bragg \cite{bragg} and Laue \cite{laue} is a beautiful manifestation of the wave nature of this high-frequency electromagnetic radiation as well as of the periodic structure of crystals, and it has played a central role in characterizing the microscopic structure of a wide range of materials, including complex proteins and DNA. A similar phenomenon involving the diffraction of neutrons by crystals was exploited to build  some of the first matter-wave interferometers and gave rise to the rich field of neutron interferometry \cite{Rauch74,rauch+werner00}. Material gratings have also been used to diffract beams of atoms and molecules \cite{cronin09}. An interesting alternative approach relies on the use of standing electromagnetic waves as phase gratings \cite{gould86,martin,rasel95} or absorption gratings \cite{fray04,haslinger13}, depending on whether resonant or nonresonant radiation is employed. These set-ups, still based on the interaction between matter and light but with their roles reversed compared to traditional optical interferometers, provide high-quality gratings with controllable properties as a consequence of our ability to manipulate light.
It should be noted that all cases mentioned so far involve static potentials (or effective potentials) for the dynamics of the matter waves, which implies that the kinetic energy (the modulus of the wave-vector) before and after scattering remains the same. Moreover, 
the duration of the interaction is determined by the transverse velocity and thickness of the (light) crystal/grating. In this respect one should distinguish between thick and thin gratings. The former, which correspond to the so-called Bragg regime, exhibits high momentum selectivity and only one non-trivial diffraction order for resonant momenta \cite{martin,Giltner95}. In contrast, for thin gratings (corresponding to the Raman-Nath regime, also known as Kapitza-Dirac for light gratings) there is little momentum selectivity and many diffraction orders are populated \cite{gould86}. An intermediate regime where velocity selectivity is somewhat relaxed but diffraction still takes place mainly to a single order, sometimes known as ``quasi-Bragg'' regime, has been studied in Ref.~\cite{quasi-Bragg} and will also be of particular interest for us.

Laser cooling to sub-Doppler temperatures \cite{Lett88,Dalibard89} made the advent of light-pulse atom interferometry \cite{borde89,kasevich91,peters99} possible. It is based on time-modulated laser pulses that drive Rabi oscillations between different momentum states (possibly entangled to different internal states) and whose duration and intensity can be adjusted to act as beam splitters ($\pi/2$ pulses) or mirrors ($\pi$ pulses). The laser beams employed are long and wide enough so that there is a fixed momentum transfer along the longitudinal direction of the beam and only the motion along this dimension matters, while transverse velocities remain unchanged.
Despite such fixed momentum transfer, a more or less narrow momentum band around
resonant states will be diffracted \cite{moler92}. This is because for pulses with finite duration differences between initial and final kinetic energies (which can be larger for shorter pulses) are allowed, as can be understood from Heisenberg's uncertainty relation for time and energy.
The role of pulse duration is then analogous to the crystal thickness mentioned above.

The first implemented scheme for light-pulse interferometry \cite{kasevich91} and widely used to date relies on two-photon Raman transitions induced by a pair of counterpropagating lasers with different frequencies. These are transitions between different internal states entangled to different momenta \cite{borde89}, and the internal-state labeling allows the read-out of the exit ports even when they spatially overlap. Moreover, the velocity selection effect can be somewhat relaxed by using shorter pulses with higher intensity. This together with the ability to deal with substantially overlapping clouds for the two exit ports, reduces the requirement of very narrow initial momentum distributions (well below recoil velocities) and cold thermal atoms obtained from optical molasses can be employed without the need for evaporative cooling.

An alternative scheme for light-pulse interferometry is based on Bragg diffraction and involves transitions between states with different momenta without changing the internal state. These are induced by a pair of counterpropagating lasers with wave numbers $k_1$ and $-k_2$, leading to a total momentum transfer of $\hbar K  \equiv \hbar k_1 + \hbar k_2$, and frequencies slightly detuned to account for the recoil energy. There is always a frame where the frequencies of the two beams are the same and one has a standing wave. In this frame only a narrow band of momentum states around the two resonant momenta $\pm \hbar K/2$ will be diffracted and there is a fixed momentum transfer $\mp \hbar K$ respectively. The process is then analogous to Bragg diffraction in crystals (except that the duration of the interaction is determined by the pulse duration). It is possible to select other resonant momenta by detuning the frequencies while keeping the total momentum transfer $\hbar K$ fixed, which can be understood as changing to a different frame where the atoms have the desired initial velocity. These kind of interferometers have been widely applied to Bose-Einstein condensates (BECs) \cite{Kozuma,torii00,Debs11,muentinga}, whose narrower momentum distribution (significantly below recoil velocity) allows good spatial separation of the exit ports and a high diffraction efficiency: For a $\pi$ pulse most atoms are diffracted and only one diffraction order is populated. (Note that contrary to the Raman case, trying to relax the effect of velocity selection by using shorter pulses with higher intensity can only be done to some extent, corresponding to the quasi-Bragg regime \cite{quasi-Bragg}, because one will otherwise start to populate other diffraction orders \cite{momentum-width}.) Using a condensate is not mandatory since one can apply initially a long velocity-selective Raman $\pi$ pulse \cite{mueller08-24hbarK}, but this reduces significantly the number of atoms available. On the other hand, an advantage of interferometry based on Bragg diffraction is that having the same internal state in both interferometer branches reduces a number of systematic effects and noise sources. Moreover, with sufficient laser intensity one can increase the effective momentum transfer to a multiple of $\hbar K$ (and with that the sensitivity of the interferometer) by adjusting the frequencies so that the resonant condition corresponds to a higher diffraction order~\cite{mueller08-24hbarK}.

\subsection{Microgravity environments and double Bragg diffraction}

The use of a retro-reflection geometry, where the two laser beams reach one side of the interferometer set-up through a common optical fiber and reflect off a mirror at the other side in such a way that the reflected beams are aligned with the incident ones (giving rise to two pairs of counterpropagating beams), is beneficial for a number of reasons. Firstly, most vibration effects on the laser phases are common to the two lasers and cancel out in a two-photon process and only mirror vibrations have an effect on processes involving pairs of counterpropagating beams. Secondly, this geometry reduces the undesired effects of wave-front distortions, which become more important for larger effective momentum transfer and longer interferometer times, since those cancel to first order provided that no additional distortions are generated while the beams propagate towards the mirror and return. 
Retro-reflection is, therefore, commonly employed in atomic fountains for high-precision interferometry \cite{peters01,dickerson13}. In that case the non-vanishing velocity of the atoms along the direction of the lasers selects one of the two pairs of counterpropagating beams, while the other remains off-resonant.
However, achieving much higher sensitivities requires longer interferometer times and sufficiently extended times are only possible in microgravity environments \cite{geiger11,muentinga,STE-quest}. Using retro-reflection in microgravity with narrow initial momentum distributions naturally leads to a double-diffraction scheme where the action of the first beam-splitter pulse on atoms initially at rest creates a superposition of two states with opposite momenta $+\hbar K$ and $-\hbar K$ since each one of the two pairs of counterpropagating beams induces a resonant transition with opposite momentum transfer. Besides doubling the total momentum transfer, this leads to a symmetric interferometer where a number of systematic effects cancel out, including those due to laser-phase noise and those involving terms proportional to $K^2$. This kind of symmetric interferometer based on a double-diffraction scheme was first implemented for Raman transitions in Ref.~\cite{Leveque}. (See Ref.~\cite{Dubetsky02} for a different proposal employing also two pairs of counterpropagating Raman beams.) In that case, the symmetric configuration led to the added benefit of a reduction of AC Stark shift effects or any other effects acting differently on internal states since the atoms are at all times in the same internal state for both interferometer branches, in contrast with the single-diffraction scheme. Note also that, although very natural under microgravity conditions, double diffraction can also be employed in a gravitational field and for non-vanishing initial velocities provided that the laser beams are transverse to the motion of the (undiffracted) atoms, as in the gyroscope set-up of Ref.~\cite{Leveque}.
Double Raman diffraction has also been employed in a gravimeter, where three different injected laser frequencies are necessary to account (via appropriate frequency chirping) for the changing Doppler shift of the accelerated atoms \cite{Malossi}.

In this article we analyze in detail the extension of the double-diffraction scheme to the case of Bragg scattering, which is particularly well suited for interferometry with BECs, as already mentioned above.
Some efforts in this direction have already been made experimentally \cite{LeCoq}, but not in connection with atom interferometry nor focusing on the special properties of the double-diffraction scheme and the much richer dynamics associated with it.
The double Bragg diffraction scheme involves a pair of slightly detuned lasers (with frequency difference corresponding to the recoil energy) retro-reflected off a mirror, and it is crucial that only two-photon processes involving counterpropagating beams (with unequal frequencies) take place, while those associated with copropagating beams should be entirely suppressed. This is achieved by injecting the two lasers with orthogonal polarizations (that can be either circular or linear 
\footnote{The injected polarizations also need to be orthogonal for the case of linear polarizations. In contrast, the linear polarizations must be parallel for Raman transitions. This difference is a consequence of the change of total angular momentum by $\Delta F = \pm 1$ in the latter case, whereas the internal state remains unaltered for Bragg diffraction.})
and inserting a quarter-wave plate in front of the mirror, which guarantees that each reflected beam has a polarization orthogonal to the incoming one. Similarly to double diffraction for Raman processes, the beam splitter creates an equal-amplitude superposition of $+\hbar K$ and $-\hbar K$ states, which leads to a symmetric interferometer configuration with similar desirable properties.
The use of Bragg scattering has certain advantages compared to set-ups based on Raman transitions, even for a double-diffraction scheme. Firstly, it is easier to implement experimentally because the small frequency detuning that is required (of the order of tens of kHz) can be achieved with a single laser and acousto-optic modulators (AOMs) rather than two phase-locked lasers as required for the Raman case, where the frequency difference is typically of several GHz. This can be particularly advantageous for compact devices. Secondly, one can use higher-order Bragg diffraction to increase the effective momentum transfer to $n \hbar K$ while keeping all the advantageous features of double diffraction and the associated symmetric interferometer configurations. Thus, in contrast to the double-diffraction scheme based on Raman transitions, where one can only do so by introducing additional $\pi$ pulses \cite{Leveque}, in the Bragg case one could use an optimized combination of both techniques to maximize the total amount of effective momentum transfer, as already done for single diffraction \cite{chiow11}.

\subsection{Special features of double Bragg diffraction}

We highlight here several important aspects that differ from single diffraction as well as new features that were absent in that case.

First of all, the resonantly connected states form a \emph{three-level system} and one has generalized Rabi oscillations between them rather than the usual Rabi oscillations for two-level systems. Nevertheless, a pulse with an appropriate duration and laser intensity can still act as a beam splitter: It evolves a zero-momentum state into an equal superposition of $+n \hbar K$ and $-n \hbar K$ momenta. Furthermore, for a double duration of the pulse it acts as a mirror, exchanging states with momentum $+n \hbar K$ to $-n \hbar K$ and vice versa.

The existence of two pairs of counterpropagating beams results in a much wider range of possibilities for higher-order processes relating any two given momentum eigenstates. In particular there are additional off-resonant transitions between resonantly connected states. This precludes the use of the standard method of adiabatic elimination of nonresonant states \cite{bernhardt,stenholm,Brion07}. Nevertheless, one can still obtain analytical results employing the so-called ``method of averaging'' described in the next subsection, which can be applied to more general situations of this kind, provided that one has two sufficiently different frequency scales, as controlled by an adiabaticity parameter $\varepsilon$. Moreover, because of the additional pair of counterpropagating beams, it is no longer possible to find a reference frame where one simply has a standing wave.

The aspects mentioned in the previous paragraph give rise to the following new features, absent in single diffraction:
\begin{itemize}
\item Fast oscillations with smaller amplitude superimposed on the slow generalized Rabi oscillations between resonant states. In contrast to single diffraction, their amplitude is of order $\varepsilon$ (when considering square pulses) rather than $\varepsilon^2$, and results from the interference between slow and fast contributions to the dynamics of resonant states.
\item A non-trivial AC Stark shift associated with higher-order processes. For single Bragg diffraction one can easily argue that the two resonant states experience the same shifts by considering the frame where one has a standing wave and where the situation is symmetrical for both momentum states. As mentioned above, this is no longer possible for two pairs of counterpropagating beams.
\end{itemize}

\subsection{The method of averaging}

The dynamics of momentum states differing by multiples of $\hbar K$ which are resonantly or nonresonantly connected by $2n$-photon transitions can be treated analytically when there is a large separation of scales between the frequencies associated with the two kinds of processes (resonant vs.\ nonresonant). This separation of scales is controlled by an adiabaticity parameter $\varepsilon$ given by the ratio of those two frequencies (in our case, the Rabi frequency for two-photon transitions and the recoil frequency). When this parameter is small, one can study in a controlled manner the slow and fast contributions to the dynamics as well as their mutual influence following the ``method of averaging'' introduced in Ref.~\cite{bogoliubov}. The fast contributions, whose dynamics is modulated by the slow part, have a small amplitude and can be organized as an expansion in powers of $\varepsilon$. In turn, the fast contributions can combine resonantly with fast oscillating terms in the Hamiltonian and give corrections to the dynamics of the slow parts. Proceeding in this way, one can write down a set of coupled equations for the slow and fast terms which are equivalent at any desired order in $\varepsilon$  to the original equations for the full solution (involving the sum of slow and fast terms).

The method just described can deal with situations where both slow and fast terms contribute to the same degree of freedom. This is a new feature of double Bragg diffraction which is not present in the single-diffraction case and cannot be dealt with using the standard adiabatic elimination technique. It is, thus, an approach that allows a controlled and systematic way of analyzing the so-called ``quasi-Bragg'' regime \cite{quasi-Bragg}, where $\varepsilon$ is sufficiently small but gives rise to non-negligible corrections, and obtaining analytic solutions. It should be noted that the basic equations describing the dynamics of momentum states in double Bragg diffraction that we have derived are valid in a more general regime (e.\,g.\ for $\varepsilon$ close to one), but then they need to be solved by other means, for instance numerically.

\subsection{Outline of the article}

Our article is organized as follows. We start by reviewing the method of averaging in Sec.~\ref{sec:method of averaging}, where we introduce a notation which is adapted to the one that we use for the examples analyzed in the present article. In that section, the method is explained up to second order, whereas arbitrary orders are discussed in Appendix~\ref{sec:averaging arbitrary}.
This technical part is the basis for the following two applications.
In Sec.~\ref{sec:Bragg diffraction} we first recall the physical process of single atomic Bragg diffraction and then discuss the connection of the method of averaging to the ordinary adiabatic elimination \cite{bernhardt,stenholm,Brion07} and the quasi-Bragg regime \cite{quasi-Bragg}.
Next, we turn in Sec.~\ref{sec:doubleBragg} to the more sophisticated case of double Bragg diffraction, where the conventional adiabatic elimination is not possible and the method of averaging needs to be applied.
Various features of this scattering process are discussed using the insights gained by the method of averaging. In that section we restrict ourselves to the case of circularly polarized light waves and square pulses. A situation with linearly polarized lasers and an underlying magnetic field is discussed in Appendix~\ref{sec:lin-lin} and leads to the same results. Furthermore, the extension of our results to time-dependent laser pulses is briefly described in Appendix~\ref{sec:pulse shapes}. Finally, we conclude in Sec.~\ref{sec:conclusions} with a discussion of our results.

\section{Method of averaging}
\label{sec:method of averaging}
We now summarize the method of averaging first introduced by Ref.~\cite{bogoliubov}. Our notation is quite different from Ref.~\cite{bogoliubov} and already adapted to the scattering situations of Sec.~\ref{sec:Bragg diffraction} and Sec.~\ref{sec:doubleBragg}. Hence, the derivation and presentation of the formalism makes a physical interpretation easier than the rather mathematical presentation in Ref.~\cite{bogoliubov}.

This method is a procedure to approximately solve coupled differential equations with time-dependent coefficients that oscillate at different frequencies. Thus, it is \textit{the} method of choice for atomic Bragg diffraction problems, since they have exactly this form.

\subsection{General idea}
Let us assume a system of $n$ coupled differential equations of the form
\ba 
\dot{\vec{g}}= \I \varepsilon \mathcal{H}\, \vec{g} \equiv \I  \varepsilon \mathcal{H}_0 \,\vec{g}+ \I \varepsilon \sum\limits_{\nu\neq 0}\e{\I \nu \omega_\text{r}  t} \mathcal{H}_\nu \,\vec{g} \, , \label{eq: dgl allgem}
\ea
with $\vec{g} \in \mathbb{C}^n$, time-independent quantities $\mathcal{H}, \mathcal{H}_0, \mathcal{H}_\nu \in \mathbb{C}^{n\times n}$, with the dimensions of a frequency, a frequency $\omega_\text{r}$, and a dimensionless parameter $\varepsilon\ll1$. We refer to this parameter as the adiabaticity parameter. When we discuss single Bragg diffraction and double Bragg diffraction, the physical meaning and relevance of this parameter becomes clearer.
In the form of Eq.~\eqref{eq: dgl allgem}, the hierarchy of increasingly faster oscillating terms with frequencies $\nu \omega_\text{r}$ is manifestly laid out. We see later that $\nu$ is associated with a certain order of a Bragg transition.

We now separate the solution into slow and fast oscillations; i.\,e., we assume a function 
\ba
		\vec{g}^{(m)}=\vec{\gamma}^{(m)} \vphantom{\sum\limits_{j=1} } + \sum\limits_{j=1}^m \varepsilon^j \vec{f}_j(\vec{\gamma}^{(m)}) \, , \label{eq: def g^{(m)}}
\ea
which satisfies the differential equation
\ba 
\dot{\vec{g}}^{(m)}=\I \varepsilon \mathcal{H}\,
		\vec{g}^{(m)}+\mathcal{O}\left(\varepsilon^{m+1}\right) \, . \label{eq: dgl g^{(m)}}
\ea
The functions $\vec{f}_j=\vec{f}_j(t,\vec{\gamma}^{(m)})$ depend not only on the on the slowly evolving term $\vec{\gamma}^{(m)}$ but are explicitly time-dependent.

This scheme is not a perturbative treatment in the conventional way. In fact, the method of averaging finds a solution $\vec{g}^{(m)}$ that satisfies the original differential equation up the order $\varepsilon^m$.

At each order, we assume that the slowly evolving part $\vec{\gamma}^{(m)}$ fulfils the equation
\ba
\dot{\vec{\gamma}}^{(m)}= \I\varepsilon \mathcal{H}_0\,\vec{\gamma}^{(m)} + \I \summe{\mu}{2}{m} \varepsilon^\mu \vec{p}_\mu(\vec{\gamma}^{(m)}) \, . \label{eq: dgl gamma^{(m)}}
\ea
Using this ansatz together with differentiation of Eq.~\eqref{eq: def g^{(m)}}, one can equate the coefficients of different orders of $\varepsilon$ on the left side of Eq.~\eqref{eq: dgl g^{(m)}} to the right side. This way, one can determine the functions $\vec{f}_j$ and $\vec{p}_j$. We now derive these conditions.

Taking the time derivative of \eq{eq: def g^{(m)}} yields
\ba
\dot{\vec{g}}^{(m)}=& \dot{\vec{\gamma}}^{(m)}+ \summe{j}{1}{m} \varepsilon^j \left. \partdifffrac{\vec{f}_j}{t}\right|_{\vec{\gamma}^{(m)}}+ \summe{j}{1}{m}\varepsilon^j \partdifffrac{\vec{f}_j(\vec{\gamma}^{(m)})}{\vec{\gamma}^{(m)}}\dot{\vec{\gamma}}^{(m)} . \nonumber
\ea
When we use the ansatz \eq{eq: dgl gamma^{(m)}} for the time derivative of $\vec{\gamma}^{(m)}$, we find the equation
\begin{widetext}
\ba
\dot{\vec{g}}^{(m)}
=&\varepsilon \left(\I \mathcal{H}_0 \, \vec{\gamma}^{(m)}+ \left.\partdifffrac{\vec{f}_1
}{t}\right|_{\vec{\gamma}^{(m)}} \right) + \varepsilon^2 \left(\I\, \vec{p}_2(\vec{\gamma}^{(m)})+ \I \partdifffrac{\vec{f}_1(\vec{\gamma}^{(m)})}{\vec{\gamma}^{(m)}}  \mathcal{H}_0 \, \vec{\gamma}^{(m)}+\left.\partdifffrac{\vec{f}_2
}{t}\right|_{\vec{\gamma}^{(m)}}\right)\nonumber \\
 &+ \summe{j}{3}{m}\varepsilon^j \left(\I\, \vec{p}_j(\vec{\gamma}^{(m)})+  \I \partdifffrac{\vec{f}_{j-1}(\vec{\gamma}^{(m)})}{\vec{\gamma}^{(m)}}  \mathcal{H}_0 \, \vec{\gamma}^{(m)}+\left.\partdifffrac{\vec{f}_j
 }{t}\right|_{\vec{\gamma}^{(m)}}+\I \summe{\mu}{1}{j-2}\partdifffrac{\vec{f}_{\mu}(\vec{\gamma}^{(m)})}{\vec{\gamma}^{(m)}}\vec{p}_{j-\mu}(\vec{\gamma}^{(m)})\right) +\mathcal{O}\left(\varepsilon^{m+1}\right) \, . \nonumber
\ea
\end{widetext}
Now we compare this expression to the one on the right-hand side of \eq{eq: dgl g^{(m)}} using the ansatz \eq{eq: def g^{(m)}}, i.\,e.\ to
\ba
\I \varepsilon \mathcal{H}\,  \vec{g}^{(m)}+\mathcal{O}&\left(\varepsilon^{m+1}\right)= \I \varepsilon \mathcal{H}\,\vec{\gamma}^{(m)}+\I \varepsilon^2 \mathcal{H}\,\vec{f}_1(\vec{\gamma}^{(m)})\nonumber \\
& + \I \summe{j}{3}{m}\varepsilon^j \mathcal{H}\vec{f}_{j-1}(\vec{\gamma}^{(m)})+\mathcal{O}\left(\varepsilon^{m+1}\right) \,, \nonumber
\ea
we find conditional equations for $\vec{f}_j$. In this way, we can determine Eq.~\eqref{eq: def g^{(m)}}.

\subsection{First-order solutions}
Up to the order of $\varepsilon$, we find with the help of $\mathcal{H}= \mathcal{H}_0 +  \sum_{\nu\neq 0}\e{\I\nu \omega_\text{r}  t} \mathcal{H}_\nu $ the condition
\ba
\left.\partdifffrac{\vec{f}_1
}{t}\right|_{\vec{\gamma}^{(m)}}= \sum\limits_{\nu \neq 0} \I \e{\I \nu \omega_\text{r} t}\mathcal{H}_\nu\, \vec{\gamma}^{(m)}. \label{eq: dgl f_1}
\ea
We can integrate this equation easily, since due to the partial derivative with respect to time the function $\vec{\gamma}^{(m)}$ can be treated as a constant, and find
\ba
\vec{f}_1(\vec{\gamma}^{(m)})=\hspace{-.15cm}\int\hspace{-.05cm}\D t\sum\limits_{\nu \neq 0}\I \e{\I \nu\omega_\text{r}  t}\mathcal{H}_\nu \,\vec{\gamma}^{(m)} = \sum\limits_{\nu \neq 0} \frac{\e{\I \nu\omega_\text{r}  t}}{\nu \omega_\text{r} }\mathcal{H}_\nu\,\vec{\gamma}^{(m)}. \label{eq: first correction}
\ea
When this integration is performed, the initial condition yields constants of integration, which may be functions of $\vec{\gamma}^{(m)}$ but are not explicitly time-dependent.
Thus, it would just yield a slowly evolving term.
In our formulation we set this constant equal to zero, since we want a clear separation of timescales.
This approach is valid if we ensure that the initial conditions are fulfilled, as we do later in this section. 

In the integration we see the link to the conventional adiabatic elimination:
Just the rapidly oscillating terms are integrated, the slowly evolving term $\vec{\gamma}^{(m)}$ is constant in time. This procedure is very close to the adiabatic elimination discussed in Sec.~\ref{sec:singe 3-recursion} in more detail.

For $m=1$, i.\,e.\ up to the first approximation, Eq.~\eqref{eq: dgl gamma^{(m)}} reads
\ba
\dot{\vec{\gamma}}^{(1)}= \I \varepsilon\mathcal{H}_0 \vec{\gamma}^{(1)}  \label{eq:dgl gamma1}
\ea
with a time-independent $\mathcal{H}_0$. Hence, we find the slowly evolving solution
\ba
\vec{\gamma}^{(1)}(t)= \ehoch{\I\varepsilon \mathcal{H}_0 t}\;\vec{\gamma}^{(1)}(0)\,, \label{eq:gamma(1)allgem}
\ea
where we have defined the matrix exponential $\ehoch{\mathcal{A}}\equiv \sum_{n=0}\mathcal{A}^n/n!$ for square matrices $\mathcal{A}$.

With \eq{eq: first correction}, we can calculate rapidly oscillating corrections to this slow solution. The full expression reads for $m=1$ with \eq{eq: def g^{(m)}}
\ba
	\vec{g}^{(1)}(t)=\left[\mathds{1}+\varepsilon\sum\limits_{\nu \neq 0} \frac{\e{\I \nu \omega_\text{r} t}}{ \nu\omega_\text{r}}\mathcal{H}_\nu\right] \vec{\gamma}^{(1)}(t) \,,\label{eq: g^(1) allgemein}
\ea
where we used the identity $\mathds{1}\equiv \delta_{n,n'}$.
Here, we see that the fast terms are suppressed with $\varepsilon\ll 1$.

To ensure that the initial conditions are fulfilled, we assume
\ba
	\vec{\gamma}^{(1)}(0)=\left[\mathds{1}+\varepsilon \sum\limits_{\nu \neq 0}\frac{\mathcal{H}_\nu}{ \nu \omega_\text{r}}\right]^{-1}\hspace{-.2cm}\vec{g}^{(1)}(0)\,. \label{eq:initial_cond}
\ea
In the following, we refer to choosing the initial condition $\vec{\gamma}^{(1)}(0)$ as the \textit{dressed-state} formulation. If we choose $\vec{g}^{(1)}(0)$ as an initial condition, we call this the \textit{bare-state} formulation. We do not discuss the interpretation of initial condition here in more detail but refer to Sec.~\ref{sec:dressed_states}, where we examine this point using the example of single Bragg diffraction.

\subsection{Second-order solutions}
To get solutions for $m\geq 1$, the next step is to compare the coefficients to the order of $\varepsilon^2$. With the solution \eq{eq: first correction} for $\vec{f}_1$ we find the differential equation
\ba
\partdifffrac{\vec{f}_2(\vec{\gamma}^{(m)})}{t}\hspace{-.075cm}=\I \hspace{-.075cm}\sum\limits_{\nu \neq 0}\hspace{-.075cm}\frac{\e{\I \nu \omega_\text{r} t} } { \nu\omega_\text{r}} \left[\mathcal{H}\mathcal{H}_\nu -\mathcal{H}_\nu\mathcal{H}_0\right] \vec{\gamma}^{(m)} \hspace{-.075cm}-\hspace{-.075cm} \I\, \vec{p}_2(\vec{\gamma}^{(m)}) .\nonumber
\ea
When defining 
\ba
\Phi_\mu^{(2)}\equiv\sum_{\nu \neq 0}\frac{\mathcal{H}_{\mu-\nu}\mathcal{H}_\nu}{\nu \omega_\text{r} }-\frac{\mathcal{H}_\mu \mathcal{H}_0}{\mu\omega_\text{r} }
\ea
for $\mu \neq 0$ and 
\ba
\Phi_0^{(2)}\equiv\sum_{\nu \neq 0}\frac{\mathcal{H}_{-\nu}\mathcal{H}_\nu}{\nu \omega_\text{r} }\, , \nonumber
\ea
this differential equation takes the form
\ba
\partdifffrac{\vec{f}_2(\vec{\gamma}^{(m)})}{t}= \I\sum\limits_{\mu}\e{\I\mu  \omega_\text{r} t} \Phi_\mu^{(2)}\, \vec{\gamma}^{(m)} -\I\, \vec{p}_2(\vec{\gamma}^{(m)})\,, \label{eq:dgl f_2}
\ea
where we have used again ${\mathcal{H}= \mathcal{H}_0 +  \sum_{\nu\neq 0}\e{\I \nu \omega_\text{r} t} \mathcal{H}_\nu} $.

We emphasize that in the sum in Eq.~\eqref{eq:dgl f_2} the term $\mu=0$ is included, i.\,e.\ a time-independent term appears where the phase factor is equal to unity. On the other side, the function $\vec{p}_2$, describing a slowly evolving term as one can see from Eq.~\eqref{eq: dgl gamma^{(m)}}, appears in the differential equation and thus has to be also independent of time. So for $\vec{p}_2(\vec{\gamma}^{(m)})\equiv \Phi_0^{(2)}\, \vec{\gamma}^{(m)}$, Eq.~\eqref{eq:dgl f_2} takes exactly the form of Eq.~\eqref{eq: dgl f_1}, namely
\ba
\partdifffrac{\vec{f}_2(\vec{\gamma}^{(m)})}{t}= \I \sum\limits_{\mu \neq 0}\e{\I \mu \omega_\text{r} t} \Phi_\mu^{(2)}\, \vec{\gamma}^{(m)}\,, \nonumber 
\ea
where just rapidly oscillating terms occur. This approach leads, in complete analogy to the previous section, to
\ba
\vec{f}_2(\vec{\gamma}^{(m)})= \sum\limits_{\mu \neq 0}\frac{\e{\I \mu \omega_\text{r} t}}{\mu \omega_\text{r}} \Phi_\mu^{(2)}\, \vec{\gamma}^{(m)} \,, \nonumber
\ea
where again the constant of integration was set to zero.
With this choice we get for $m=2$ from Eq.~\eqref{eq: dgl gamma^{(m)}} the differential equation
\ba 
\dot{\vec{\gamma}}^{(2)}=&\I \varepsilon \mathcal{H}_0 \vec{\gamma}^{(2)}+\I \varepsilon^2  \Phi_0^{(2)} \vec{\gamma}^{(2)}\, , \nonumber
\ea
where we have time-independent coefficients and hence find 
\ba
\vec{\gamma}^{(2)}(t)=&\ehoch{\I\varepsilon \mathcal{H}_0 t+ \I\varepsilon^2 \Phi_0^{(2)} t}\;\vec{\gamma}_2(0) \nonumber\\
	=&\ehoch{\I \varepsilon \left(\mathcal{H}_0+ \varepsilon \sum_{\nu \neq 0}\frac{\mathcal{H}_{-\nu}\mathcal{H}_\nu}{\nu \omega_\text{r}} \right) t}\;\vec{\gamma}^{(2)}(0)\, . \label{eq: gamma_2}
\ea
Since we have already determined $\vec{f}_2$, we could also obtain the rapidly oscillating corrections. Because we discuss in the following sections the second approximation without fast corrections, we refrain from presenting this cumbersome expression.

The procedure described in this section can be extended step by step to arbitrary orders of $\varepsilon$, as we show in Appendix~\ref{sec:averaging arbitrary}.

\section{Bragg diffraction}
\label{sec:Bragg diffraction}
The method of averaging presented so far can be applied to a large class of problems. In particular, it gives us the opportunity to gain more insight into Bragg diffraction processes.
Even though it is necessary for the double-diffraction case, one can also apply it to single diffraction.
In this approach, the distinction between the deep Bragg regime and the so-called quasi-Bragg regime as investigated in Ref.~\cite{quasi-Bragg} comes out clearly. Moreover, the methods and approximations used in this chapter can be generalized to the much more complex double Bragg diffraction case.
As an introduction we now discuss the familiar Bragg diffraction from this point of view.

\subsection{Model}
\begin{figure}[htb]
\centering
\includegraphics[scale=1]{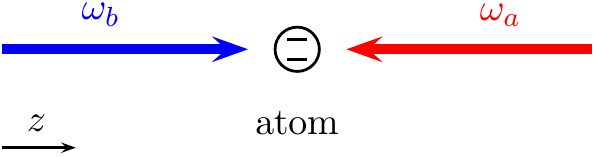}
\caption{Schematic set-up of Bragg diffraction. A two-level atom is interacting with two counterpropagating light fields of frequency $\omega_a$ and $\omega_b$, which are far detuned from the atomic transition.}
\label{fig: schematic setup single bragg}
\end{figure}

We assume a two-level atom with an energy separation $\hbar \omega_{eg}$ interacting with two counterpropagating light waves, one with frequency $\omega_a$, and the other with $\omega_b$. Both light fields are aligned parallel to the $z$-direction. This set-up is depicted in Fig.~\ref{fig: schematic setup single bragg}. 

Thus, the atom is interacting with the electric field
\ba
 \hat{\vec{\mathcal{E}}}=  \vec{E}_b \e{\I (k_b \hat{z}-\omega_bt)} +\vec{E}_a \e{\I (-k_a \hat{z}-\omega_at)}+~ \text{h.c.} \, ,\label{eq: el field single}
\ea
with the amplitudes $\vec{E}_{a}$ and $\vec{E}_b$ and the absolute values $k_a$ and $k_b$ of the wave vectors. The sign in front of $k_a$ in the second exponential reflects the fact that the light is travelling in the opposite direction.

\begin{figure}[htb]
\centering
\includegraphics[scale=1]{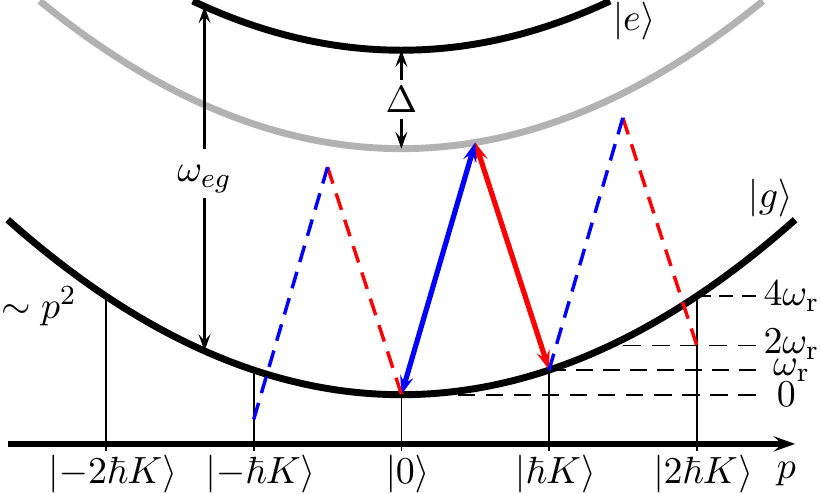}
\caption{Momentum transfer and resonances in Bragg diffraction. The absorption of a photon from one field and the emission of a photon into the field travelling in the opposite direction results in a recoil of the value $\hbar K \equiv \hbar (k_b+k_a)$. This process is resonant, if the difference between the two lasers $\Delta \omega= \omega_b-\omega_a$ is equal to the change of kinetic energy, which is proportional to $p^2$. The dashed transitions are off-resonant and suppressed; the gray line denotes the sum of potential and kinetic energy of the virtual excited state. $\omega_\text{r} =\hbar K^2/(2M)$ denotes the recoil frequency.}
\label{fig: parabula single}
\end{figure}

The field operator $\hat{\vec{\mathcal{E}}}$ describes two classical fields, but accounts for the mechanical action of the light on the atom, i.\,e.\ the recoil. Indeed, the operator 
\ba
\e{\pm \I k \hat{z}}= \int\D p \ket{p\pm\hbar k}\bra{p} \nonumber
\ea
shifts the momentum by $\pm \hbar k$.
Hence, we can change the momentum of the atom by applying these lasers.
This can be understood in terms of momentum conservation: The atom in the ground state $\ket{g}$ absorbs one photon of momentum $\hbar k_b$.
The atom is now in the excited state $\ket{e}$ and has picked up the recoil $\hbar k_b$. An emission of a photon with momentum $-\hbar k_a$ may be stimulated by the other laser and lead to a total transfer of the atomic momentum by $\hbar K \equiv \hbar (k_b+k_a)$, where $K$ is called the effective wave vector of the Bragg pulse.
This process is shown in Fig.~\ref{fig: parabula single}.

Since in typical Bragg diffraction a superposition of different momenta in the ground state is created, the population of the excited state needs to be virtual.
As a result, both light fields have to be far detuned from the atomic resonance $\omega_{eg}$. The detuning is defined as $\Delta\equiv \omega_{eg}-\omega_b\approx \omega_{eg}-\omega_a $, as shown in Fig.~\ref{fig: parabula single}. This condition is essential for the adiabatic elimination of the excited state that is performed in the next section and leads to a two-photon process.

Figure~\ref{fig: parabula single} shows an additional feature of the Bragg scattering process.
Only if the difference $\Delta \omega\equiv \omega_b-\omega_a$ of the two frequencies of the light waves is equal to the difference of the kinetic energy gained by the momentum transfer, this process is resonant, i.\,e.\ each resonant process has to start and end on the kinetic parabola $p^2/(2 M)$, where $M$ denotes the mass of the atom and $p$ the momentum.
Other processes are of higher order and thus suppressed.
For example, the dashed lines in the figure represent such diffraction orders.
Usually, all off-resonant momentum states can be adiabatically eliminated as well. We now demonstrate that the method of averaging provides a convenient way to calculate the populations of these levels.

The resonance condition can be written in a general form: For the $j$-th resonance, we have $j$ two-photon processes, i.\,e.\ the gain of energy due to the lasers is $j  \Delta \omega$. If this energy coincides with the difference of the kinetic energy caused by the momentum transfer $j \hbar K$, the process is on resonance. This condition reads
\ba
\Delta \omega = \frac{1}{j}\frac{\hbar (j K)^2}{2M}= j \frac{\hbar K^2}{2M}\equiv j \omega_\text{r}\, , \label{eq:res_cond}
\ea
where $\omega_\text{r}$ is called the recoil frequency.

\subsection{Three-term recurrence relation in adiabatic approximation}
\label{sec:singe 3-recursion}
In this section, we derive a three-term recurrence relation for the momentum distribution of atoms in the ground state. For this, we perform the conventional adiabatic approximation, as e.\,g.\ explained in Refs.~\cite{bernhardt,stenholm}.
We use in rotating wave approximation, e.\,g.\ see Ref.~\cite{schleich}, the Hamiltonian
\ba
	\hat{H}&= \frac{\hat{p}^2}{2M}+ \hbar \omega_{eg} \ket{e}\bra{e}  \nonumber \\
		 &+\hbar\left\lbrace  \left[\Omega_b \e{\I (k_b \hat{z}-\omega_b t)} +\Omega_a\e{\I (-k_a \hat{z}-\omega_a t)}\right] \ket{e}\bra{g}+~\text{h.c.} \right\rbrace , \nonumber
\ea
where we have assumed a dipole interaction of the atomic dipole moment $\hat{\vec{d}}$ with the electric field $\hat{\vec{\mathcal{E}}}$ determined by \eq{eq: el field single}. The Rabi frequencies are defined as ${\Omega_j \equiv-\bra{e}\hat{\vec{d}}\cdot\vec{E}_j\ket{g}/\hbar}$ and the atom has the mass $M$.

We emphasize that we consider throughout this article a regime where we have a large detuning and the effects of spontaneous emission can be neglected. An approach to Bragg diffraction taking the effects of recoil due to spontaneous emission into account can be found e.\,g.\ in Refs.~\cite{Jex,chudesnikovBragg,YakovlevBook}.

We now allow this Hamiltonian to act on an arbitrary state
\ba
  \ket{\psi}\equiv \int \D p \left[ g(p)\ket{g,p}+ e(p) \ket{e,p} \right] \, ,\nonumber
\ea
apply the Schr\"odinger equation 
\ba
\I \hbar\frac{\D}{\D t}\ket{\psi}= \hat{H}\ket{\psi} \label{eq:schroedinger}
\ea
to equate the coefficients in front of $\ket{g,p}$ and $\ket{e,p}$, and find the coupled differential equations
\ba
\I \dot{e}(p)=&\Omega_b \e{\I(\omega_{eg}-\omega_b)t} \e{-\I(\omega_{-k_b}+\nu_{-k_b})t}g(p-\hbar k_b)\nonumber \\
&+\Omega_a \e{\I(\omega_{eg}-\omega_a)t} \e{-\I (\omega_{k_a}+\nu_{k_a})t}g(p+\hbar k_a) \label{eq: single e(p)} \\
\I \dot{g}(p)=&\Omega_a^* \e{-\I(\omega_{eg}-\omega_a)t} \e{-\I (\omega_{-k_a}+\nu_{-k_a})t}e(p-\hbar k_a)\nonumber \\
&+\,\Omega_b^* \e{-\I(\omega_{eg}-\omega_b)t} \e{-\I (\omega_{k_b}+\nu_{k_b})t}e(p+\hbar k_b)\, . \label{eq: single g(p)}
\ea
Here, we have already transformed into the interaction picture in which $e(p)$ has to be multiplied by $\ehoch{\I (p^2/(2M\hbar)+\omega_{eg})t}$ and $g(p)$ by $\ehoch{\I p^2/(2M\hbar)t}$. In the equations above, we have introduced the frequency $\omega_k\equiv\hbar k^2/(2M)$, which corresponds to the kinetic energy of an atom with momentum $\hbar k$ and the frequency $\nu_k\equiv pk/M$, which accounts for the Doppler effect.

In the following, we simplify this set of differential equations by eliminating the excited state adiabatically. We assume the change of the population amplitude of the ground state $g(p)$ and the frequencies $\omega_k+\nu_k$ to be much smaller than the detuning ${\Delta=\omega_{eg}-\omega_{a,b}}$ and thus find by only integrating the exponentials $\ehoch{\I(\omega_{eg}-\omega_{a,b})t}$ in \eq{eq: single e(p)} the approximate expression 
\ba
e(p)\cong&-\frac{\Omega_b}{\Delta}\e{\I(\omega_{eg}-\omega_b)t}\e{-\I  (\omega_{-k_b}+\nu_{-k_b})t}g(p-\hbar k_b)\nonumber \\
&-\frac{\Omega_a}{\Delta} \e{\I(\omega_{eg}-\omega_a)t} \e{-\I  (\omega_{k_a}+\nu_{k_a})t}g(p+\hbar k_a) \label{eq:adiabat.e(p)single}
\ea
for the excited state.
Of course, this approximation is only valid for $\Omega_{a,b}\ll \Delta$.

At this point, we see the connection of the method of averaging to the adiabatic elimination:
In \eq{eq: first correction} we also perform an integration assuming that the slow solution $\vec{\gamma}^{(m)}$ is constant in time in comparison to the oscillating terms.
In general, the adiabatic elimination is a delicate interplay of fast and slow variables.
The interpretation in this case is simple:
If the detuning is high, the Rabi oscillations from $\ket{g}$ to $\ket{e}$ are very fast and the population of the excited state is suppressed by $\Omega_{a,b}/\Delta$. In this case the fraction $\Omega_{a,b}/\Delta$ corresponds to the adiabaticity parameter. We assume it to be so small that a further treatment through the method of averaging is not necessary.

We can now use the solution Eq.~\eqref{eq:adiabat.e(p)single} and substitute it into \eq{eq: single g(p)} to find the effective equation
\ba
\I \dot{g}(p)=&-\frac{\Omega_a\Omega_b^*}{\Delta} \e{\I \Delta \omega t} \e{-\I  (\omega_{K}+\nu_{K})t} g(p+\hbar K)\nonumber \\
&-\frac{\Omega_a^*\Omega_b}{\Delta} \e{-\I \Delta \omega t} \e{-\I (\omega_{-K}+\nu_{-K})t} g(p-\hbar K) \nonumber \\
=&- \Omega \; \e{\I (\Delta \omega - \omega_r)t} \e{-\I \nu_\text{D} t}g(p+\hbar K)\nonumber \\
&-\Omega^* \;\e{-\I (\Delta \omega + \omega_r)t} \e{\I \nu_\text{D} t}g(p-\hbar K), \label{eq: single dgl g(p) lang}
\ea
where we have defined the effective frequency $\Omega\equiv\Omega_a \Omega_b^*/\Delta$, the difference $\Delta \omega\equiv \omega_b-\omega_a$ of the laser frequencies, and the effective wave vector $K= k_b+k_a$. The recoil frequency is defined as
\ba
\omega_\text{r}\equiv \frac{\hbar K^2}{2M} \label{eq:rec freq}
\ea
and the Doppler frequency as
\ba
\nu_\text{D}\equiv \frac{p K}{M}\, . \label{eq:doppler freq}
\ea
The coupling to the same momentum has been removed by multiplying $g(p)$ with $\ehoch{\I(|\Omega_a|^2+|\Omega_b|^2)t/\Delta}$.

At this point we want to emphasize that we have defined our effective frequency $\Omega$ in such a way that it is the frequency of the probability amplitude, rather than the Rabi frequency $\Omega_\text{Rabi}$ at which the population of a state oscillates. To link our definition to the latter--maybe more familiar--definition, we make use of the relation
\ba
\Omega \equiv \frac{\Omega_\text{Rabi}}{2} \nonumber\, .
\ea
Indeed, a $\pi/2$ pulse, i.\,e.\ ${\Omega_\text{Rabi}\, t = \pi/2}$, is in our description $\Omega\, t= \pi/4$. On the other side, when we face double diffraction, we are not dealing with a two-level system anymore, but with three levels. Nevertheless, effective Rabi oscillations occur and with our definition of the frequency we find here a more intuitive connection.

The value of $\Delta \omega$ now determines the resonances of the Bragg diffraction process.
According to \eq{eq:res_cond}, if we set $\Delta \omega=j\omega_\text{r}$, the transition from $\ket{0}$ to $\ket{j \hbar K}$ is on resonance, i.\,e.\ the time-dependent phase factors vanish for these states as we see in Eq.~\eqref{eq: single dgl g(p) lang}.

We now focus on the first-order Bragg diffraction process, which is the process shown in Fig.~\ref{fig: parabula single}. For that, we set $\Delta \omega=\omega_\text{r}$ and assume $\Omega$ to be real. The last assumption is not necessary, but simplifies the notation and calculation. To keep track of the laser phases, one can associate with $\Delta \omega$ the respective difference of laser phases.

With these assumptions we thus get from \eq{eq: single dgl g(p) lang} the system of coupled differential equations
 \ba
\dot{g}(p+n\hbar K)=& \I \Omega \e{-\I 2 n \omega_\text{r}t} \e{-\I \nu_\text{D}t}g(p+\hbar (n+1)K) \nonumber \\
 &+\I \Omega \e{\I 2 (n-1) \omega_\text{r}t} \e{\I \nu_\text{D} t}g(p+\hbar (n-1)K) \label{eq:3term recurrence single}
\ea
for $n\in \mathbb{Z}$, which corresponds to a time-dependent three-term recurrence relation.

\subsection{Application of the method of averaging}
\label{sec:single-appl-averaging}
In order to cast this equation into the form of \eq{eq: dgl allgem}, we first have to introduce the dimensionless adiabaticity parameter $\varepsilon$. Equation~\eqref{eq:3term recurrence single} implies that there are two timescales: The Rabi frequency $\Omega$ and the recoil frequency $\omega_\text{r}$. In the spirit of the rotating wave or the adiabatic approximation, we assume the coupling to higher momentum states, i.\,e.\ to higher $n$, to be suppressed due to fast oscillating terms. This condition implies that the recoil frequency has to be large in comparison with the Rabi frequency, and hence we define the adiabaticity parameter
\ba
\varepsilon \equiv \frac{\Omega}{\omega_\text{r}} \label{eq:adiabaticity}
\ea
as the comparison of these two timescales.

 In the Bragg regime, the population of higher momentum states is suppressed and hence we find $\Omega \ll \omega_\text{r}$, or $\varepsilon\ll1$, and the method of averaging can be applied. Thus, the parameter $\varepsilon$ describes the adiabaticity of the diffraction process.

We now define $g(p+\hbar n K)\equiv g_n$ as components of a vector and arrive at
\ba
\dot{\vec{g}}= \I \varepsilon \left(\mathcal{H}_0+ \sum\limits_{\nu\neq 0}\e{\I \nu \omega_\text{r}  t} \mathcal{H}_\nu \right)\vec{g}\, , \nonumber
\ea
which has the form of Eq.~\eqref{eq: dgl allgem} and where the matrices are defined as
\ba 
 \left(\mathcal{H}_\nu\right)_{n,n'}=\omega_\text{r} &\left(\e{-\I \nu_\text{D} t} \delta_{n+1,n'}\delta_{2n,-\nu}  \right. \nonumber \\
 &\left.+ \e{\I \nu_\text{D} t} \delta_{n-1,n'} \delta_{2(n-1),\nu} \right)\, ,\nonumber
\ea
with the Kronecker delta $\delta_{m,n}$. 
The coupling strength in Eq.~\eqref{eq:3term recurrence single} was $\Omega$ and it still is, but in order to have a dimensionless expansion parameter we have introduced $\Omega\equiv\varepsilon \,\omega_\text{r}$, which is why the Hamilton matrix has the dimensions of a frequency.

For the moment, we assume $p=0$, i.\,e.\ $\nu_\text{D}=0$. In this case, $\mathcal{H}_\nu$ becomes time-independent, namely
\ba 
 \left(\mathcal{H}_\nu\right)_{n,n'}=&\omega_\text{r} \left(\delta_{n+1,n'}\delta_{2n,-\nu} +\delta_{n-1,n'}\delta_{2(n-1),\nu}\right)\, , \label{eq: single def H_nu}
\ea
and the differential equation is exactly of the form of Eq.~\eqref{eq: dgl allgem}. However, this equation just describes the time evolution of $g(\hbar n K)\equiv g_n$ for the resonant momenta. We discuss the role of a deviation from resonance in the context of double Bragg diffraction in Sec.~\ref{sec:velocity selectivity}.

\subsubsection{Rabi oscillations between momentum states}
We now apply the method of averaging described in Sec.~\ref{sec:method of averaging} to find a slow solution $\vec{\gamma}$ with the matrices $\mathcal{H}_\nu$ from Eq.~\eqref{eq: single def H_nu}.
According to Eq.~\eqref{eq:dgl gamma1}, the differential equation for the first approximation ($m=1$) of the slowly oscillating terms is with $\vec{\gamma}^{(1)}= (\gamma_0^{(1)},\gamma_1^{(1)})^\text{T}$, where the superscript $\text{T}$ denotes the transpose,
\ba
\dot{\vec{\gamma}}^{(1)}=\I \varepsilon \mathcal{H}_0 \vec{\gamma}^{(1)}= \I \Omega \begin{pmatrix}
 0&1\\
 1&0
\end{pmatrix} \vec{\gamma}^{(1)}\,. \nonumber 
\ea
Taking more than two states into account is possible, but since we chose $\Delta \omega= \omega_\text{r}$, just the two neighboring states are resonant.

Solving this differential equation by taking the matrix exponential as in Eq.~\eqref{eq:gamma(1)allgem}, we find
\ba
\vec{\gamma}^{(1)}(t)= \left[ \cos \left(\Omega t \right)\begin{pmatrix}
 1&0\\
 0&1
\end{pmatrix}  + \I \sin \left(\Omega t \right) \begin{pmatrix}
 0&1\\
 1&0
\end{pmatrix} \right]  \vec{\gamma}^{(1)}(0)\,. \label{eq: Rabi single Bragg}
\ea
This result contains the well-known Rabi oscillations \cite{bernhardt} between the states $\ket{0}$ and $\ket{\hbar K}$ with the effective Rabi frequency $\Omega$.
With this result, we see how Bragg pulses can be used to create superpositions of different momentum states. A beam splitter, or $\pi/2$ pulse, is achieved for $\Omega\, t= \pi/4$, a mirror, or $\pi$ pulse, for $\Omega\, t= \pi/2$.
\vspace{12pt}

\subsubsection{The quasi-Bragg regime}

These Rabi oscillations have been derived many times before, see e.\,g.\ \cite{bernhardt,stenholm,Brion07}, by adiabatically eliminating all momentum states but $\ket{0}$ and $\ket{\hbar K}$.
We see now how the method of averaging corresponds to the technique of adiabatic elimination.
Of course this is an approximation,
but our method allows us to easily derive corrections to these Rabi oscillations, and thus to calculate populations of the levels that would have been eliminated if we had pursued the conventional approach.

We note that this solution is completely independent of the adiabaticity parameter $\varepsilon$. Hence, the slowly evolving solution $\vec{\gamma}^{(1)}$ is correct to lowest order of $\varepsilon$. We call the regime where this description is sufficient the \textit{deep Bragg regime}. If we want to calculate corrections up to the next higher order, which now leads into the \textit{quasi-Bragg regime} explored by Ref.~\cite{quasi-Bragg}, we can use Eq.~\eqref{eq: g^(1) allgemein} to get
\begin{widetext}
\ba
	\vec{g}^{(1)}(t)&=\left[\mathds{1}+\varepsilon\sum\limits_{\nu \neq 0} \frac{\e{\I\nu \omega_\text{r}  t}}{ \nu \omega_{r}}\mathcal{H}_\nu\right]\;\vec{\gamma}^{(1)} (t)\nonumber \\
	=&\left[  
	\begin{pmatrix}
 1& \frac{\varepsilon }{2} \e{\I 2  \omega_\text{r}t}\cos \left(\Omega t \right)&  \I \frac{\varepsilon }{2}\e{\I 2  \omega_\text{r}t}\sin \left(\Omega t \right) & 0\\
- \frac{\varepsilon}{2}\e{-\I 2  \omega_\text{r}t}& 0 & 0 & 0\\
 0& 0 &  0& - \frac{\varepsilon}{2}\e{-\I 2  \omega_\text{r}t}\\ 
  0 &  \I \frac{\varepsilon}{2}\e{\I 2  \omega_\text{r}t}\sin \left(\Omega t \right) &\frac{\varepsilon}{2}\e{\I 2  \omega_\text{r}t} \cos \left(\Omega t \right)  & 1
\end{pmatrix} + \cos \left(\Omega t \right) \begin{pmatrix}
 1&0\\
 0&1
\end{pmatrix}  + \I \sin \left(\Omega t \right) \begin{pmatrix}
 0&1\\
 1&0
\end{pmatrix}  \right]\vec{\gamma}^{(1)}(0)  \,.\label{eq:single Rabi corrections}
\ea
\end{widetext}
To be able to calculate these correction terms and the population of the states $\ket{- \hbar K}$ and $\ket{2 \hbar  K}$, we have enlarged the matrices to $4\times 4$ and considered the vector ${\vec{g}^{(1)}=(g_{-1}^{(1)},g_0^{(1)},g_1^{(1)},g_2^{(1)})^\text{T}}$.
This correction does not change the Rabi oscillations between $\ket{0}$ and $\ket{\hbar K}$ but creates populations in the coefficients $g_{-1}^{(1)}$ and $g_2^{(1)}$.
The population of these states is suppressed by a factor $\varepsilon$ and thus not relevant in the deep Bragg regime.

In Eq.~\eqref{eq:single Rabi corrections}, the coefficients oscillate with a frequency of $2 \omega_\text{r}$ which is identical to the frequency that a second-order process deviates from the resonant kinetic energy, as shown in Fig.~\ref{fig: parabula single}.
For processes of this kind, energy conservation is not ensured, if such a dashed off-resonant transition is made, but due to the energy-time uncertainty this is possible.
\begin{figure}[htb]

\includegraphics{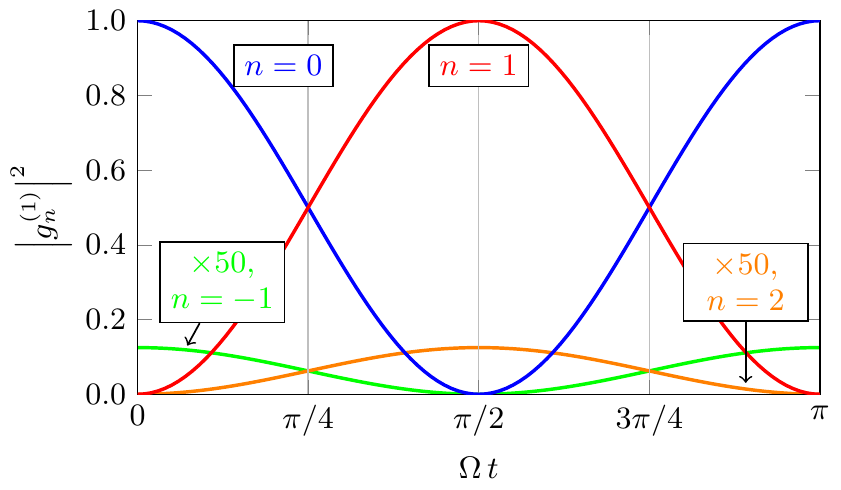}
 \caption{Populations $|g_n^{(1)}|^2$ of momentum states for single Bragg diffraction in the quasi-Bragg regime. The initial condition ${\vec{\gamma}^{(1)}(0)=(0,1,0,0)^\text{T}}$ corresponds to the dressed-state formulation, and the adiabaticity parameter is $\varepsilon=0.1$. The two bottom lines are magnified by a factor of 50 and show the small populations in the off-resonant states $n=-1,2$.}
 \label{fig:single_oscis}
\end{figure}

The solutions Eq.~\eqref{eq:single Rabi corrections} are depicted in Fig.~\ref{fig:single_oscis} for the initial condition ${\vec{\gamma}^{(1)}(0)=(0,1,0,0)^\text{T}}$.
The two bottom lines show the small-amplitude off-resonant states ${n=-1,2}$ magnified by a factor of 50.
The off-resonant state $n=-1$ behaves exactly like its adjacent resonant state with $n=0$, but with a suppressed amplitude. The same is true for the other two states. Since the fast oscillations of the off-resonant terms are contained in a phase factor $\ehoch{\pm \I 2 \omega_\text{r}t}$, we do not see them in the corresponding population, given by the modulus square. In Sec.~\ref{sec:doubleBragg} we show that in double Bragg diffraction effects from these fast oscillations do occur even for the resonant states.

For Fig.~\ref{fig:single_oscis}, we have chosen an initial condition for $\vec{\gamma}^{(1)}(0)$ and are thus in a dressed-state formulation instead of choosing $\vec{g}^{(1)}(0)$ and defining $\vec{\gamma}^{(1)}(0)$ according to \eq{eq:initial_cond} in the bare-state formulation.
Therefore, $\vec{\gamma}^{(1)}(0)$ cannot be interpreted as the initial momentum distribution. 
For this reason, the initial conditions for $|g_n^{(1)}|^2$ in Fig.~\ref{fig:single_oscis} do not coincide with the intuitive ones. We discuss in Sec.~\ref{sec:dressed_states} the bare-state formulation.

The discussion of this section predicts features of the diffraction process when leaving the deep Bragg regime and serves as an application of the method of averaging. However, for the case of double Bragg diffraction, this method is necessary. We focus on this point in Sec.~\ref{sec:doubleBragg}.

\subsection{Dressed versus bare states}
\label{sec:dressed_states}
\begin{figure}[htb]

\includegraphics{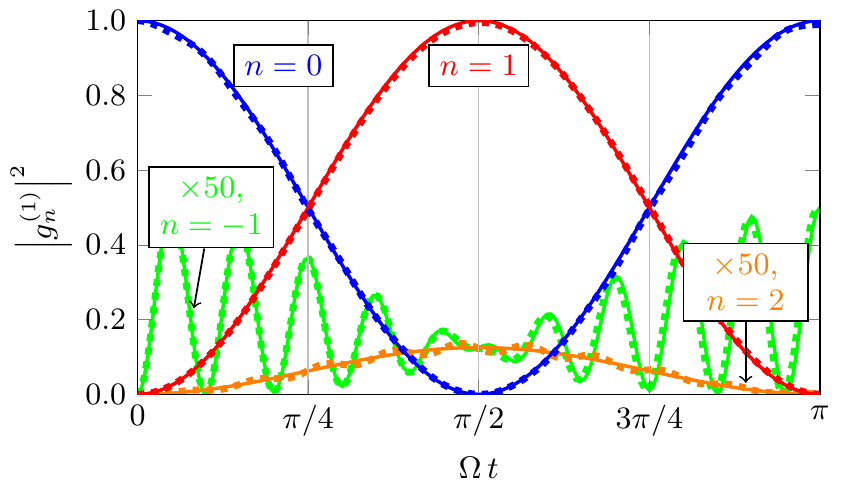}
 \caption{Populations $|g_n^{(1)}|^2$ of momentum states for single Bragg diffraction in the quasi-Bragg regime. In contrast to Fig.~\ref{fig:single_oscis} now the initial condition is ${\vec{g}^{(1)}(0)=(0,1,0,0)^\text{T}}$, which corresponds to the bare-state formulation. The adiabaticity parameter is again $\varepsilon=0.1$. The two bottom lines are magnified by a factor of 50 and show the small populations in the off-resonant states $n=-1,2$. The dotted lines represent numerical results.}
 \label{fig:single_bare}
\end{figure}

In the previous section, we have discussed the dressed-state formulation of single Bragg diffraction in the quasi-Bragg regime. We have used Eq.~\eqref{eq:single Rabi corrections} and displayed the time evolution of the dressed state, i.\,e.\ $\vec{\gamma}^{(1)}(0)= (0,1,0,0)^\text{T}$, in Fig.~\ref{fig:single_oscis}. For square-shaped light pulses, the dressed state might seem unphysical. So in Fig.~\ref{fig:single_bare} we depict the dynamics of a bare state with the initial condition $\vec{g}^{(1)}(0)= (0,1,0,0)^\text{T}$. This initial condition translates with Eq.~\eqref{eq:initial_cond} into a sophisticated superposition of initial states $\vec{\gamma}^{(1)}(0)$. In Fig.~\ref{fig:single_bare} we just display terms where the amplitudes scale up to the order of $\varepsilon$ to be consistent with the first-order treatment.

We note that the population of the state $\ket{-\hbar K}$ now oscillates on a fast timescale with the frequency $2\omega_\text{r}$. The populations of the other states do not change significantly. When compared to the dotted numerical results, small-amplitude deviations from the analytical solutions appear. To find these analytically, higher orders of $\varepsilon$ would have to be taken into account. 

For square-shaped Bragg pulses, for which the treatment based on the method of averaging is very well-suited, the bare-state formulation is easier to interpret since the initial condition corresponds to the physical situation. Nevertheless, one has to keep in mind that most applications involve adiabatically turned-on light waves. In this case, the adiabaticity parameter $\varepsilon$ becomes time-dependent, as discussed in Appendix~\ref{sec:pulse shapes}. In Eq.~\eqref{eq:initial_cond} this translates into $\varepsilon(0)=0$, and thus the dressed-state corresponds to the bare-state formulation.

\section{Double Bragg diffraction}
\label{sec:doubleBragg}
To introduce symmetric Bragg pulses analogously to the symmetric Raman pulses in the double-diffraction scheme by Ref.~\cite{Leveque}, we now extend our model to an interaction with four light waves. The method of averaging outlined in Sec.~\ref{sec:method of averaging} and applied in Sec.~\ref{sec:Bragg diffraction} is needed to find a theoretical description of this process, since the elimination of nonresonant momentum states is much more subtle.

\subsection{Model}
\label{sec:double-model}
\begin{figure}[htb]
\centering
\includegraphics[scale=1]{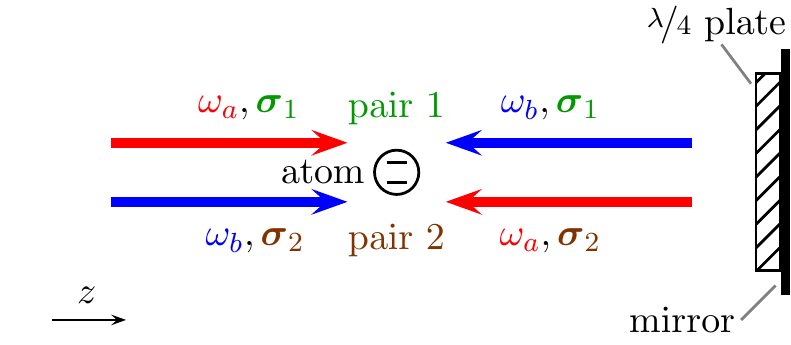}
\caption{Schematic set-up of double Bragg diffraction. Two pairs of counterpropagating light waves (pair 1 and pair 2) induce a Bragg scattering process in opposite directions. In order to distinguish both pairs, they have orthogonal polarizations $\vec{\sigma}_1$ and $\vec{\sigma}_2$. This orthogonality can be achieved by a retroreflecting geometry using a mirror and a $\lambda/4$ wave plate.}
\label{fig: schematic setup double bragg}
\end{figure}

In contrast to single Bragg diffraction, where two counterpropagating light waves interact with a two-level atom, we consider the following problem: One atom interacts with two pairs of light fields. Each pair induces a Bragg diffraction process, but in opposite directions. A possible set-up is shown in Fig.~\ref{fig: schematic setup double bragg}.

To suppress stimulated emission induced by {pair 1} if the atom was excited by {pair 2}, the polarizations of both pairs are chosen to be orthogonal, i.\,e.\ $\vec{\sigma}_i \cdot \vec{\sigma}_j= \delta_{i,j}$.
Otherwise, spurious scattering processes might occur:
The atom in Fig.~\ref{fig: schematic setup double bragg} could, for example, absorb a `red' photon from {pair 1} and thus gain the momentum $\hbar k_a$.
If now an emission of a `blue' photon from {pair 2} into the same direction occurs, the atom loses the momentum $\hbar k_b$, which yields the total momentum transfer $\hbar (k_a-k_b)$.
In addition, the desired processes with momentum transfers of ${\pm \hbar (k_a + k_b)}$ take place.
Taking into account all possible processes, one momentum state is therefore coupled to eight different momenta.

In the main part of this article, we perform the calculation for circularly polarized light fields, i.\,e., $\vec{\sigma}_1=\vec{\sigma}_+$ and $\vec{\sigma}_2=\vec{\sigma}_-$.
However, other choices of orthogonal polarizations are possible as well.
In Appendix~\ref{sec:lin-lin} we extend our model to orthogonal linear polarizations with a magnetic field in an arbitrary direction causing a Zeeman splitting.
\begin{figure}[htb]
\centering
\includegraphics[scale=1]{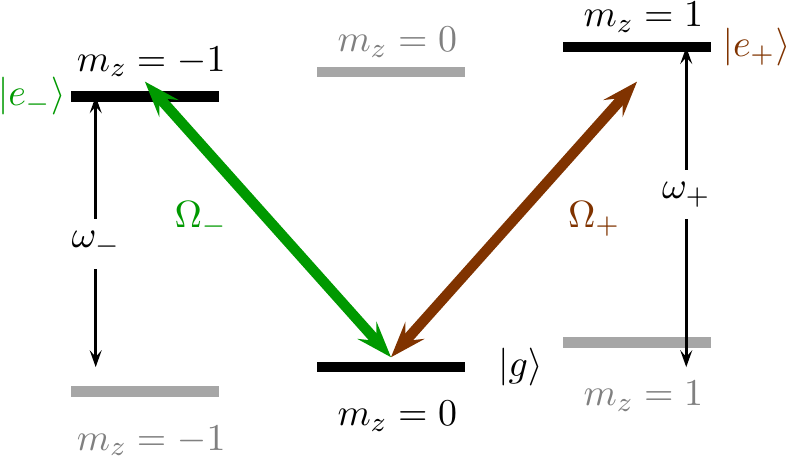}
\caption{Transitions between magnetic sublevels due to circular polarizations. Laser pair 1 with polarization $\vec{\sigma}_-$ drives the transitions between $\ket{g}$ and $\ket{e_-}$, which have the energy difference $\hbar \omega_-$, with a Rabi frequency $\Omega_-$; laser pair~2 with polarization $\vec{\sigma}_+$ causes the transitions between $\ket{g}$ and $\ket{e_+}$, which have the energy difference $\hbar \omega_+$, with a Rabi frequency $\Omega_+$. The kinetic energy is neglected in this figure and $m_z$ denotes the magnetic quantum number.}
\label{fig: atomic structure}
\end{figure}

We assume that the atom can be excited to two different states $\ket{e_+}$ and $\ket{e_-}$, corresponding to each polarization.
Depending on the specific species of atoms used in the experiment, one can identify magnetic sub-levels as in Fig.~\ref{fig: atomic structure}.
Although even atoms with ground state magnetic quantum numbers $m_z=\pm 2$ are possible, it is sufficient to consider just these three states if a magnetic field is applied such that the transitions to the $m_z=\pm 2$ levels are far detuned. For now, we use the configuration of Fig.~\ref{fig: atomic structure}.
\begin{figure}[htb]
\centering
\includegraphics[scale=1]{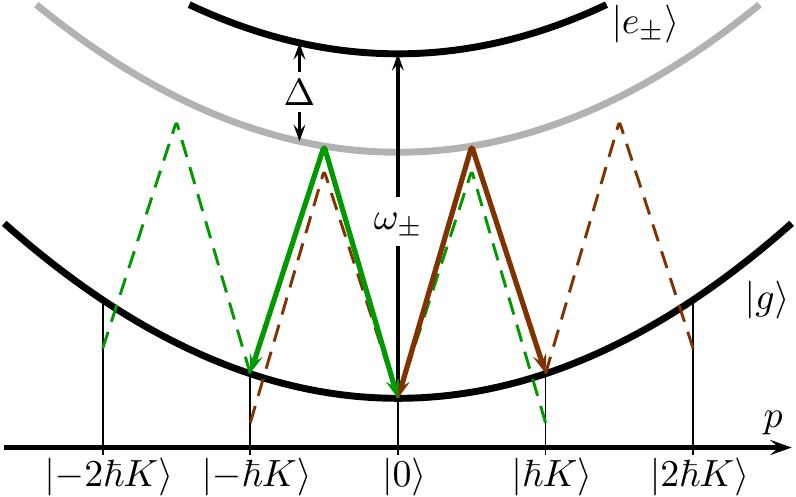}
\caption{Momentum transfer and resonances in double Bragg diffraction. The Bragg scattering processes induced by each pair of lasers correspond to the one of Fig.~\ref{fig: parabula single}, but are of opposite directions. The frequencies of the excited states are $\omega_+$ and $\omega_-$. The dashed transitions are off-resonant and get suppressed.}
\label{fig: parabula double}
\end{figure}

For this specific configuration, it is not necessary to make any restrictions on the Zeeman splitting $\omega_+-\omega_-$. However, for different polarizations (e.\,g., linear polarizations as discussed in Appendix~\ref{sec:lin-lin}) it is necessary that $|{\omega_+-\omega_-|\ll \Delta}$. So we assume ${\omega_+\cong \omega_-\equiv \omega_{eg}}$.

We show the momentum transfer for double Bragg diffraction in Fig.~\ref{fig: parabula double}.
Here, each pair of light fields induces independently a Bragg scattering process; {pair 1} to the left and {pair 2} to the right.
The dashed lines are the off-resonant processes of each pair.
We discuss their meaning and influence in Sec.~\ref{sec:double-beam-spitter}.

\subsection{Three-term recurrence relation}
\label{sec:double recurrenece}
We now derive analogously to Sec.~\ref{sec:singe 3-recursion} a three-term recurrence relation for the double Bragg diffraction process of the model described in the preceding section. The interaction of a two-level atom with four light waves shown in Fig.~\ref{fig: schematic setup double bragg} yields in rotating wave approximation \cite{schleich} the Hamiltonian
\ba
	\hat{H}=& \frac{\hat{p}^2}{2M}+ \hbar \omega_{eg} \left(\ket{e_+}\bra{e_+} + \ket{e_-}\bra{e_-}\right) \nonumber \\
		 &+\hbar\left\lbrace \Omega_+ \left[ \e{\I (k_a \hat{z}-\omega_a t)} +\e{\I (-k_b \hat{z}-\omega_b t)}\right] \ket{e_+}\bra{g} \right. \nonumber \\
		 &+\left.\Omega_- \left[ \e{\I (-k_a \hat{z}-\omega_a t)} +\e{\I (k_b\hat{z}-\omega_b t)}\right]\ket{e_-}\bra{g}+ ~\text{h.c.} \right\rbrace, \label{eq:Hamiltonian double}
\ea
where we have introduced the Rabi frequencies $\Omega_\pm\equiv- |E|\bra{e_\pm}\hat{\vec{d}}\cdot \vec{\sigma}_\pm\ket{g} /\hbar$ and used $\bra{e_\pm} \hat{\vec{d}} \cdot \vec{\sigma}_\mp\ket{g}=0$.
For the sake of simplicity we assume all the light fields to have the same amplitude $|E|$, even though this assumption is not necessary for the further treatment.
In particular, different amplitudes are assumed in Appendix~\ref{sec:lin-lin}.

The Hamiltonian $\hat{H}$ given by Eq.~\eqref{eq:Hamiltonian double} describes exactly the situation shown in Fig.~\ref{fig: atomic structure} and Fig.~\ref{fig: parabula double}.
Acting with it on the state
\ba
	\ket{\psi}= \int \D p \left[ g(p)\ket{p,g}+ e_+(p) \ket{e_+,p}+ e_-(p)\ket{e_-,p} \right] \nonumber
\ea
and applying the Schr\"odinger Eq.~\eqref{eq:schroedinger} to equate the coefficients of $\ket{g,p}$ and $\ket{e_\pm,p}$ to the time derivative of this state, we find a system of coupled differential equations for $g(p)$ and $e_\pm(p)$.
The latter ones can be eliminated adiabatically in the case of large detuning $\Delta\equiv\omega_{eg}-\omega_{a,b}$ in comparison to the Rabi frequencies $\Omega_\pm$, i.\,e.\ $\left|\Omega_\pm/ \Delta\right|\ll 1$, analogous to the approach of Sec.~\ref{sec:singe 3-recursion}.
In this way, we find the system of differential equations
\ba
\I \dot{g}(p)=&- g(p+\hbar K) \e{-\I \frac{p K}{m}t} \nonumber \\
&\times\left[ \frac{|\Omega_+|^2}{\Delta}\e{-\I (\Delta\omega+\omega_\text{r})t} + \frac{|\Omega_-|^2}{\Delta}\e{\I (\Delta\omega-\omega_\text{r})t}\right]	\nonumber \\
	&- g(p-\hbar K) \e{\I \frac{p K}{m}t} \nonumber \\
	&\times \left[\frac{|\Omega_+|^2}{\Delta}\e{\I (\Delta\omega-\omega_\text{r})t} + \frac{|\Omega_-|^2}{\Delta}\e{-\I (\Delta\omega+\omega_\text{r})t}\right]	 \nonumber
\ea
for the probability amplitude of the state $\ket{g}$. As in Sec.~\ref{sec:singe 3-recursion}, we have transformed again into the interaction picture.
Here, $\Delta \omega \equiv \omega_b-\omega_a$ denotes the difference of the light frequencies and $\omega_\text{r}\equiv \hbar K^2/(2M)$ the recoil frequency, respectively, where the effective wave vector is again defined as $K\equiv k_a+k_b$. 

With the notation $g_n\equiv g(p+\hbar n K)$ and defining the effective transition frequency $\Omega\equiv|\Omega_\pm|^2/\Delta$, the time-dependent three-term recurrence relation reads 
\begin{widetext}
\ba
\I \dot{g}_n=&-\Omega \e{-\I \frac{p K}{m}t}  g_{n+1}\left[\e{-\I [\Delta \omega + (2n+1)\omega_\text{r}]t} +\e{\I [\Delta \omega -(2n+1)\omega_\text{r}]t}\right]-\Omega  \e{\I \frac{p K}{m}t} g_{n-1} \left[\e{\I [\Delta \omega +(2n-1)\omega_\text{r}]t} + \e{-\I [\Delta \omega -(2n-1)\omega_\text{r}]t}\right]. \label{eq: dgl with delta omega}
\ea
\end{widetext}
When we compare this expression to the conventional Bragg diffraction case, Eq.~\eqref{eq:3term recurrence single}, we identify two terms instead of one within each of the brackets. The first term comes from the light fields of {pair 1}, the second one from {pair 2}.
Since each term oscillates at a different frequency, there are always two couplings between neighboring momentum levels.
Hence, we cannot perform the adiabatic elimination of off-resonant momentum levels in the familiar way, since some levels are resonantly and off-resonantly coupled at the same time.
For example, Fig.~\ref{fig: parabula double} shows that the state $\ket{-\hbar K}$ is coupled to $\ket{0}$ resonantly by {pair~1} with a solid green line, and off-resonantly by {pair~2} with a dashed brown line.
Nevertheless, the method of averaging described in Sec.~\ref{sec:method of averaging} can be applied and automatically deals with this problem.

\subsection{First-order diffraction}
In this section, we concentrate on first-order diffraction where $\Delta \omega= \omega_\text{r}$. For this case, Eq.~\eqref{eq: dgl with delta omega} reduces to
\ba
\I \dot{g}_n=&-\Omega \e{-\I \frac{p K}{m}t}  g_{n+1}\left[\e{-\I  2 (n+1)\omega_\text{r}t} +\e{-\I 2n\omega_\text{r}t}\right]	\nonumber \\
	&-\Omega  \e{\I \frac{p K}{m}t} g_{n-1} \left[\e{\I 2n\omega_\text{r}t} + \e{\I 2(n-1)\omega_\text{r}t}\right]. \nonumber
\ea
With the notation from Sec.~\ref{sec:singe 3-recursion}, the matrices have the form
\ba 
 \left(\mathcal{H}_\nu\right)_{n,n'}= &\omega_\text{r}\left[\e{\I \nu_\text{D} t}\delta_{n-1,n'} (\delta_{2n,\nu}+\delta_{2(n-1),\nu}) \right.\nonumber\\
		&\left.+\e{-\I \nu_\text{D} t} \delta_{n+1,n'}(\delta_{2(n+1),-\nu}+\delta_{2n,-\nu})\right]\,, \label{eq: def H_nu}
\ea
where we have recalled the definition of the Doppler frequency $\nu_\text{D}$ from Eq.~\eqref{eq:doppler freq}.
Note that we have not yet chosen $p=0$ as in \eq{eq: single def H_nu}. In Sec.~\ref{sec:double-beam-spitter} we consider $p=0$, whereas in Sec.~\ref{sec:velocity selectivity} we allow for $p\neq 0$.

\subsubsection{Beam splitters and mirrors}
\label{sec:double-beam-spitter}
To see if we can realize beam splitters and mirrors based on double Bragg diffraction, we turn to the deep Bragg regime, where the slow solution $\vec{\gamma}^{(1)}$ is sufficient.

We consider first the case of $p=0$, which also means that $\nu_\text{D}=0$, and the matrices $ \mathcal{H}_\nu$ become time-independent. According to Eq.~\eqref{eq:dgl gamma1} we have to solve the differential equation
\ba
\dot{\vec{\gamma}}^{(1)}= \I\varepsilon\mathcal{H}_0 \vec{\gamma}^{(1)}=\I \Omega \begin{pmatrix}
 0&\,1\,&0\\
 1&0&1\\
 0&1&0
\end{pmatrix}
\vec{\gamma}^{(1)}\,, \nonumber
\ea
with $\vec{\gamma}^{(1)}=(\gamma_{-1}^{(1)},\gamma_0^{(1)},\gamma_1^{(1)})^\text{T}$.

\begin{figure}[htb]
\includegraphics{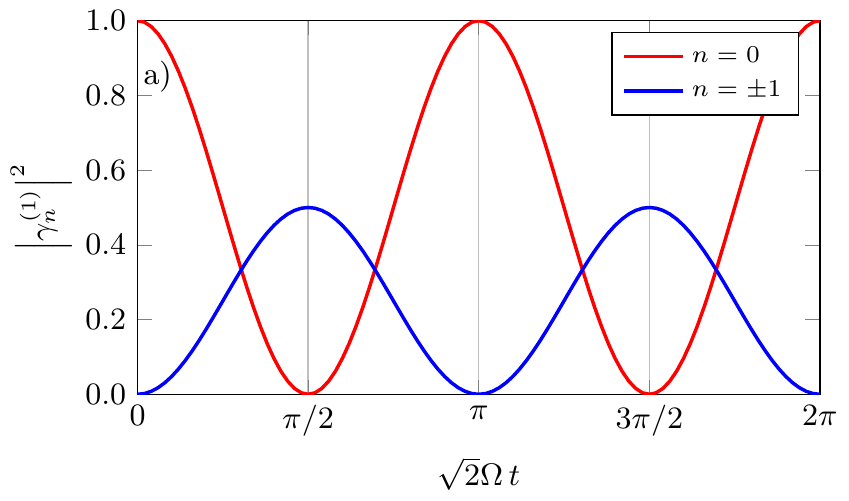}

\includegraphics{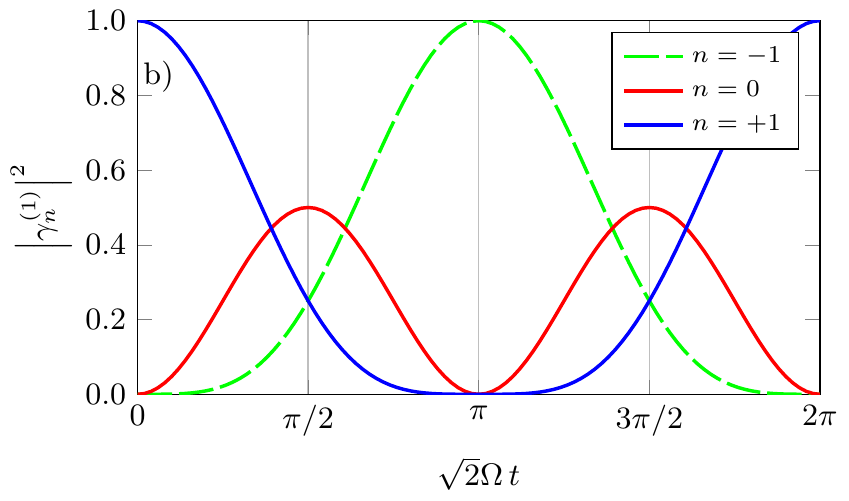}

\caption{Rabi oscillations to realize a beam splitter (a) and mirror (b) in the deep Bragg regime. For the adiabaticity parameter $\varepsilon\ll1$ the quantity $|\gamma^{(1)}_n|^2$ corresponds to the population of a momentum state $\ket{n\hbar K}$. For the demonstration of a beam splitter (a) the initial condition is $(0,1,0)^\text{T}$. Starting in  $\ket{0}$, a superposition $(\ket{-\hbar K}+\ket{\hbar K})/ \sqrt{2}$ is generated for $\sqrt{2}\Omega \, t=\pi/2$. For the demonstration of a mirror (b) the initial condition is $(1,0,0)^\text{T}$. Starting in $\ket{\hbar K}$, the whole population is transferred to $\ket{-\hbar K}$ at the time $\sqrt{2}\Omega\, t=\pi$. }
\label{fig: mirror and beam splitter}
\end{figure}

According to Eq.~\eqref{eq:gamma(1)allgem}, the solution of this system is given by the matrix exponential which can be carried out by any computer algebra system. In this case, we find
\ba
\vec{\gamma}^{(1)}(t)=& \left[ \frac{1}{2} \begin{pmatrix}
\,1\,&0&-1\,\\
 0&0&0\\
 -1\,&0&\,1\,
\end{pmatrix}+ \frac{1}{2}\cos\left( \sqrt{2}\Omega t \right) \,\begin{pmatrix}
\,1\,&0&\,1\,\\
 0&2&0\\
\,1\,&0&\,1\,
\end{pmatrix} \right. \nonumber \\
&\left. +  \frac{\I}{\sqrt{2}} \sin \left( \sqrt{2}\Omega t \right)\, \begin{pmatrix}
 0&\,1\,&0\\
 1&0&1\\
 0&1&0
\end{pmatrix}\right]\vec{\gamma}^{(1)}(0)\, . \label{eq: dbd gamma^1}
\ea
which describes quasi-Rabi oscillations of a three-level system. The populations $|\gamma_n^{(1)}|^2$ of the momentum states are shown in Fig.~\ref{fig: mirror and beam splitter} for the initial conditions $(0,1,0)^\text{T}$ and $(1,0,0)^\text{T}$. Indeed, $\pi/2$ and $\pi$ pulses can be performed and the effective Rabi frequency is $\sqrt{2}\Omega$. Thus, beam splitters and mirrors can be realized.

This result is not obvious, since these oscillations occur in an effective three-level system.
In contrast to single Bragg diffraction, where a $\pi/2$ pulse was generating an equal superposition of the initial state and the other state, we now use the term ``$\pi/2$ pulse'' when an equal superposition of the two states $\ket{\pm \hbar K}$ is created and the initial state is completely depopulated. In the same spirit, a ``$\pi$ pulse'' now exchanges the population of the two states $\ket{- \hbar K }$ and $\ket{\hbar K}$.

\begin{figure}[htb]

\includegraphics{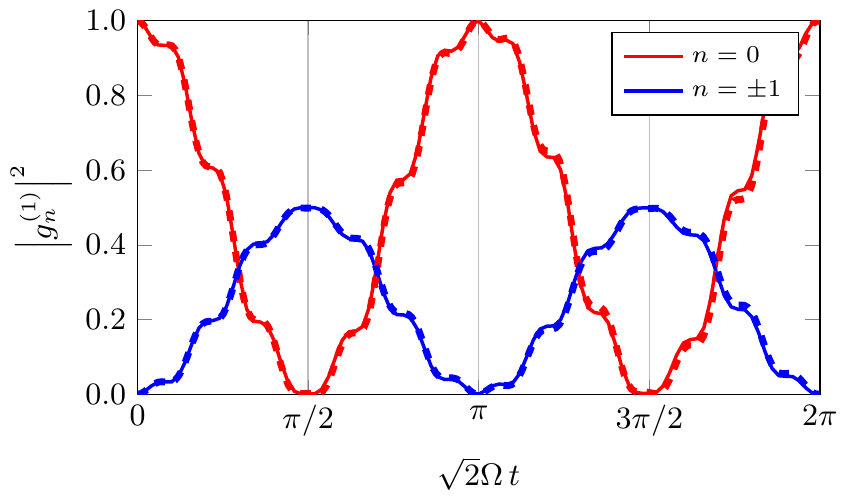}

 \caption{Beam splitter in the quasi-Bragg regime for the adiabaticity parameter $\varepsilon=0.1$ represented by the population $|g_n^{(1)}|^2$ of the momentum state $\ket{n \hbar K}$. The initial condition is $\vec{g}^{(1)}(0)=(0,1,0)^\text{T}$. The dashed lines are the result of a numerical simulation, the solid lines are the approximate analytical solutions. In this regime, fast oscillations with small amplitude modify the Rabi oscillations from Fig.~\ref{fig: mirror and beam splitter}.}
 \label{fig: first order corrections}
\end{figure}

The definition of the matrices $\mathcal{H}_\nu$ in \eq{eq: def H_nu} shows that there are adjustments to the momentum states $\ket{0}$ and $\ket{\pm \hbar K}$ if we leave the deep Bragg regime and take rapidly oscillating corrections of order $\varepsilon$ into account.
In addition to that, fast oscillations of the momentum states $\ket{\pm  2\hbar K}$ with small amplitudes can be found analogously to the single-diffraction case discussed in Sec.~\ref{sec:single-appl-averaging}.
But since we are only interested in the change of the dynamics of the states $\ket{0}$ and $\ket{\pm \hbar K}$ we just look at the innermost $3\times 3$ matrix, even though we have used a larger matrix to calculate it. In this case \eq{eq: g^(1) allgemein} reads
\ba
\vec{g}^{(1)}=\vec{\gamma}^{(1)}+\frac{\varepsilon}{2} \begin{pmatrix}
 0& \e{\I 2 \omega_\text{r}t}&0\\
 -\e{-\I 2 \omega_\text{r}t}&0&-\e{-\I 2 \omega_\text{r}t}\\
 0&\e{\I 2 \omega_\text{r}t}&0
\end{pmatrix}\vec{\gamma}^{(1)}. \nonumber
\ea
The influence of off-resonant processes is taken into account by a fast oscillation on top of the slow Rabi oscillation.
The fast frequency is again $2\omega_\text{r}$, which corresponds to the energy that the off-resonant process deviates from the resonance.
The populations $|g_n^{(1)}|^2$ are plotted in Fig.~\ref{fig: first order corrections} and compared to a numerical simulation. For the choice of $\varepsilon=0.1$, we see excellent agreement. For the plot, we used the bare state initial condition $g^{(1)}(0)=(0,1,0)^\text{T}$ from \eq{eq:initial_cond}.

These small-amplitude fast oscillations at the order of $\varepsilon$ for dressed states are a new feature of double Bragg diffraction, namely that we have simultaneously a resonant and off-resonant coupling of a momentum state.

\subsubsection{Small deviations from resonant momenta: Velocity selectivity}
\label{sec:velocity selectivity}
So far, we have just considered first-order double Bragg diffraction with corrections of order $\varepsilon$ for the case $p=0$.
In this case, the momenta $\ket{0}$ and $\ket{\pm \hbar K}$ are resonant, but all other states $\ket{n \hbar K}$ are suppressed.
But if we now allow $p\neq 0$, we can still find approximate analytic solutions which are a double Bragg generalization of the single Bragg case as discussed in Ref.~\cite{momentum-width}.
We dedicate the current section to investigate this feature in more detail.

For $p \neq 0$, the matrices $\mathcal{H}_\nu$ are, according to \eq{eq: def H_nu}, proportional to $\ehoch{\pm\I \nu_\text{D} t}$ and thus time-dependent.
Since the method of averaging delicately plays with the combination of fast and slowly oscillating terms, these factors are of importance.

In Sec.~\ref{sec:method of averaging}, we have integrated in \eq{eq: first correction} by neglecting any time-dependence of $\mathcal{H}_\nu$.
This approximation is only true if these matrices vary slower than the exponents $\ehoch{\I\nu \omega_\text{r} t}$.
This requirement is fulfilled for $|\nu_\text{D}|\ll \omega_\text{r}$ or $|p|\ll \hbar K/2$. Our solution is only a good approximation if we fulfill this condition, so we just consider small deviations from the resonant momenta.

In this case $\vec{\gamma}^{(1)}$ can be found easily. If we multiply $\gamma_{\pm1}$ by $\ehoch{\mp \I \nu_\text{D} t}$, we find the new system
\ba
\dot{\vec{\gamma}}^{(1)}= \I  \begin{pmatrix}
 \nu_\text{D} & \Omega &0\\
 \Omega &0& \Omega\\
 0&\Omega &-\nu_\text{D}
\end{pmatrix}
\vec{\gamma}^{(1)} \label{eq:dgl p neq 0}
\ea
of differential equations for these coefficients.
\begin{figure}

\includegraphics{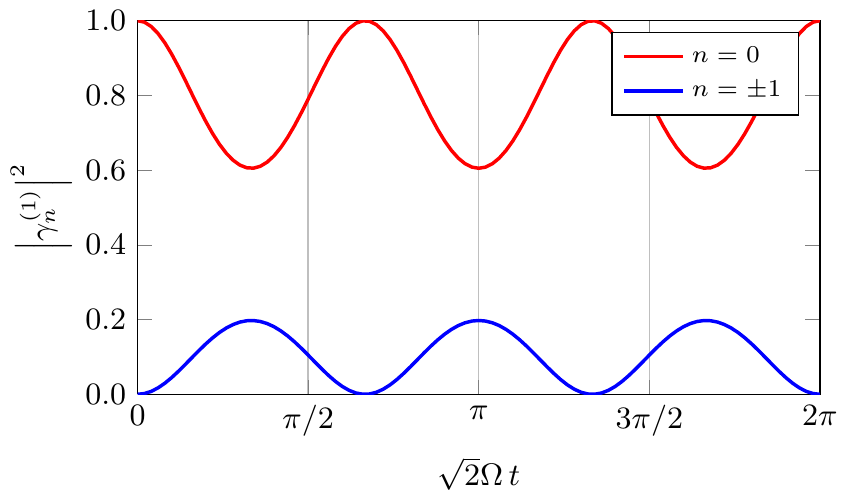}

 \caption{Rabi oscillations of the populations $|\gamma_n^{(1)}|^2$ of the momentum states $\ket{(n+0.01)\hbar K}$ in the deep Bragg regime (${\varepsilon=0.01}$) for the off-resonant momentum $p= 0.01 \hbar K$. The Rabi oscillations are detuned and suppressed, which leads to velocity selectivity. The initial condition is $(0,1,0)^\text{T}$ and the time is in units of the resonant frequency $\sqrt{2}\Omega$.}
 \label{fig: p neq 0}
\end{figure}

This equation shows what happens if the momentum deviates from a multiple of $\hbar K$: 
Two elements $\pm \nu_\text{D}$ show up on the main diagonal of the matrix, which differ from the element in the center of the matrix. This is why they act as a detuning or AC Stark shift \cite{James} on the Rabi oscillations. For increasing deviations from the resonant momenta the interaction is therefore suppressed. This effect is well-known and called velocity selectivity \cite{Kozuma, momentum-width}. This feature becomes more obvious when we look at the solution
\begin{widetext}
\ba
\vec{\gamma}^{(1)}(t)=& \left[\frac{1}{\Omega_\text{eff}^2} \begin{pmatrix}
\Omega^2 & -\nu_\text{D}\Omega&-\Omega^2\\
  -\nu_\text{D}\Omega &\nu_\text{D}^2 &\nu_\text{D} \Omega\\
 -\Omega^2 & \nu_\text{D} \Omega &\Omega^2
\end{pmatrix}\hspace{-.1cm}
+ \frac{\cos\left( \Omega_\text{eff} t \right)}{\Omega_\text{eff}^2}\begin{pmatrix}
\nu_\text{D}^2 \hspace{-.1cm}+\hspace{-.1cm}\Omega^2 & \nu_\text{D} \Omega& \Omega^2\\
 \nu_\text{D}\Omega & 2\Omega^2 & -\nu_\text{D}\Omega\\
\Omega^2 & -\nu_\text{D}\Omega &\nu_\text{D}^2\hspace{-.1cm} +\hspace{-.1cm}\Omega^2
\end{pmatrix} \hspace{-.1cm}
+  \frac{\I \sin \left( \Omega_\text{eff} t \right)}{\Omega_\text{eff}} \begin{pmatrix}
 \nu_\text{D}&\Omega & 0\\
 \Omega & 0 & \Omega\\
 0& \Omega &-\nu_\text{D}
\end{pmatrix}\right]\vec{\gamma}^{(1)}(0) \label{eq:velocity selectivity}
\ea
\end{widetext}
of Eq.~\eqref{eq:dgl p neq 0}, where we have defined the effective Rabi frequency ${\Omega_\text{eff}\equiv\sqrt{2\Omega^2+\nu_\text{D}^2}}$. The matrix exponential defined by Eq.~\eqref{eq:gamma(1)allgem} was calculated with the help of a computer algebra system.
Indeed, Eq.~\eqref{eq:velocity selectivity} reduces to \eq{eq: dbd gamma^1} for $p=0$, i.\,e.\ if the Doppler frequency $\nu_\text{D}$ vanishes. The effective Rabi frequency clearly demonstrates that the Doppler frequency $\nu_\text{D}$ acts as a detuning.

Figure~\ref{fig: p neq 0} shows in comparison with Fig.~\ref{fig: mirror and beam splitter} the effect of detuning and suppression for a momentum ${p=0.01 \hbar K}$, which corresponds to $\nu_\text{D}= 0.02 \omega_\text{r}$.
We note that the frequency of the oscillation as well as the amplitude are changed.

\begin{figure}[htb]

\includegraphics{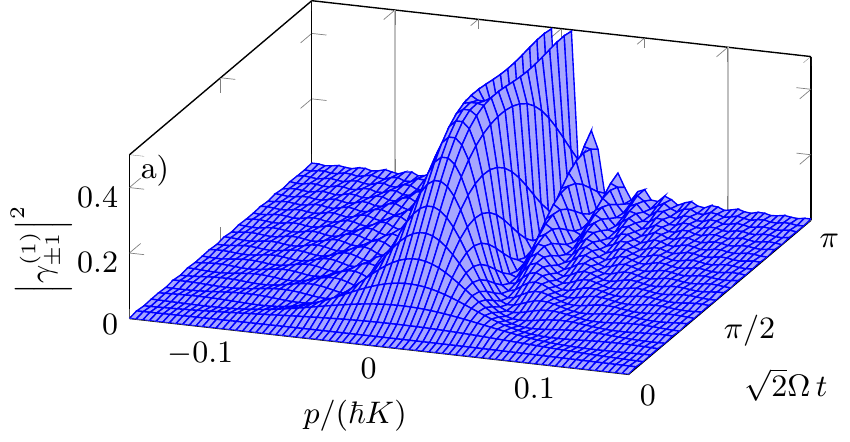}

\includegraphics[scale=.98]{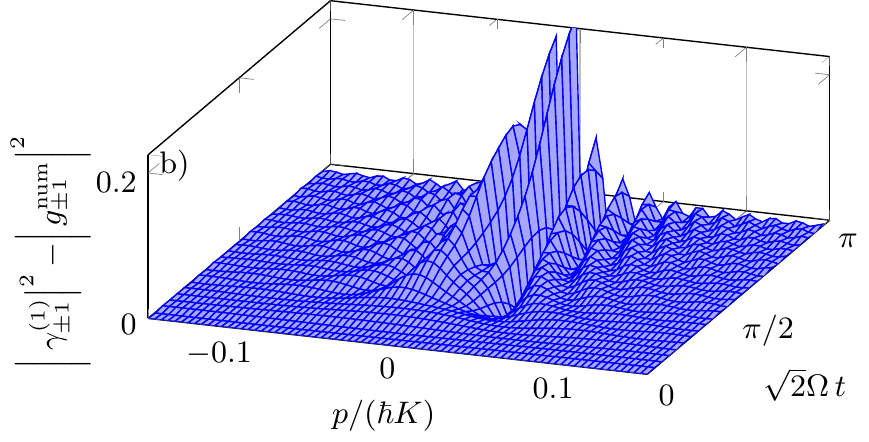}

 \caption{Velocity selectivity of double Bragg diffractionexpressed by the Rabi oscillation of the population $|\gamma^{(1)}_{\pm1}|^2$ in their dependence on the initial momentum (a). For $p=0$, i.\,e.\ for the resonant momenta $\pm \hbar K$, we get full Rabi oscillations, but as we leave this resonance this process is suppressed and detuned.
The difference between analytical and numerical solutions (b) shows a large deviation between both solutions. This deviation is mainly due to different frequencies, not to different amplitudes. The width of the resonance is comparable in both solutions. Here we have chosen $\varepsilon=0.01$.}
\label{fig:3d velocity}
\end{figure}

Figure~\ref{fig:3d velocity} shows a three-dimensional plot of the Rabi oscillations for different momenta.
The analytical solution, i.\,e.\ Eq.~\eqref{eq:velocity selectivity}, for the diffracted state populations $|\gamma^{(1)}_{\pm 1}|^2$ is shown in the top part of the figure.
As already discussed above, for large deviations from the resonance $p=0$ the interaction is suppressed, and almost no population occurs, which very clearly demonstrates the velocity selectivity.

In the bottom part of the figure the difference between this analytical solution and a numerical simulation is shown.
As expected, they coincide only in the immediate neighborhood of $p=0$ but deviate for larger values of $p$. 

Even though the exact frequencies and amplitudes are not very well approximated for increasing momenta, the width of the resonance is similar and thus the approximation made above surprisingly good.
%
\begin{figure}

\includegraphics{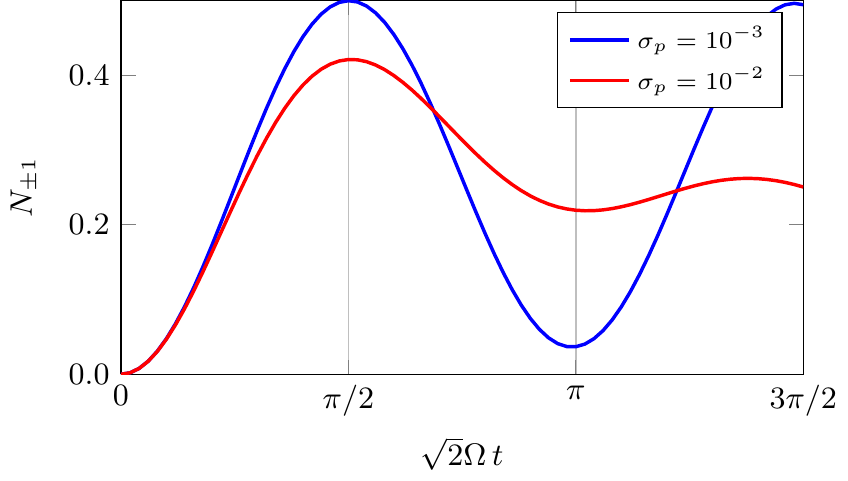} 

 \caption{Damped Rabi oscillations in the numerically integrated population $N_{\pm1}$ when the adiabaticity parameter is $\varepsilon=0.01$. For a broad momentum distribution, the oscillations get damped out due to the average over different frequencies and amplitudes. This effect is less pronounced for smaller values of the momentum width $\sigma_p$. }
 \label{fig:damping}
\end{figure}

Since in experiments the atoms do have a distribution over many momenta, one has to average the Rabi oscillations shown in Fig.~\ref{fig:3d velocity} over the momentum and weigh them with their initial momentum distribution. In our simulation we use an initial Gaussian distribution
\ba
\gamma^{(1)}_0(0,p)= \mathcal{N}\ehoch{-\frac{(p/\hbar K)^2}{4 \sigma_p^2}} \, ,\nonumber
\ea
with a momentum width of $\sigma_p$ in units of $\hbar K$, and a normalization constant
\ba
\mathcal{N} \equiv\left(\sqrt{2\pi}\,\sigma_p\, \hbar K\right)^{-\frac{1}{2}}\, , \nonumber
\ea
and calculate numerically the integrated population 
\ba
N_{\pm 1}= \int\limits_{-\hbar K /2}^{\hbar K/2} \D p\, \left|\gamma^{(1)}_{\pm1}(t,p)\right|^2 \, , \nonumber
\ea
of atoms with momenta around $\pm \hbar K$. In time-of-flight experiments these can be identified by measuring the clouds drifting away from the center.

The results of this numerical integration for different momentum widths shown in Fig.~\ref{fig:damping} display a damping of the Rabi oscillation. This feature is only an artefact of the interaction times; for very long times the oscillation revives. Since in atom interferometry the interest lies on $\pi/2$ or $\pi$ pulses, the interaction time is limited and revivals are not observable. The damping can be explained very easily: As we already discussed, the Rabi frequency depends on the initial momentum and thus gets washed out when averaged over a momentum distribution. 
This dephasing effect is an intrinsic problem of the diffraction process, which can be improved by using narrower momentum distributions, generated, e.\,g., by techniques such as delta-kick cooling \cite{muentinga}.

The frequency of the effective Rabi oscillation is now $\sqrt{2} \Omega$ instead of $\Omega$. This property might lead to the conclusion that the interaction times in double diffraction would be shorter.
But to compare the velocity selectivity of single and double Bragg diffraction, it is important to look not at the interaction times, but at the arguments of the trigonometric functions in Eq.~\eqref{eq:velocity selectivity}. From this point of view, a beam splitter and a mirror take twice as long, e.\,g.\ at $\sqrt{2} \Omega\, t = \pi/2$ instead of $\Omega\, t = \pi/4$ for a $\pi/2$ pulse. Hence, the higher velocity selectivity in comparison with single Bragg diffraction has its origin in the three-level behavior of double diffraction. An extensive numerical discussion of the importance of the momentum width for single Bragg diffraction for different diffraction orders is given in Ref.~\cite{momentum-width}.

Of course, the velocity selectivity decreases with increasing light field intensity, i.\,e.\ in the end with increasing adiabaticity parameter $\varepsilon$. This feature can be seen from Eq.~\eqref{eq:velocity selectivity}, where the states are suppressed by the fraction $\Omega/\sqrt{2 \Omega^2+ \nu_\text{D}^2}= \varepsilon/\sqrt{2 \varepsilon^2 + [2p/(\hbar K)]^2 }$. For increasing $\varepsilon$, the significance of $p\neq 0$ decreases. 

On the other side, an increasing $\varepsilon$ leads to the quasi-Bragg regime. So higher-order excitations become more and more significant.

\subsection{Second-order double Bragg diffraction}
\label{sec:2nd order double}
\begin{figure}[htb]
\centering
\includegraphics[scale=1]{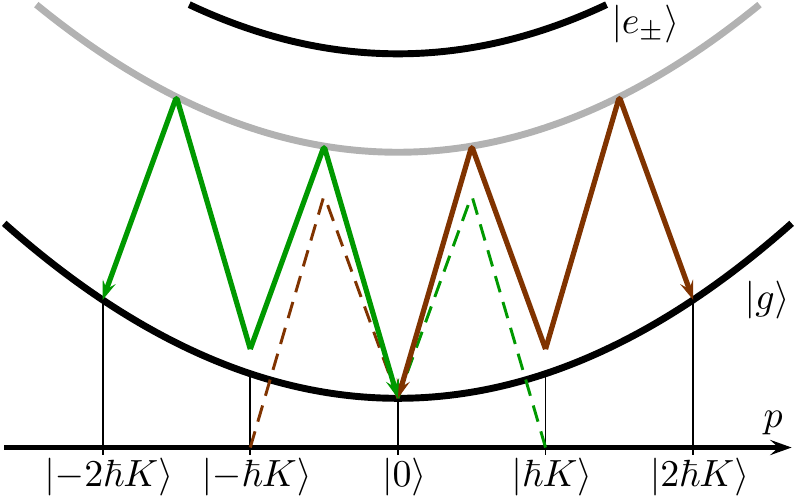}
\caption{Two processes in second-order double Bragg diffraction ($\Delta \omega= 2 \omega_\text{r}$). A four-photon transition in each direction occurs. For an initial population of the momentum states $\ket{0}$ and $\ket{\pm 2 \hbar K}$, the states $\ket{\pm \hbar K}$ are just virtually occupied. Different initial conditions are sketched in Fig.~\ref{fig: parabula 2nd pm1}. The dashed lines show off-resonant processes.}
\label{fig: parabula 2nd}
\end{figure}

The formalism introduced in our article makes it possible to also describe second-order double Bragg diffraction.
For that, we choose according to \eq{eq:res_cond} the resonance condition $\Delta \omega = 2 \omega_\text{r}$.
The resulting second-order process is depicted in Fig.~\ref{fig: parabula 2nd}.

The first scattering process, a two-photon transition, is off-resonant by $\omega_\text{r}$.
But adding a subsequent two-photon transition, one reaches resonantly the states $\ket{\pm 2 \hbar K}$.
For certain parameters, the intermediate state is just occupied virtually. This is a process similar to the adiabatic elimination of the excited state described in Sec.~\ref{sec:singe 3-recursion}.
Now, the transition frequency $\Omega$ has to be considerably smaller than the detuning, i.\,e.\ the deviation from resonance, which is $\omega_\text{r}$.

On the other hand, in the deep Bragg regime, where off-resonant processes are strongly suppressed, nothing would happen.
Therefore, we have to leave the deep Bragg regime in order to allow this process, as we shall see now.

When we use the resonance condition $\Delta \omega = 2 \omega_\text{r}$ for the second-order Bragg scattering process, Eq.~\eqref{eq: dgl with delta omega} can be written as 
\ba
\I \dot{g}_n=&-\Omega \e{-\I \nu_\text{D} t}  g_{n+1}\left[\e{-\I  (2n+3)\omega_\text{r}t} +\e{-\I (2n-1)\omega_\text{r}t}\right]	\nonumber \\
	&-\Omega  \e{\I\nu_\text{D} t} g_{n-1} \left[\e{\I (2n+1)\omega_\text{r}t} + \e{\I (2n-3)\omega_\text{r}t}\right]. \nonumber
\ea

\subsubsection{Emergence of the asymmetric AC Stark effect}
For the sake of simplicity we look at $p=0$, i.\,e.\ $\nu_\text{D}=0$, which leads us to the matrices
\ba 
 \left(\mathcal{H}_\nu\right)_{n,n'}= \omega_\text{r}&\left[
 \delta_{n-1,n'} (\delta_{2n+1,\nu}+\delta_{2n-3,\nu}) \right.\nonumber\\
		&\left.+
		 \delta_{n+1,n'}(\delta_{2n+3,-\nu}+\delta_{2n-1,-\nu})\right]. \label{eq: H_nu 2nd order}
\ea
Hence, $\mathcal{H}_\nu$ is zero for even $\nu$; in particular, $\mathcal{H}_0=0$.

Since the differential equation for the first slow approximation reads $\dot{\vec{\gamma}}^{(1)}=\vec{0}$ we find the trivial solution $\vec{\gamma}^{(1)}(t)=\vec{\gamma}^{(1)}(0)$.
This result is not surprising, since in this approximation we can only describe first-order processes and they are all suppressed, in agreement with Fig.~\ref{fig: parabula 2nd}.

Next, we turn to the second-order approximation; i.\,e., we go to the next higher order in $\varepsilon$, that is $m=2$. 
According to Sec.~\ref{sec:method of averaging}, the solution given by Eq.~\eqref{eq: gamma_2} reads
\ba
\vec{\gamma}^{(2)}(t)=&\ehoch{\I\varepsilon^2 \sum_{\nu \neq 0}\frac{\mathcal{H}_{-\nu}\mathcal{H}_\nu}{\nu \omega_\text{r}} t}\;\vec{\gamma}^{(2)}(0). \nonumber
\ea
With the specific form of $\mathcal{H}_\nu$ given by \eq{eq: H_nu 2nd order} we find
\ba
\sum_{\nu \neq 0}\frac{\mathcal{H}_{-\nu}\mathcal{H}_\nu}{\nu \omega_\text{r}}= \omega_\text{r}
\begin{pmatrix}
-\frac{76}{105}	&	0	&	-1	&	0	&	0 \\
0	&	\mathbf{\frac{28}{15}}	&	0	&	\mathbf{\frac{2}{3}}	&	0 \\
-1	&	0	&	-\frac{4}{3}	&	0	&	-1 \\
0	&	\mathbf{\frac{2}{3}}	&	0	&	\mathbf{  \frac{28}{15}}	&	0 \\
0	&	0	&	-1	&	0	&	-\frac{76}{105}
\end{pmatrix}, \label{eq: 2nd order Phi0}
\ea
where we have focused on the inner $5\times5$ matrix and used ${\vec{\gamma}^{(2)}\equiv(\gamma_{-2}^{(2)},\gamma_{-1}^{(2)},\gamma_0^{(2)},\gamma_1^{(2)},\gamma_2^{(2)})^\text{T}}$.

We can already recognize the structure of the resulting oscillations.
Similar to the case of $p\neq 0$ in \eq{eq:dgl p neq 0} for first-order diffraction in the previous section, we now see entries on the main diagonal of the matrix.
As already analyzed in Sec.~\ref{sec:velocity selectivity}, values on the main diagonal that differ from the central element account for an AC Stark shift.
For a second-order process, this effect is not surprising.
But in contrast to the case in the context of velocity selectivity, the deviations do have the same sign.
We come back to this point after we have discussed the solutions
\ba
\begin{pmatrix}
\gamma_{-2}^{(2)}(t) \\
\gamma_0^{(2)}(t)\\
\gamma_{+2}^{(2)}(t)
\end{pmatrix}= \mathcal{M }_{\pm2} (t) \cdot
\begin{pmatrix}
\gamma_{-2} ^{(2)}(0)\\
\gamma_0 ^{(2)}(0)\\
\gamma_{+2}^{(2)}(0)
\end{pmatrix}\label{eq: 2nd order osci}
\ea
and
\ba\begin{pmatrix}
\gamma_{-1}^{(2)}(t) \\
\gamma_{+1}^{(2)}(t)
\end{pmatrix} = \mathcal{M}_{\pm1} (t)\cdot
\begin{pmatrix}
\gamma_{-1}^{(2)} (0)\\
\gamma_{+1}^{(2)}(0)
\end{pmatrix}   \label{eq: 2nd order pm1}
\ea
for the slowly evolving part in second approximation. Here we have separated the oscillations between the states $\ket{-2\hbar K}, \ket{0},\text{ and }\ket{2 \hbar K}$ from the ones between $\ket{-\hbar K}$ and $\ket{\hbar K}$.

To show that this separation follows directly from the structure of the differential equation, we boldfaced those entries in \eq{eq: 2nd order Phi0} that are responsible for the transitions between $\ket{-\hbar K}$ and $\ket{\hbar K}$. The dynamics of the states is determined by the matrices
\begin{widetext}
\ba
&\mathcal{M}_{\pm2}=\e{-\I \frac{76}{108}\varepsilon^2\omega_\text{r}t} \frac{1}{2} \begin{pmatrix}
1	&	0	&	-1 \\
0	&	0	&	0 \\
-1	&	0	&	1
\end{pmatrix}  + \e{-\I \frac{108}{105}\varepsilon^2\omega_\text{r}t} \left[\frac{\cos \left( \Omega_{\pm2} t \right)}{2} 
\begin{pmatrix}
1	&	0	&	1 \\
0	&	2	&	0 \\
1	&	0	&	1
\end{pmatrix} 
-\frac{\I\sin \left( \Omega_{\pm2} t \right)}{\sqrt{23074}}  
\begin{pmatrix}
-16	&	105	&	-16 \\
105	&	32	&	105 \\
-16	&	105	&	-16
\end{pmatrix} 
\right]\nonumber
\ea
\end{widetext}
and
\ba
\mathcal{M}_{\pm1}\hspace{-.1cm}=&\e{-\I \frac{28}{15}\varepsilon^2\omega_\text{r}t} \left[\cos  \Omega_{\pm1} t
\begin{pmatrix}
1	&	0 \\
0	&	1
\end{pmatrix} 
+\I \sin \Omega_{\pm1} t 
\begin{pmatrix}
0	&	1	 \\
1	&	0	 
\end{pmatrix} 
\right], \label{eq:M_pm1}
\ea
where the Rabi oscillation between the states $\ket{\pm \hbar K}$ has the frequency $\Omega_{\pm 1}\equiv (2/3)\,\varepsilon^2\omega_\text{r}=(2/3)\,\varepsilon\Omega$, determined by one of the boldfaced numbers in \eq{eq: 2nd order Phi0}, and the quasi-Rabi oscillation between the states $\ket{\pm 2 \hbar K}$ has the frequency  $\Omega_{\pm 2}\equiv (\sqrt{23074}/105)\varepsilon^2\omega_\text{r}=(\sqrt{23074}/105) \varepsilon \Omega$.

Analogously to the result of the adiabatic elimination of the excited state Eq.~\eqref{eq: single dgl g(p) lang}, where we found the Rabi frequency $\Omega= |\Omega_\pm|^2/\Delta$, the frequency of this solution is now proportional to $\Omega^2/\omega_\text{r}= \Omega \varepsilon$. The same structure makes the connection to the adiabatic approximation obvious. The adiabaticity parameter $\varepsilon$ now decreases the frequency and makes the second-order oscillation slower in comparison to a first-order process. 
\begin{figure}[htb]
\centering
\includegraphics[scale=1]{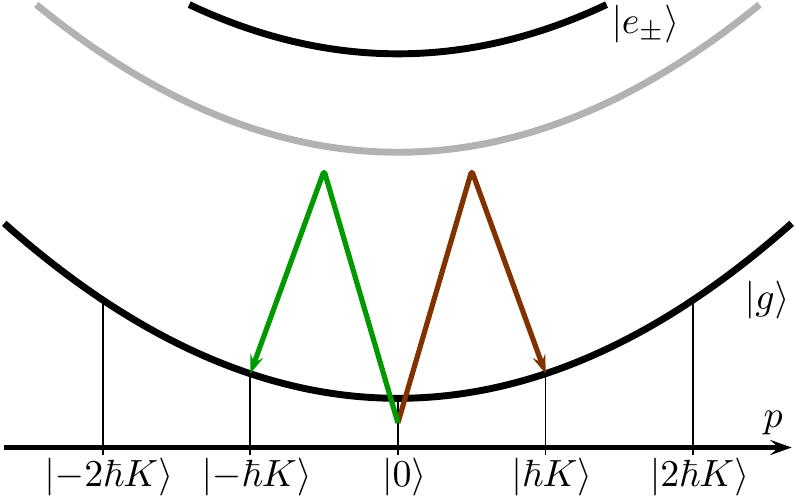}
\caption{Scattering process between the states $\ket{\pm \hbar K}$ according to Eq.~\eqref{eq: 2nd order pm1}. If the atoms are initially in one of these states, a resonant Rabi oscillation between them takes place in second-order Bragg diffraction. This phenomenon can be explained by the resonant four-photon process shown above. The transitions for other initial conditions are shown in Fig.~\ref{fig: parabula 2nd}.}
\label{fig: parabula 2nd pm1}
\end{figure}

We first discuss the oscillation between the states $\ket{\pm \hbar K}$. According to Eq.~\eqref{eq:M_pm1} they do have the same form as the ones in \eq{eq: Rabi single Bragg}, which are just Rabi oscillations between these two states.
There is no suppression and no involvement of other states. The resonant second-order process that connects these two states is depicted in Fig.~\ref{fig: parabula 2nd pm1}.
\begin{figure}

\includegraphics{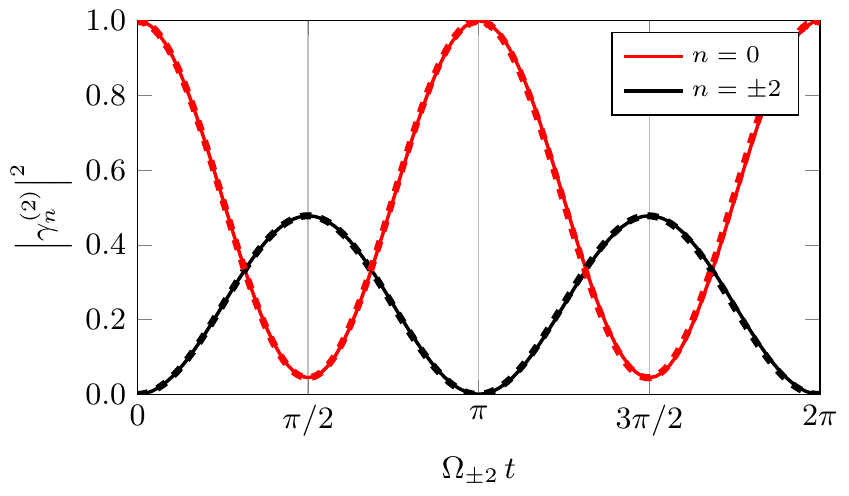} 

 \caption{Rabi oscillations in second-order double Bragg diffraction where $|\gamma_n^{(2)}|^2$ corresponds to the population of the momentum state $\ket{n\hbar K}$ for $\varepsilon\ll1$ .The initial condition $(0,0,1,0,0)^\text{T}$ was chosen to focus on the diffraction process $\ket{0}\leftrightarrow \ket{\pm2\hbar K}$ where the states $\ket{\pm \hbar K}$ are not involved. The oscillations are slightly suppressed due to an asymmetric AC Stark shift. The dashed lines show the numerical solution for $\varepsilon=0.01$.}
 \label{fig:2nd order}
\end{figure}

However, we are more interested in the oscillations from Eq.~\eqref{eq: 2nd order osci}. They correspond to the processes shown in Fig.~\ref{fig: parabula 2nd}.
Indeed, these oscillations are the quasi-Rabi oscillations of a three-level system between the relevant momentum states $\ket{\pm 2 \hbar K}$ and $\ket{0}$.
The oscillations are suppressed by a factor smaller than unity in front of the sine-function and plotted in Fig.~\ref{fig:2nd order}.
The states cannot be completely depopulated with a $\pi/2$ or $\pi$ pulse, i.\,e.\ perfect beam splitters and mirrors are not possible.
This feature is very different from single second-order Bragg scattering.
Qualitatively, this difference is a consequence of the form of the matrix Eq.~\eqref{eq: 2nd order Phi0} where we have already highlighted the origin of this AC Stark shift due to the `asymmetry' of the coupling to off-resonant states.

We also want to emphasize that the two-photon light shift occurring in Raman transitions \cite{clade, gauguet} is related to this asymmetric AC Stark shift, but there are subtle differences. First of all, the two-photon light shift is a frequency shift due to off-resonant states occurring in a double Raman set-up, where an asymmetry is introduced by giving the atom an initial velocity, in contrast to the double Raman diffraction of Ref.~\cite{Leveque}. However, frequency shifts due to off-resonant transitions occur even in single Bragg and Raman diffraction. The asymmetric AC Stark shift in second-order double Bragg diffraction is a consequence of the special form of Eq.~\eqref{eq: 2nd order Phi0} and not just due to higher off-resonant states.

Since the possibility of higher-order diffraction is one of the advantages of Bragg diffraction over Raman diffraction, it seems as if this advantage is lost in double diffraction, since no perfect beam splitters and mirrors in higher orders can be realized. However, we show in the following section that this effect can be compensated. In addition to that, we believe that the asymmetric AC Stark shift is a small effect in comparison to the damping and the imperfections due to the width of the momentum distribution discussed above.

\subsubsection{Compensation for the asymmetric AC Stark shift}
Fortunately, the asymmetric AC Stark shift can be compensated for. To understand this claim analytically, we recall from Sec.~\ref{sec:velocity selectivity} the case of momenta $p\neq 0$ and study the effect of a slightly violated resonance condition; i.\,e., we choose $\Delta \omega= (2 +\delta)\omega_{r}$ with $\delta \ll 1$.

When we include the phase factor $\ehoch{\I2\delta\omega_\text{r}t}$ in the states $\gamma_{\pm2}$, the matrix for the new system of differential equations reads
\ba
\sum_{\nu \neq 0} \hspace{-.075cm}\frac{\mathcal{H}_{-\nu}\mathcal{H}_\nu}{\nu \omega_\text{r}} \hspace{-.05cm}=\hspace{-.1cm} \omega_\text{r} \hspace{-.1cm} 
\begin{pmatrix}
-\frac{76}{105}	+ 2 \frac{\delta}{\varepsilon^2}&	0	&	-1	&	0	&	0 \\
0	&	\frac{28}{15}	&	0	&	\frac{2}{3}	&	0 \\
-1	&	0	&	-\frac{4}{3}	&	0	&	1 \\
0	&	\frac{2}{3}	&	0	&	\frac{28}{15}	&	0 \\
0	&	0	&	-1	&	0	&	-\frac{76}{105}	+ 2  \frac{\delta}{\varepsilon^2}
\end{pmatrix}\hspace{-.1cm} . \nonumber
\ea
With the choice $\delta\equiv -(32/105) \, \varepsilon^2$ this matrix reduces to
\ba
\sum_{\nu \neq 0}\frac{\mathcal{H}_{-\nu}\mathcal{H}_\nu}{\nu \omega_\text{r}}= \omega_\text{r}
\begin{pmatrix}
-\frac{4}{3}&	0	&	-1	&	0	&	0 \\
0	&	\frac{28}{15}	&	0	&	\frac{2}{3}	&	0 \\
-1	&	0	&	-\frac{4}{3}	&	0	&	-1 \\
0	&	\frac{2}{3}	&	0	&	\frac{28}{15}	&	0 \\
0	&	0	&	-1	&	0	&	-\frac{4}{3}&
\end{pmatrix}, \nonumber
\ea
where now the elements on the main diagonal that belong to the states $\ket{\pm 2 \hbar K}$ do not differ from the central element, and thus no asymmetric AC Stark shift occurs.

The Rabi oscillation Eq.~\eqref{eq: 2nd order pm1} between the states $\ket{\pm \hbar K}$ does not change at all, but the matrix of \eq{eq: 2nd order osci} reads now
\begin{widetext}
\ba
\mathcal{M}_{\pm2}=&\e{-\I \frac{4}{3}\Omega\varepsilon t} \left[ \frac{1}{2} \begin{pmatrix}
1	&	0	&	-1 \\
0	&	0	&	0 \\
-1	&	0	&	1
\end{pmatrix}  
+\frac{\cos\left( \sqrt{2}\Omega\varepsilon t \right) }{2} 
\begin{pmatrix}
1	&	0	&	1 \\
0	&	2	&	0 \\
1	&	0	&	1
\end{pmatrix}-\frac{\I\sin \left( \sqrt{2}\Omega\varepsilon t \right)}{\sqrt{2}}  
\begin{pmatrix}
0	&	1	&	0 \\
1	&	0	&	1 \\
0	&	1	&	0
\end{pmatrix} 
\right]\nonumber
\ea
\end{widetext}
and a resonant effective Rabi oscillation as in \eq{eq: dbd gamma^1} with a frequency $\sqrt{2} \Omega\varepsilon$ occurs. Hence, the frequency of the first-order diffraction is just multiplied by $\varepsilon$.
\begin{figure}

\includegraphics{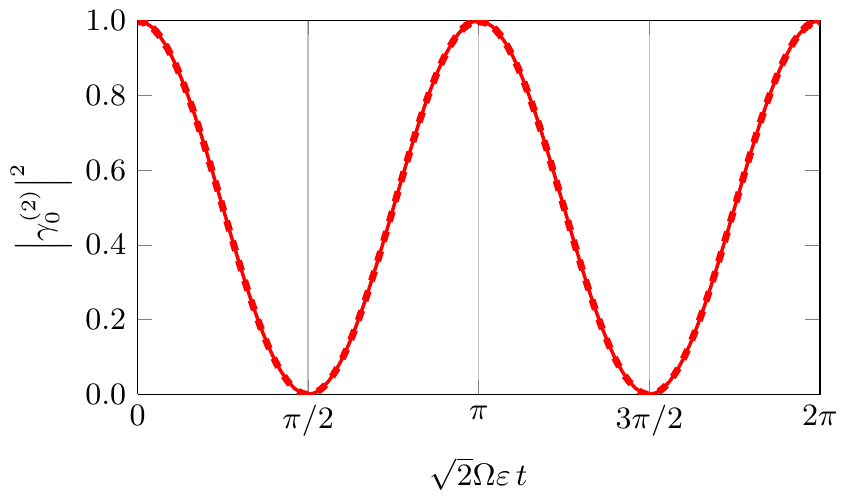} 

 \caption{Rabi oscillations in second-order double Bragg diffraction after elimination of the asymmetric AC Stark shift. The elimination was achieved by choosing $\Delta \omega=(2-32/105 \,\varepsilon^2)\omega_\text{r}$ (solid line). The initial condition was again $(0,0,1,0,0)^\text{T}$. The oscillations are not detuned anymore. The dashed line which represents the numerical solution for $\varepsilon=0.01$ and the same choice of $\Delta \omega$ is in agreement with the approximate analytical result.}
 \label{fig:2nd order corrected}
\end{figure}

In general, the effect of the asymmetric AC Stark shift leading to suppressed Rabi oscillations in second-order double diffraction can be eliminated by just changing the frequencies of the lasers.
This result is numerically verified in Fig.~\ref{fig:2nd order corrected}.

\subsection{Quasi-resonances}
\label{sec:quasi resonances}

\begin{figure}[htb]
\centering
\includegraphics[scale=1]{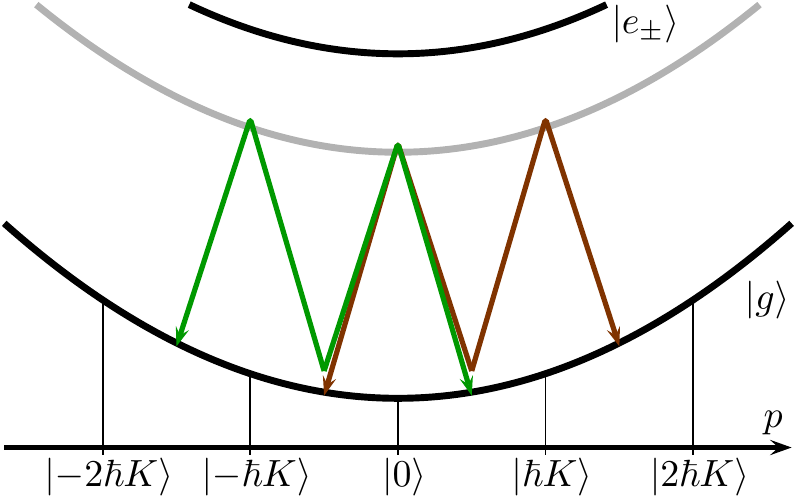}
\caption{Quasi-resonances in first-order double Bragg diffraction for $\Delta \omega= \omega_\text{r}$. Even though we consider first-order diffraction, a second-order process between momenta $\ket{\pm \hbar K/2}$ and $\ket{\pm \hbar 3 K/2}$ is resonant.}
\label{fig: parabula quasi resonance}
\end{figure}

In second-order approximation a feature for the first-order Bragg condition occurs which we call quasi-resonances. In order to demonstrate this phenomenon we again choose $\Delta \omega=\omega_\text{r}$, but now concentrate on momenta close to half-integers of $\hbar K$, e.\,g.\ of $\pm\hbar K/2$ and so on. As Fig.~\ref{fig: parabula quasi resonance} shows, a second-order process now becomes resonant. 

For this reason we set $p= \hbar K/2$, i.\,e., $\nu_\text{D}=\omega_\text{r} $, and insert it into Eq.~\eqref{eq: dgl with delta omega} which yields via the relation
\ba
\I \dot{g}_n=&-\Omega\,  g_{n+1}\left[\e{-\I  (2n+3)\omega_\text{r}t} +\e{-\I (2n+1)\omega_\text{r}t}\right]	\nonumber \\
	&-\Omega\,  g_{n-1}\left[\e{\I (2n+1)\omega_\text{r}t} + \e{\I (2n-1)\omega_\text{r}t}\right]\nonumber
\ea 
the matrices 
\ba 
 \left(\mathcal{H}_\nu\right)_{n,n'}=\omega_\text{r} &\left[\delta_{n-1,n'} (\delta_{2n+1,\nu}+\delta_{2n-1,\nu}) \right.\nonumber\\
		&\left.+ \delta_{n+1,n'}(\delta_{2n+3,-\nu}+\delta_{2n+1,-\nu})\right]\,. \nonumber
\ea
Since $\mathcal{H}_0\equiv0$ we directly turn to the second approximation and find
\ba
\sum_{\nu \neq 0}\frac{\mathcal{H}_{-\nu}\mathcal{H}_\nu}{\nu \omega_\text{r}}= \omega_\text{r}
\begin{pmatrix}
\mathbf{-\frac{4}{5}}	&	0	&	\mathbf{-1}	&	0	\\
0	&	\frac{4}{3}	&	0	&	-1	\\
\mathbf{-1}	&	0	&	\mathbf{\frac{4}{3}}	&	0	\\
0	&	-1	&	0	&	-\frac{4}{5}	
\end{pmatrix}\,, \nonumber
\ea
for $\vec{\gamma}^{(2)}=(\gamma_{-\frac{3}{2}}^{(2)},\gamma_{-\frac{1}{2}}^{(2)},\gamma_{\frac{1}{2}}^{(2)},\gamma_{\frac{3}{2}}^{(2)})^\text{T}$, where we have defined ${\gamma^{(2)}(p+n \hbar K)=\gamma^{(2)}(\hbar K/2+n \hbar K)\equiv \gamma_{n+\frac{1}{2}}}^{(2)}$.

Since there are deviations on the main diagonal, the matrix exponential yields again detuned oscillations. The solution reads
\ba
\vec{\gamma}^{(2)}(t)=\e{-\I \frac{4}{15}\varepsilon^2\omega_\text{r}} \mathcal{M}_{\pm\frac{1}{2}}(t)\, \vec{\gamma}^{(2)}(0)\nonumber
\ea
with
\ba
\mathcal{M}_{\pm\frac{1}{2}}\hspace{-.075cm}=\hspace{-.075cm}
 \cos (\Omega_{\pm\frac{1}{2}} t )\;\mathds{1}
- \frac{\I\sin (\Omega_{\pm\frac{1}{2}} t)}{\sqrt{481}}
\begin{pmatrix}
\mathbf{16} &	0	&	\mathbf{15}	&	0	\\
0	&	-16	&	0	&	15	\\
\mathbf{15}	&	0	&	\mathbf{-16}	&	0	\\
0	&	15	&	0	&	16	
\end{pmatrix} \nonumber
\ea
and the effective Rabi frequency ${\Omega_{\pm\frac{1}{2}}\equiv (\sqrt{481}/15)\,\varepsilon\Omega}$. Here, $\mathds{1}\equiv \delta_{n,n'}$ denotes the identity operator.

\begin{figure}

\includegraphics{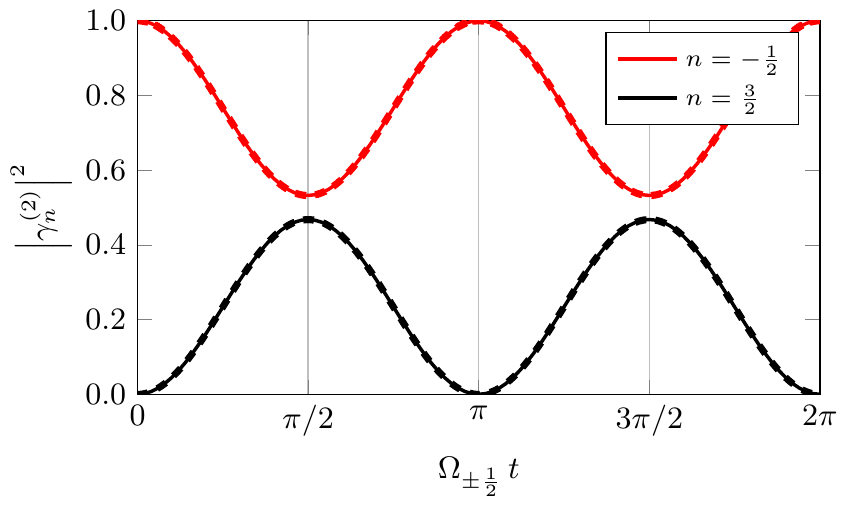} 

 \caption{Rabi oscillations between the quasi-resonances $\ket{-\hbar K/2}\leftrightarrow\ket{3\hbar K/2}$ for $\Delta \omega=\omega_\text{r}$ (solid lines). The oscillations are suppressed by approximately one half. The dashed lines show a numerical solution of the Rabi equations for $\varepsilon=0.01$ and confirm our analytical result.}
 \label{fig:quasi resonances}
\end{figure}
The suppressed Rabi oscillations are separately oscillating between the states $\ket{-\hbar K/2}$ and $\ket{3 \hbar K/2}$ as well as between $\ket{\hbar K/2}$ and $\ket{-3 \hbar K/2}$. To show the emergence of these two separated diffraction processes, the elements corresponding to the latter are boldfaced in the matrices above.
These oscillations are shown in Fig.~\ref{fig:quasi resonances} and coincide with the dotted numerical solution as well.

In fact, these quasi-resonances are no special feature of double Bragg diffraction, as one sees easily from Fig.~\ref{fig: parabula quasi resonance}. We included them for completeness and want to emphasize that this processes are in double diffraction asymmetric, too. So, in general the symmetry of double diffraction is lost if there are non-vanishing initial momenta. On the other side, the width of these resonances in momentum space is much narrower and the oscillation occurs on a much larger timescale than the first-order resonance, as one sees from the discussion above. 

\section{Conclusions}
\label{sec:conclusions}

In this paper we have introduced and analyzed double Bragg diffraction in the context of light-pulse atom interferometry, which has a number of appealing properties for precision interferometry, particularly in microgravity environments, as summarized in the introduction.
In the first place, we have derived the basic equations governing the dynamics of momentum eigenstates as driven by two-photon processes that result from eliminating an excited state largely detuned from single-photon transitions.
This system of coupled linear ordinary differential equations with time-dependent coefficients can be solved numerically, but it is, of course, convenient to have simple analytical solutions, from which deeper insight can be gained, whenever possible. 
The existence of additional nonresonant transitions between resonantly connected states precludes the use of the standard adiabatic elimination procedure. Nevertheless, the method of averaging introduced in Ref.~\cite{bogoliubov} provides a systematic treatment for this kind of situations in terms of slow and fast contributions (corresponding in our case to the Rabi frequency associated with two-photon processes and the recoil frequency) as long as the adiabaticity parameter $\varepsilon$ given by the ratio of those two frequency scales is small.

For the deep Bragg regime, with $\varepsilon \ll 1$, we find generalized Rabi oscillations within an effective three-level system comprising states with momenta $0,\ \hbar K$ and $-\hbar K$. By an appropriate choice of pulse duration and laser intensity one can generate the analogue of $\pi/2$ and $\pi$ pulses, acting, respectively, as beam splitters and mirrors.
In addition, we have also considered nonvanishing initial momenta, for which the resonance condition is no longer fulfilled exactly. Analytical results have been obtained for small momenta, whose effect is analogous to that of a detuning and which lead to damped Rabi oscillations for initial states with nonvanishing momentum width. It is, however, worth pointing out that for larger initial momenta the states with opposite momentum transfer ($+n\hbar K$ and $-n\hbar K$) are no longer equally populated, as illustrated by the exact result on ``quasi-resonances'' of Sec.~\ref{sec:quasi resonances}.
This can be intuitively understood by considering the transformation to the reference frame where the initial momentum vanishes and noticing that in this case the frequency detuning for the two pairs of counterpropagating beams is not the same.

With our approach we have also been able to study analytically the quasi-Bragg regime, with $\varepsilon < 1$ but not too small (e.\,g.\ $\varepsilon \sim 0.1$). This is a regime of particular interest because it relaxes somewhat the effect of velocity selectivity and it corresponds to a parameter range typically accessible in experiments; in particular it makes possible Bragg scattering of relatively high order without excessively long pulse durations. This regime was studied for single Bragg diffraction in Ref.~\cite{quasi-Bragg}. For square pulses our analysis of the double diffraction case reveals the existence of fast oscillations with smaller amplitude superimposed on the slow generalized Rabi oscillations between resonant states. The amplitude of these oscillations for double diffraction is of order $\varepsilon$, in contrast with the single diffraction case, where the amplitude is of order $\varepsilon^2$ and, hence, much more suppressed. Our analytical results for these oscillations are in excellent agreement with numerical solutions, as shown in Fig.~\ref{fig: first order corrections}.

It should be stressed that besides being well suited to problems where the conventional adiabatic elimination procedure cannot be applied, such as double Bragg diffraction, the method of averaging can also be valuable for studying certain aspects of single diffraction since it provides a simpler and mathematically more transparent description. An example of that are the superimposed fast oscillations for square pulses in single Bragg diffraction. These have been found \cite{quasi-Bragg} when solving exactly the Mathieu equation for a pulse of constant amplitude \cite{laemmerzahl95} and matching the solution to the free solution (in absence of external electromagnetic field) before and after the square pulse. A satisfactory description when employing adiabatic expansions requires a proper understanding of the subtle connection between ``dressed'' states and the ``bare'' initial states typically accessible in experiments, as discussed in Sec.~\ref{sec:dressed_states}.

Our treatment demonstrates that higher-order Bragg processes are also possible for double diffraction. Moreover, the existence of an asymmetric AC Stark shift for the two resonant states (which can be easily compensated in practice) has been established. This feature is absent for single diffraction. In that case there is a frame where the two counterpropagating beams give rise to a standing wave and where the two resonant states are symmetric. This cannot be done simultaneously for the two pairs of counterpropagating beams employed in double diffraction.

Having orthogonal polarizations for the copropagating beams and for the equal-frequency counterpropagating ones is crucial in order to avoid spurious transitions, as emphasized in the introduction. While we have restricted our attention to circularly polarized light beams in the main body of the paper, the case of linear polarizations and an arbitrary direction of the magnetic field defining the quantization axis and determining the Zeeman splitting is analyzed in detail in Appendix~\ref{sec:lin-lin}.

Throughout most of the article we have focused on the case of square pulses, but many of our results can be extended to smooth time-dependent pulses, as explained in Appendix~\ref{sec:pulse shapes}. The advantages of using smooth pulses were already pointed out for single diffraction in Ref.~\cite{quasi-Bragg}. Similar advantages apply to the case of double diffraction. Furthermore, the dependence on the laser phases in the contributions that give rise to the fast superimposed oscillations is potentially rather problematic for precision interferometry, and for double diffraction they appear already at order $\varepsilon$ rather than $\varepsilon^2$. Fortunately, for smooth pulses, such as a Gaussian profile, these contributions are instead exponentially suppressed like $\exp(-1/\varepsilon^2)$ (which falls off faster then any power as $\varepsilon \to 0$) as shown in Appendix~\ref{sec:adiabatic_pulse_shape}.
We plan to address these issues in more detail in a future analysis of the response of an interferometer based on double diffraction.

\begin{acknowledgments}
We thank H.~Ahlers, S.~Kleinert, M.~Meister, V.~Tamma, and W.~Zeller for their support and many fruitful discussions.
A.~R., E.~M.~R.\ and W.~P.~S.\ acknowledge financial support by the German Space Agency (DLR) with funds provided by the Federal Ministry of Economics and Technology (BMWi) due to an enactment of the German Bundestag under Grant No.~DLR 50WM1131-1137 (project QUANTUS-III). G.~T. thanks the Max-Planck-Gesellschaft for financial support.
\end{acknowledgments}

\appendix

\section{Method of averaging to arbitrary orders}
\label{sec:averaging arbitrary}
In Sec.~\ref{sec:method of averaging}, we introduced the method of averaging \citep{bogoliubov} and the two lowest approximations. 
In this appendix, we now show that this method is valid to arbitrary orders and we derive the corresponding expressions.

For this purpose, we recall from Sec.~\ref{sec:method of averaging} that we have to solve the equation
\ba 
\dot{\vec{g}}= \I  \varepsilon \mathcal{H}_0 \,\vec{g}+\I \varepsilon \sum\limits_{\nu\neq 0}\e{\I\nu \omega_\text{r}  t} \mathcal{H}_\nu \,\vec{g}\,. \nonumber
\ea
The ansatz 
\ba
	\vec{g}^{(m)}=\vec{\gamma}^{(m)} \vphantom{\sum\limits_{j=1} } + \sum\limits_{j=1}^m \varepsilon^j \vec{f}_j(\vec{\gamma}^{(m)})\, , \label{eq:app_g^m}
\ea
separating the slow from the fast oscillations, satisfies the differential equation
\ba 
\dot{\vec{g}}^{(m)}=\I\varepsilon \mathcal{H}\,
		\vec{g}^{(m)}+\mathcal{O}\left(\varepsilon^{m+1}\right) \nonumber
\ea
up to the $m$-th order in $\varepsilon$.

At each order, we assume that the slowly evolving part $\vec{\gamma}^{(m)}$ fulfills the equation
\ba
\dot{\vec{\gamma}}^{(m)}=\I \varepsilon \mathcal{H}_0\,\vec{\gamma}^{(m)} +\I \summe{\mu}{2}{m} \varepsilon^\mu \vec{p}_\mu(\vec{\gamma}^{(m)})\,. \label{eq:app_gamma-dot}
\ea

The task is now to find the $\vec{f}_j$ and $\vec{p}_j$. For this reason, we use Eqs.~\eqref{eq:app_g^m} and \eqref{eq:app_gamma-dot} above and equate the coefficients
\begin{widetext}
\ba
\dot{\vec{g}}^{(m)}
=&\varepsilon \left(\I \mathcal{H}_0 \, \vec{\gamma}^{(m)}+ \left. \partdifffrac{\vec{f}_1
}{t}\right|_{\vec{\gamma}^{(m)}} \right) + \varepsilon^2 \left(\I\vec{p}_2(\vec{\gamma}^{(m)})+ \I \partdifffrac{\vec{f}_1(\vec{\gamma}^{(m)})}{\vec{\gamma}^{(m)}}  \mathcal{H}_0 \, \vec{\gamma}^{(m)}+\left.\partdifffrac{\vec{f}_2
}{t}\right|_{\vec{\gamma}^{(m)}}\right)\nonumber \\
 &+ \summe{j}{3}{m}\varepsilon^j \left(\vphantom{\summe{j}{3}{m}}\I \vec{p}_j(\vec{\gamma}^{(m)})+ \I \partdifffrac{\vec{f}_{j-1}(\vec{\gamma}^{(m)})}{\vec{\gamma}^{(m)}}  \mathcal{H}_0 \, \vec{\gamma}^{(m)}+\left.\partdifffrac{\vec{f}_j
 }{t}\right|_{\vec{\gamma}^{(m)}}+\I \summe{\mu}{1}{j-2}\partdifffrac{\vec{f}_{\mu}(\vec{\gamma}^{(m)})}{\vec{\gamma}^{(m)}}\vec{p}_{j-\mu}(\vec{\gamma}^{(m)})\right) +\mathcal{O}\left(\varepsilon^{m+1}\right). \nonumber
\ea
\end{widetext}
to the ones on the right-hand side of \eq{eq: dgl g^{(m)}}, i.\,e.\ with
\ba
\I\varepsilon \mathcal{H}\,  \vec{g}^{(m)}+\mathcal{O}&\left(\varepsilon^{m+1}\right)= \I\varepsilon \mathcal{H}\,\vec{\gamma}^{(m)}+\I\varepsilon^2 \mathcal{H}\,\vec{f}_1(\vec{\gamma}^{(m)})\nonumber \\
&+\I\summe{j}{3}{m}\varepsilon^j \mathcal{H} \vec{f}_{j-1}(\vec{\gamma}^{(m)})+\mathcal{O}\left(\Omega^{m+1}\right). \nonumber
\ea

The terms proportional to $\varepsilon$ and $\varepsilon^2$ were dealt with in Sec.~\ref{sec:method of averaging}, where we have determined $\vec{f}_1$ and $\vec{f}_2$ for a specific choice of $\vec{p}_1$ and $\vec{p}_2$. We can extend this treatment iteratively to arbitrary orders of $\varepsilon$.

In order to illustrate this technique we now consider the case of $\varepsilon^k$. 
The functions $\vec{f}_j$ can be written as
\ba
\vec{f}_j(\vec{\gamma}^{(m)})=\sum\limits_{\mu \neq 0}\frac{\e{\I \mu \omega_\text{r} t}}{ \mu \omega_\text{r}} \Phi_\mu^{(j)}\, \vec{\gamma}^{(m)} \nonumber
\ea
for $j<k$ and each of the slow corrections $\vec{p}_j$ is chosen to be of the from ${\vec{p}_j(\vec{\gamma}^{(m)})=\Phi_0^{(j)}\, \vec{\gamma}^{(m)}}$.
Then, the differential equation for $\vec{f}_k$ reads
\begin{widetext}
\ba
\left.\partdifffrac{\vec{f}_k
}{t}\right|_{\vec{\gamma}^{(m)}}=&\I \mathcal{H}\sum\limits_{\mu\neq0}\frac{\e{\I  \omega_\text{r}\mu t}}{ \mu \omega_\text{r}} \Phi_\mu^{(k-1)}\, \vec{\gamma}^{(m)}-\I\sum\limits_{\mu\neq0}\frac{\e{\I  \omega_\text{r}\mu t}}{ \mu \omega_\text{r}} \Phi_\mu^{(k-1)}\mathcal{H}_0\, \vec{\gamma}^{(m)} -\I\sum\limits_{l=1}^{k-2}\sum\limits_{\mu\neq0}\frac{\e{\I \mu \omega_\text{r} t}}{ \mu \omega_\text{r}} \Phi_\mu^{(l)}\Phi_0^{(k-l)}\, \vec{\gamma}^{(m)}- \I \vec{p}_k(\vec{\gamma}^{(m)})\,.\nonumber 
\ea
\end{widetext}
We now define
\ba
\Phi_\mu^{(k)}\equiv\sum_{\nu \neq 0}\frac{\mathcal{H}_{\mu-\nu}\Phi^{(k-1)}_\nu}{\nu \omega_\text{r} }-\frac{\Phi^{(k-1)}_\mu \mathcal{H}_0}{\mu \omega_\text{r}}-\sum\limits_{l=1}^{k-2}\frac{\Phi_\mu^{(l)}\Phi_0^{(k-l)}}{ \mu \omega_\text{r}} \nonumber
\ea
for $\mu \neq 0$ and
\ba
\Phi_0^{(k)}\equiv\sum_{\nu \neq 0}\frac{\mathcal{H}_{-\nu}\Phi^{(k-1)}_\nu}{\nu \omega_\text{r} } \nonumber
\ea
so that with $\vec{p}_k(\vec{\gamma}^{(m)})= \Phi_0^{(k)}\,\vec{\gamma}^{(m)}$ we get the differential equation
\ba
\left.\partdifffrac{\vec{f}_k
}{t}\right|_{\vec{\gamma}^{(m)}}= \I\sum\limits_{\mu \neq 0}\e{\I \mu \omega_\text{r} t} \Phi_\mu^{(k)}\, \vec{\gamma}^{(m)} \nonumber
\ea
which can again be integrated. In this way, all orders up to $m$ can be iteratively determined.

\section{Double Bragg diffraction with orthogonal linear polarizations}
\label{sec:lin-lin}
In Sec.~\ref{sec:double-model} we considered circularly polarized laser beams.
In this appendix we show that even for different orthogonal polarizations and an arbitrary magnetic field we arrive at the same differential equations used throughout this article, provided certain requirements are satisfied.
\begin{figure}[htb]
\centering
\includegraphics[scale=1]{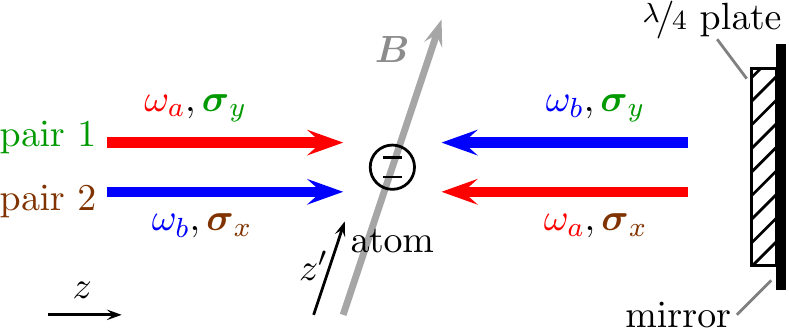}
\caption{Linear-linear configuration for double Bragg diffraction. The magnetic field $\vec{B}$ is not in the $z$-direction of the pair of light fields, but in a different one denoted by $z^\prime$. The polarizations of the two pairs are linear and orthogonal to the $z$-direction of propagation, i.\,e., $\vec{\sigma}_x$ and $\vec{\sigma}_y$.}
\label{fig: setup lin-lin}
\end{figure}

We now assume an experimental set-up as shown in Fig.~\ref{fig: setup lin-lin}.
The linear polarizations $\vec{\sigma}_x$ and $\vec{\sigma}_y$ are orthogonal and named with respect to the direction of propagation $z$ of the light fields.
The magnetic field $\vec{B}$ responsible for a Zeeman splitting and determining the quantization axis points in a different direction $z^\prime$.

With respect to the propagation direction of the light, the electric field reads
\ba
\vec{\mathcal{\hat{E}}}=&E_b \left[\vec{\sigma}_x \e{\I(k_b \hat{z}-\omega_b t)}+\vec{\sigma}_y \e{\I(-k_b \hat{z}-\omega_b t)}\right]\nonumber \\
&+E_a \left[\vec{\sigma}_x \e{\I(-k_a \hat{z}-\omega_a t)}+\vec{\sigma}_y \e{\I(k_a \hat{z}-\omega_a t)}\right ]+ ~ \text{h.c.} \label{eq:app.el.field}
\ea
Writing the polarizations as a linear combination of polarizations $\vec{\sigma}_\pm^\prime$ and $\vec{\sigma}_z^\prime$, which are now with respect to the $z^\prime$-direction of the quantization axis of the atom, we get
\ba
\vec{\sigma}_k=a_+^{(k)}\vec{\sigma}_+^\prime + a_-^{(k)} \vec{\sigma}_-^\prime + a_z^{(k)}\vec{\sigma}_z^\prime = \sum\limits_{j=\pm,z}a_j^{(k)} \vec{\sigma}_j^\prime \nonumber
\ea
for $k=x,y$. Here, $\vec{\sigma}_\pm^\prime$ are circular polarizations orthogonal to the $z^\prime$-direction and $\vec{\sigma}_z^\prime$ is the unit vector along the $z^\prime$-direction.

Since in the unprimed frame the polarizations were orthogonal, we find with the use of the orthogonality in the primed frame the relation
\ba 
\delta_{k,l}=&\vec{\sigma}_k\cdot\vec{\sigma}_l^* =\sum\limits_{j,i=\pm,z}a_j^{(k)}a_i^{*(l)} 
 \vec{\sigma}_j^\prime\cdot\vec{\sigma}_i^{\prime\, *} \nonumber \\
 =&\sum\limits_{j,i=\pm,z}a_j^{(k)}a_i^{*(l)} \delta_{j,i}=\sum\limits_{j=\pm,z}a_j^{(k)}a_j^{*(l)}. \label{eq: orthogonality}
\ea
This orthogonality relation will prove to be crucial later.

\begin{figure}[htb]
\centering
\includegraphics[scale=1]{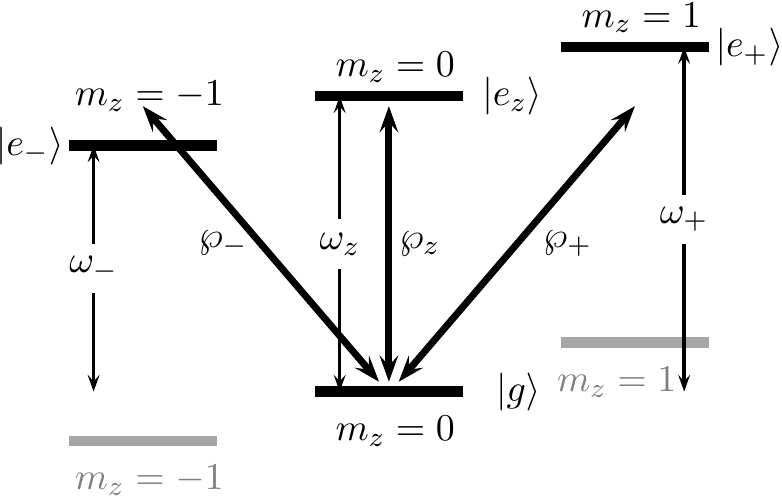}
\caption{Atomic level structure. We consider three excited states $\ket{e_{\pm}}$ and $\ket{e_z}$ with the frequencies $\omega_{\pm}$ and $\omega_z$, respectively, and one ground state $\ket{g}$. The dipole moments $\wp_{\pm}$ and $\wp_z$ initiate transitions between sublevels and correspond to the polarizations $\vec{\sigma}_{\pm}^\prime$ and $\vec{\sigma}_z^\prime$ in the primed frame.}
\label{fig: atomic structure lin-lin}
\end{figure}

We now assume the atomic level structure shown in Fig.~\ref{fig: atomic structure lin-lin}.
Since the magnetic field in $z^\prime$-direction is responsible for the Zeeman splitting, we consider all polarizations with respect to this quantization axis.

The dipole transitions induced by the electric field from Eq.~\eqref{eq:app.el.field}, exciting the states $\ket{e_{\pm}}$ and $\ket{e_z}$ from the ground state $\ket{g}$, that is from the magnetic quantum number $m_z=0$, correspond to each of the three polarizations in the primed frame. The frequencies of the excited states are $\omega_\pm$ and $\omega_z$, and the dipole moments are defined as 
\ba
\wp_{\pm,z}\equiv-\frac{1}{\hbar}\bra{e_{\pm,z}}\vec{\hat{d}}\cdot \vec{\sigma}_{\pm,z}^\prime\ket{g} \, ,\nonumber
\ea
where $\hat{\vec{d}}$ denotes again the dipole operator.
Note that the amplitudes of the electric fields are not included in this definition.

Electric field and atom couple through dipole interaction, which leads to the interaction
\ba
\hat{H}_\text{I}=&\hbar \sum\limits_{j=\pm,z}\wp_j^* \left[ \e{-\I\omega_b t}E_b\left(a_j^{(x)}\e{\I k_b\hat{z}}+a_j^{(y)}\e{-\I k_b\hat{z}}\right)\right. \nonumber \\
 &\hspace{.25cm}+\left.\e{-\I\omega_a t}E_a\left(a_j^{(x)}\e{-\I k_a\hat{z}}+a_j^{(y)}\e{\I k_a\hat{z}}\right)\right]\ket{e_j}\bra{g} \nonumber \\
 &+\text{h.c.}\, \nonumber
\ea
which is much more complicated than Eq.~\eqref{eq:Hamiltonian double}. Nevertheless, we can derive the differential equations for the coefficients in the same way as in Sec.~\ref{sec:singe 3-recursion} and find, when changing into the interaction picture,
\begin{widetext}
\ba 
\I \dot{e}_j(p)=&\wp_j^*  \e{-\I(\omega_b-\omega_j)t}E_b \left( \e{-\I (\omega_{-k_b}+\nu_{-k_b})t}a^{(x)}_j g(p-\hbar k_b)+\e{-\I (\omega_{k_b}+\nu_{k_b})t}a^{(y)}_j g(p+\hbar k_b)\right) \nonumber \\
&+\wp_j^*  \e{-\I(\omega_a-\omega_j)t}E_a \left( \e{-\I (\omega_{k_a}+\nu_{k_a})t}a^{(x)}_j g(p+\hbar k_a)+\e{-\I(\omega_{-k_a}+\nu_{-k_a})t}a^{(y)}_j g(p-\hbar k_a)\right) \nonumber
\ea
and
\ba 
\I \dot{g}(p)=\sum\limits_{j=\pm,z}\wp_j  \left[\e{\I(\omega_b-\omega_j)t} E_b^* \left( \e{-\I (\omega_{k_b}+\nu_{k_b})t}a^{*(x)}_j e_j(p+\hbar k_b)\right.
+\e{-\I (\omega_{-k_b}+\nu_{-k_b})t}a^{*(y)}_j e_j(p-\hbar k_b)\right) \nonumber \\
+ \e{\I(\omega_a-\omega_j)t}E_a^* \left( \e{-\I (\omega_{-k_a}+\nu_{-k_a})t}a^{*(x)}_j e_j(p-\hbar k_a)+\left.\e{-\I (\omega_{k_a}+\nu_{k_a})t}a^{*(y)}_j e_j(p+\hbar k_a)\right)\right] ,\nonumber
\ea
where we have recalled the definitions $\omega_k\equiv\hbar k^2/(2M)$ and $\nu_k\equiv pk/M$ from Sec.~\ref{sec:singe 3-recursion}.

We can again eliminate the excited states $e_j(p)$ adiabatically and find the resulting differential equations for the ground state
\ba
\I \dot{g}(p)=&g(p)\sum\limits_{j=\pm,z}\left[\frac{|\wp_j E_b|^2}{\omega_b-\omega_j} \left( \left|a^{(x)}_j\right|^2+\left|a^{(y)}_j \right|^2 \right)+\frac{|\wp_j E_a|^2}{\omega_a-\omega_j} \left( \left|a^{(x)}_j\right|^2+\left|a^{(y)}_j \right|^2 \right)\right] \nonumber \\
+&g(p+\hbar K) \sum\limits_{j=\pm,z} \left|\wp_j\right|^2 \e{-\I (\omega_{K}+\nu_{K})t} \left(\frac{ \e{\I(\omega_b-\omega_a)t}}{\omega_a-\omega_j} E_aE_b^*\left|a^{(x)}_j\right|^2 + \frac{ \e{\I(\omega_a-\omega_b)t}}{\omega_b-\omega_j}E_bE_a^*\left|a^{(y)}_j\right|^2 \right) \nonumber \\
+&g(p-\hbar K) \sum\limits_{j=\pm,z} \left|\wp_j\right|^2 \e{-\I (\omega_{-K}+\nu_{-K})t} \left(\frac{ \e{\I(\omega_b-\omega_a)t}}{\omega_a-\omega_j}E_aE_b^*\left|a^{(y)}_j\right|^2 + \frac{ \e{\I(\omega_a-\omega_b)t}}{\omega_b-\omega_j}E_bE_a^*\left|a^{(x)}_j\right|^2 \right) \nonumber \\
+&g(p+2\hbar k_b) \sum\limits_{j=\pm,z} \frac{ \left|\wp_jE_b\right|^2 }{\omega_b-\omega_j} \e{-\I (\omega_{2k_b}+\nu_{2k_b})t} a^{*(x)}_j a^{(y)}_j
+g(p-2\hbar k_b) \sum\limits_{j=\pm,z} \frac{ \left|\wp_jE_b\right|^2 }{\omega_b-\omega_j} \e{-\I (\omega_{-2k_b}+\nu_{-2k_b})t} a^{*(y)}_j a^{(x)}_j\nonumber \\
+&g(p+2\hbar k_a) \sum\limits_{j=\pm,z} \frac{ \left|\wp_jE_a\right|^2 }{\omega_a-\omega_j} \e{-\I (\omega_{2k_a}+\nu_{2k_a})t} a^{*(y)}_j a^{(x)}_j
+g(p-2\hbar k_a) \sum\limits_{j=\pm,z} \frac{ \left|\wp_jE_a\right|^2 }{\omega_b-\omega_j} \e{-\I (\omega_{-2k_a}+\nu_{-2k_a})t} a^{*(x)}_j a^{(y)}_j\nonumber \\
+&g(p+\hbar \Delta k) \sum\limits_{j=\pm,z} \left|\wp_j\right|^2 \e{-\I (\omega_{\Delta k}+\nu_{\Delta k})t} \left(\frac{ \e{\I(\omega_b-\omega_a)t}}{\omega_a-\omega_j} E_aE_b^*+ \frac{ \e{\I(\omega_a-\omega_b)t}}{\omega_b-\omega_j}E_bE_a^*\right)a^{(y)}_j a^{*(x)}_j \nonumber \\
+&g(p-\hbar \Delta k) \sum\limits_{j=\pm,z} \left|\wp_j\right|^2 \e{-\I (\omega_{-\Delta k}+\nu_{-\Delta k})t} \left(\frac{ \e{\I(\omega_b-\omega_a)t}}{\omega_a-\omega_j} E_aE_b^*+ \frac{ \e{\I(\omega_a-\omega_b)t}}{\omega_b-\omega_j}E_bE_a^*\right)a^{(x)}_j a^{*(y)}_j\, ,\label{eq:gpunkt-lin-lin} 
\ea
\end{widetext}
where we defined $\Delta k \equiv k_b-k_a$.

Equation \eqref{eq:gpunkt-lin-lin} is quite cumbersome and, in general, cannot be solved with our method of averaging since it couples to eight different momenta of different timescales.
These are the spurious couplings mentioned in Sec.~\ref{sec:double-model}.
Fortunately, we can neglect all terms except the coupling to $g(p\pm \hbar K)$ under the following conditions:

If we assume all $\wp_j $ to be identical, which is a reasonable argument since the light fields $E_{a,b}$ are not included in the definition, and demand that the detuning $\Delta$ is much larger than the Zeeman splitting, i.\,e.\
\ba 
\omega_j-\omega_{a,b}\cong \Delta \nonumber
\ea
for $j= \pm, z$, which is fulfilled under the condition $\omega_+-\omega_- \ll\Delta$, we can use the orthogonality relation of \eq{eq: orthogonality} since all other terms are independent of $j$ and perform the sums.
The only surviving terms of Eq.~\eqref{eq:gpunkt-lin-lin} lead to Eq.~\eqref{eq: dgl with delta omega}. Indeed, the sums proportional to the spurious terms vanish.

Hence, we can conclude that for a linear-linear configuration we can use Eq.~\eqref{eq: dgl with delta omega} , as long as the magnetic field is not too strong, i.\,e.\ the Zeeman splitting $\omega_+-\omega_-$ is not too large.
Very similar arguments hold true for circular polarizations with a magnetic field not aligned with the light propagation.

\section{Pulse shapes}
\label{sec:pulse shapes}
Throughout this paper, we have focused on box-shaped pulses, where we have a time-dependent pulse in the form of a step function. We refer at certain stages of the discussion of the time-evolution to $\pi/2$ and $\pi$ pulses, depending on the duration of the pulses. They create an equal superposition of two momentum states, or an exchange of their populations, respectively.

This approximation does not correspond to most experimental situations, where the lasers are turned on and off according to a certain time-dependent function that differs from a step function. In fact, a smooth pulse results in more convenient properties of the populated states. Of course, this has subtle implications for the method of averaging. This section of the appendix deals with the modification of our method for smooth time-dependent pulse shapes. 
In this case, the Rabi frequencies are proportional to the amplitudes of the electric fields; i.\,e., we get a time-dependent frequency $\tilde{\Omega}\, \tilde{\kappa}(t) $, where the function $\tilde{\kappa}=\tilde{\kappa}(t)$ corresponds to the pulse shape.

\subsection{Adiabaticity condition}
For the adiabatic elimination of the excited state, we integrate a differential equation of the same form as Eq.~\eqref{eq: single e(p)}, namely
\ba
\dot{e}(t)=-\I\tilde{\Omega}\, \tilde{\kappa}(t) \e{\I \Delta t} g(t)\, ,\nonumber
\ea
where we now have included the pulse shape. Integrating this equation by parts we find
\ba
e(t)=&-\I\int\limits_{t_0}^t \D t'\, \tilde{\Omega}\,\tilde{\kappa}(t') \e{\I \Delta t'} g(t')\nonumber  \\
 =& -\frac{\tilde{\Omega}\,\tilde{\kappa}(t)}{\Delta}\e{\I \Delta t} g(t)+\int\limits_{t_0}^t \D t' \frac{\tilde{\Omega}\,\dot{\tilde{\kappa}}(t')}{\Delta}\e{\I \Delta t} g(t') \nonumber \\
 =& -\frac{\tilde{\Omega}\,\tilde{\kappa}(t)}{\Delta}\e{\I \Delta t} g(t)+\frac{\tilde{\Omega}\,\dot{\tilde{\kappa}}(t)}{\I\Delta^2}\e{\I \Delta t}g(t)-\int\limits_{t_0}^t \D t' \ldots
 \,, \label{eq:int_by_parts}
\ea
where we have again assumed $g(t)$ to be slowly varying in comparison to $\Delta$, as in Sec.~\ref{sec:singe 3-recursion} and used the fact that $\tilde{\kappa}(t_0)= 0$.
If now ${|\dot{\tilde{\kappa}}(t)/\Delta^2|\ll|\tilde{\kappa}(t)/\Delta|}$, which can be written as ${|\dot{\tilde{\kappa}}/\tilde{\kappa}|\cong 1/ T\ll \Delta }$ with a characteristic pulse time $T$, we can neglect the contribution coming from the first integration by parts. This condition can be seen as an adiabaticity condition for the pulse shape compared to the high-frequency oscillation of the highly detuned excited state. Neglecting also the higher-order terms, we recover the already known expression for the adiabatic elimination.
In this way, we get the very same equation for the ground state, but now with a time-dependent effective Rabi frequency.

\subsection{Method of averaging with time-dependent pulses}
We can incorporate time-dependent pules shapes in our equations just by replacing $\mathcal{H}$ by $ \kappa(t)\,\mathcal{H} $. After the adiabatic elimination we get a Rabi frequency $\Omega=\tilde{\Omega}^2/\Delta$, and the pulse shape of this Rabi frequency reads ${\kappa(t)=\tilde{\kappa}^2(t)}$. Hence, $\varepsilon$ has to be replaced by $\varepsilon \kappa(t)$. According to Eq.~\eqref{eq: dgl gamma^{(m)}}, the differential equations for the slowly oscillating terms take, in general, the form
\ba
\dot{\vec{\gamma}}=\I\mathcal{A}(t) \vec{\gamma}\, , \nonumber
\ea
where $\mathcal{A}(t)$ is now a time-dependent square matrix.
\subsubsection{Leading-order approximation}
To first order, we find $\mathcal{A}(t)\equiv \varepsilon \kappa(t) \mathcal{H}_0$ for the first-order Bragg resonance ($\Delta\omega= \omega_\text{r}$), where the time-dependence can be factored out and the solution of this differential equation reads
\ba
\vec{\gamma}^{(1)}(t)=\ehoch{\I\varepsilon \int\limits_{t_0}^t \D t'\kappa( t') \mathcal{H}_0}\vec{\gamma}^{(1)} (t_0)\,. \nonumber
\ea
Hence, in the solution of the slowly oscillating term we just have to replace $\Omega$ with $  \Omega \int \D t\kappa( t) $. This substitution coincides with other discussions, e.\,g., the one of Ref.~\cite{quasi-Bragg}.

For higher orders, the solution is not that elementary, since the time-dependence of $\mathcal{A}(t)$ cannot be factored out anymore. So, in general, the quasi-Bragg regime with time-dependent pulses cannot be treated with our method up to higher than leading order in the adiabaticity parameter $\varepsilon$.

On the bright side, we showed in Sec.~\ref{sec:2nd order double} that for second-order Bragg diffraction the matrices simplify and we again can factor out the complete time-dependence. In this sense, finding slowly oscillating solutions with higher-order accuracy than the leading order is not possible for time-dependent pulses, but the leading order with rapidly oscillating corrections can be calculated. 
In second-order double Bragg diffraction we thus can replace $\Omega^2$ with $ \Omega^2 \int \D t \,\kappa^2(t)$, which is in complete agreement with Ref.~\cite{quasi-Bragg}.

But unfortunately, an analytic treatment of $p \neq 0$ is much more complicated, since time-independent terms $\nu_\text{D}$ in the matrices appear and hence even in leading order the time-dependence cannot be factored out. In the treatment of Ref.~\cite{quasi-Bragg} the quasi-Bragg regime for time-dependent pulses is also described just for vanishing momentum width.

\subsubsection{Adiabatic pulse shapes in the method of averaging}
\label{sec:adiabatic_pulse_shape}
When solving the partial differential equation for the rapidly oscillating terms $\vec{f}_j$, one can perform an integration by parts analogously to \eq{eq:int_by_parts}. In the same way, we find the condition $|\dot{\kappa}/\kappa|\cong 1 / T \ll \omega_\text{r}$, which clearly shows that the role of the detuning is now played by the recoil frequency.

This condition corresponds in the limit to the case where the lasers are turned on and off adiabatically. On the other hand, the pulse duration cannot be arbitrarily long if we want to perform $\pi/2$ or $\pi$ pulses, so the characteristic time is of the order of $\Omega\, T \sim \pi$ which limits the pulse duration. Together with this condition above we find
\ba
\frac{1}{\omega_\text{r} T} \sim \frac{\Omega}{\omega_\text{r}\pi}\ll 1\, , \nonumber
\ea 
which is fulfilled only in the Bragg regime $ \Omega/\omega_\text{r}\ll 1 $.\\
This requirement is a consequence of the differential equation satisfied by $\vec{f}_j$, for example  Eq.~\eqref{eq: dgl f_1} for $\vec{f}_1$
\ba
\left.\partdifffrac{\vec{f}_1
}{t}\right|_{\vec{\gamma}^{(m)}}= \kappa(t) \sum\limits_{\nu \neq 0} \I \e{\I \nu \omega_\text{r} t}\mathcal{H}_\nu\, \vec{\gamma}^{(m)}\,, \label{eq:dgl-f1-pulse}
\ea
where we have now included the time dependence $\kappa(t)$ of the pulse. As one can see from Eq.~\eqref{eq:dgl-f1-pulse}, $\vec{\gamma}^{(m)}$ is treated as a constant. This makes it possible to perform the integration for certain pulse shapes even analytically. For instance, if we assume a Gaussian pulse
\ba
\kappa(t)= \ehoch{- t^2/T^2} \,,\nonumber
\ea 
where the characteristic time $T= \sqrt{\pi}/(2 \Omega)$ corresponds to a $\pi/2$ pulse and $T=\sqrt{\pi}/\Omega$ to a $\pi$ pulse. In the following we discuss $\pi$ pulses, but the difference in the $\pi/2$ pulse is just a factor $1/2$.

We can now integrate Eq.~\eqref{eq:dgl-f1-pulse} easily over all times (i.\,e., neglecting the experimental truncation of the Gaussian pulse) 
and find
\ba
\vec{f}_1=\sqrt{\pi} T \sum\limits_{\nu \neq 0} \e{-(\nu \omega_\text{r} T)^2 /4 } \mathcal{H}_\nu\,\vec{\gamma}^{(m)}\,. \nonumber
\ea
Since $\mathcal{H}_\nu$ is proportional to $\omega_\text{r}$, we find for $\pi$ pulses the correction
\ba
\varepsilon \vec{f}_1= & \frac{\Omega}{\omega_\text{r}}\sqrt{\pi} T \sum\limits_{\nu \neq 0} \e{-(\nu \omega_\text{r} T)^2 /4 }  \omega_\text{r}\tilde{\mathcal{H}}_\nu\,\vec{\gamma}^{(m)} \nonumber \\
=& \pi \sum\limits_{\nu \neq 0} \ehoch{-\left(\frac{\sqrt{\pi} \nu}{2 \varepsilon} \right)^2 } \tilde{\mathcal{H}}_\nu\, \vec{\gamma}^{(m)}\,, \nonumber
\ea
where $\tilde{\mathcal{H}}_\nu \equiv \mathcal{H}_\nu/\omega_\text{r}$ are now the dimensionless matrices. We see now that in the deep Bragg regime these corrections are suppressed by a Gaussian $\ehoch{-1/ \varepsilon^{2}}$, which falls off faster than any power of $\varepsilon$ as $\varepsilon$ approaches $0$.  Thus, the scaling is better in comparison with the box-shaped pulses where the correction is suppressed by $\varepsilon$.

\subsection{Dressed states}
Furthermore, time-dependent pulses make the pictures of dressed or bare states obsolete, because in this case both coincide, which can be seen from the initial condition Eq.~\eqref{eq:initial_cond}. Indeed, for time-dependent pulses we get
\ba
	\vec{\gamma}^{(1)}(t_0)=\left[\mathds{1}+\varepsilon\, \kappa(t_0) \sum\limits_{\nu \neq 0}\frac{\mathcal{H}_\nu}{ \nu}\right]^{-1}\;\vec{g}^{(1)}(t_0) \,. \nonumber
\ea
Since the light fields are initially switched off, $\kappa(t_0)$ vanishes and thus
 \ba
	\vec{\gamma}^{(1)}(t_0)=\vec{g}^{(1)}(t_0) \,. \nonumber
\ea
As a consequence the fast oscillatory behavior in single Bragg diffraction found in Fig.~\ref{fig:single_bare}, which is an artefact of the bare-state formulation, vanishes for time-dependent pulses and the population of the states is better described by the smooth oscillations seen in Fig.~\ref{fig:single_oscis}.

\bibliography{BraggScattering}

\begin{thebibliography}{54}%
\makeatletter
\providecommand \@ifxundefined [1]{%
 \@ifx{#1\undefined}
}%
\providecommand \@ifnum [1]{%
 \ifnum #1\expandafter \@firstoftwo
 \else \expandafter \@secondoftwo
 \fi
}%
\providecommand \@ifx [1]{%
 \ifx #1\expandafter \@firstoftwo
 \else \expandafter \@secondoftwo
 \fi
}%
\providecommand \natexlab [1]{#1}%
\providecommand \enquote  [1]{``#1''}%
\providecommand \bibnamefont  [1]{#1}%
\providecommand \bibfnamefont [1]{#1}%
\providecommand \citenamefont [1]{#1}%
\providecommand \href@noop [0]{\@secondoftwo}%
\providecommand \href [0]{\begingroup \@sanitize@url \@href}%
\providecommand \@href[1]{\@@startlink{#1}\@@href}%
\providecommand \@@href[1]{\endgroup#1\@@endlink}%
\providecommand \@sanitize@url [0]{\catcode `\\12\catcode `\$12\catcode
  `\&12\catcode `\#12\catcode `\^12\catcode `\_12\catcode `\%12\relax}%
\providecommand \@@startlink[1]{}%
\providecommand \@@endlink[0]{}%
\providecommand \url  [0]{\begingroup\@sanitize@url \@url }%
\providecommand \@url [1]{\endgroup\@href {#1}{\urlprefix }}%
\providecommand \urlprefix  [0]{URL }%
\providecommand \Eprint [0]{\href }%
\providecommand \doibase [0]{http://dx.doi.org/}%
\providecommand \selectlanguage [0]{\@gobble}%
\providecommand \bibinfo  [0]{\@secondoftwo}%
\providecommand \bibfield  [0]{\@secondoftwo}%
\providecommand \translation [1]{[#1]}%
\providecommand \BibitemOpen [0]{}%
\providecommand \bibitemStop [0]{}%
\providecommand \bibitemNoStop [0]{.\EOS\space}%
\providecommand \EOS [0]{\spacefactor3000\relax}%
\providecommand \BibitemShut  [1]{\csname bibitem#1\endcsname}%
\let\auto@bib@innerbib\@empty
\bibitem [{\citenamefont {van Zoest}\ \emph {et~al.}(2010)\citenamefont {van
  Zoest}, \citenamefont {Gaaloul}, \citenamefont {Singh}, \citenamefont
  {Ahlers}, \citenamefont {Herr}, \citenamefont {Seidel}, \citenamefont
  {Ertmer}, \citenamefont {Rasel}, \citenamefont {Eckart}, \citenamefont
  {Kajari} \emph {et~al.}}]{van-zoest10}%
  \BibitemOpen
  \bibfield  {author} {\bibinfo {author} {\bibfnamefont {T.}~\bibnamefont {van
  Zoest}}, \bibinfo {author} {\bibfnamefont {N.}~\bibnamefont {Gaaloul}},
  \bibinfo {author} {\bibfnamefont {Y.}~\bibnamefont {Singh}}, \bibinfo
  {author} {\bibfnamefont {H.}~\bibnamefont {Ahlers}}, \bibinfo {author}
  {\bibfnamefont {W.}~\bibnamefont {Herr}}, \bibinfo {author} {\bibfnamefont
  {S.}~\bibnamefont {Seidel}}, \bibinfo {author} {\bibfnamefont
  {W.}~\bibnamefont {Ertmer}}, \bibinfo {author} {\bibfnamefont
  {E.}~\bibnamefont {Rasel}}, \bibinfo {author} {\bibfnamefont
  {M.}~\bibnamefont {Eckart}}, \bibinfo {author} {\bibfnamefont
  {E.}~\bibnamefont {Kajari}},  \emph {et~al.},\ }\href
  {http://www.sciencemag.org/content/328/5985/1540.short} {\bibfield  {journal}
  {\bibinfo  {journal} {Science}\ }\textbf {\bibinfo {volume} {328}},\ \bibinfo
  {pages} {1540} (\bibinfo {year} {2010})}\BibitemShut {NoStop}%
\bibitem [{\citenamefont {Geiger}\ \emph {et~al.}(2011)\citenamefont {Geiger},
  \citenamefont {M{\'e}noret}, \citenamefont {Stern}, \citenamefont {Zahzam},
  \citenamefont {Cheinet}, \citenamefont {Battelier}, \citenamefont {Villing},
  \citenamefont {Moron}, \citenamefont {Lours}, \citenamefont {Bidel} \emph
  {et~al.}}]{geiger11}%
  \BibitemOpen
  \bibfield  {author} {\bibinfo {author} {\bibfnamefont {R.}~\bibnamefont
  {Geiger}}, \bibinfo {author} {\bibfnamefont {V.}~\bibnamefont {M{\'e}noret}},
  \bibinfo {author} {\bibfnamefont {G.}~\bibnamefont {Stern}}, \bibinfo
  {author} {\bibfnamefont {N.}~\bibnamefont {Zahzam}}, \bibinfo {author}
  {\bibfnamefont {P.}~\bibnamefont {Cheinet}}, \bibinfo {author} {\bibfnamefont
  {B.}~\bibnamefont {Battelier}}, \bibinfo {author} {\bibfnamefont
  {A.}~\bibnamefont {Villing}}, \bibinfo {author} {\bibfnamefont
  {F.}~\bibnamefont {Moron}}, \bibinfo {author} {\bibfnamefont
  {M.}~\bibnamefont {Lours}}, \bibinfo {author} {\bibfnamefont
  {Y.}~\bibnamefont {Bidel}},  \emph {et~al.},\ }\href@noop {} {\bibfield
  {journal} {\bibinfo  {journal} {Nature communications}\ }\textbf {\bibinfo
  {volume} {2}},\ \bibinfo {pages} {474} (\bibinfo {year} {2011})}\BibitemShut
  {NoStop}%
\bibitem [{\citenamefont {M\"untinga}\ \emph {et~al.}(2013)\citenamefont
  {M\"untinga} \emph {et~al.}}]{muentinga}%
  \BibitemOpen
  \bibfield  {author} {\bibinfo {author} {\bibfnamefont {H.}~\bibnamefont
  {M\"untinga}} \emph {et~al.},\ }\href {\doibase
  10.1103/PhysRevLett.110.093602} {\bibfield  {journal} {\bibinfo  {journal}
  {Phys. Rev. Lett.}\ }\textbf {\bibinfo {volume} {110}},\ \bibinfo {pages}
  {093602} (\bibinfo {year} {2013})}\BibitemShut {NoStop}%
\bibitem [{STE()}]{STE-quest}%
  \BibitemOpen
  \href@noop {} {}\bibinfo {note} {STE-QUEST space mission in ESA’s Cosmic
  Vision 2020-22, \url{http://sci.esa.int/ste-quest}}\BibitemShut {NoStop}%
\bibitem [{\citenamefont {Bodart}\ \emph {et~al.}(2010)\citenamefont {Bodart},
  \citenamefont {Merlet}, \citenamefont {Malossi}, \citenamefont {Pereira~dos
  Santos}, \citenamefont {Bouyer},\ and\ \citenamefont {Landragin}}]{Bodart10}%
  \BibitemOpen
  \bibfield  {author} {\bibinfo {author} {\bibfnamefont {Q.}~\bibnamefont
  {Bodart}}, \bibinfo {author} {\bibfnamefont {S.}~\bibnamefont {Merlet}},
  \bibinfo {author} {\bibfnamefont {N.}~\bibnamefont {Malossi}}, \bibinfo
  {author} {\bibfnamefont {F.}~\bibnamefont {Pereira~dos Santos}}, \bibinfo
  {author} {\bibfnamefont {P.}~\bibnamefont {Bouyer}}, \ and\ \bibinfo {author}
  {\bibfnamefont {A.}~\bibnamefont {Landragin}},\ }\href {\doibase
  10.1063/1.3373917} {\bibfield  {journal} {\bibinfo  {journal} {Applied
  Physics Letters}\ }\textbf {\bibinfo {volume} {96}},\ \bibinfo {pages}
  {134101} (\bibinfo {year} {2010})}\BibitemShut {NoStop}%
\bibitem [{\citenamefont {Hauth}\ \emph {et~al.}(2013)\citenamefont {Hauth},
  \citenamefont {Freier}, \citenamefont {Schkolnik}, \citenamefont {Senger},
  \citenamefont {Schmidt},\ and\ \citenamefont {Peters}}]{Hauth13}%
  \BibitemOpen
  \bibfield  {author} {\bibinfo {author} {\bibfnamefont {M.}~\bibnamefont
  {Hauth}}, \bibinfo {author} {\bibfnamefont {C.}~\bibnamefont {Freier}},
  \bibinfo {author} {\bibfnamefont {V.}~\bibnamefont {Schkolnik}}, \bibinfo
  {author} {\bibfnamefont {A.}~\bibnamefont {Senger}}, \bibinfo {author}
  {\bibfnamefont {M.}~\bibnamefont {Schmidt}}, \ and\ \bibinfo {author}
  {\bibfnamefont {A.}~\bibnamefont {Peters}},\ }\href {\doibase
  10.1007/s00340-013-5413-6} {\bibfield  {journal} {\bibinfo  {journal} {Appl.
  Phys. B}\ }\textbf {\bibinfo {volume} {113}},\ \bibinfo {pages} {49}
  (\bibinfo {year} {2013})}\BibitemShut {NoStop}%
\bibitem [{muq()}]{muquans}%
  \BibitemOpen
  \href@noop {} {}\bibinfo {note} {$\mu$QuanS Precision Quantum Sensors,
  \url{http://www.muquans.com}}\BibitemShut {NoStop}%
\bibitem [{\citenamefont {Herr}\ \emph {et~al.}()\citenamefont {Herr} \emph
  {et~al.}}]{herr}%
  \BibitemOpen
  \bibfield  {author} {\bibinfo {author} {\bibfnamefont {W.}~\bibnamefont
  {Herr}} \emph {et~al.},\ }\href@noop {} {}\bibinfo {howpublished} {in
  preparation}\BibitemShut {NoStop}%
\bibitem [{\citenamefont {Kasevich}\ and\ \citenamefont
  {Dubetsky}()}]{kasevich-patent}%
  \BibitemOpen
  \bibfield  {author} {\bibinfo {author} {\bibfnamefont {M.~A.}\ \bibnamefont
  {Kasevich}}\ and\ \bibinfo {author} {\bibfnamefont {B.}~\bibnamefont
  {Dubetsky}},\ }\href@noop {} {\emph {\bibinfo {title} {Kinematic sensors
  employing atom interferometer phases}}}\ (\bibinfo {address} {United States
  Patent 7317184})\BibitemShut {NoStop}%
\bibitem [{\citenamefont {Durfee}\ \emph {et~al.}(2006)\citenamefont {Durfee},
  \citenamefont {Shaham},\ and\ \citenamefont {Kasevich}}]{durfee06}%
  \BibitemOpen
  \bibfield  {author} {\bibinfo {author} {\bibfnamefont {D.~S.}\ \bibnamefont
  {Durfee}}, \bibinfo {author} {\bibfnamefont {Y.~K.}\ \bibnamefont {Shaham}},
  \ and\ \bibinfo {author} {\bibfnamefont {M.~A.}\ \bibnamefont {Kasevich}},\
  }\href {\doibase 10.1103/PhysRevLett.97.240801} {\bibfield  {journal}
  {\bibinfo  {journal} {Phys. Rev. Lett.}\ }\textbf {\bibinfo {volume} {97}},\
  \bibinfo {pages} {240801} (\bibinfo {year} {2006})}\BibitemShut {NoStop}%
\bibitem [{\citenamefont {Schmidt}\ \emph {et~al.}(2011)\citenamefont
  {Schmidt}, \citenamefont {Senger}, \citenamefont {Hauth}, \citenamefont
  {Freier}, \citenamefont {Schkolnik},\ and\ \citenamefont
  {Peters}}]{Schmidt11}%
  \BibitemOpen
  \bibfield  {author} {\bibinfo {author} {\bibfnamefont {M.}~\bibnamefont
  {Schmidt}}, \bibinfo {author} {\bibfnamefont {A.}~\bibnamefont {Senger}},
  \bibinfo {author} {\bibfnamefont {M.}~\bibnamefont {Hauth}}, \bibinfo
  {author} {\bibfnamefont {C.}~\bibnamefont {Freier}}, \bibinfo {author}
  {\bibfnamefont {V.}~\bibnamefont {Schkolnik}}, \ and\ \bibinfo {author}
  {\bibfnamefont {A.}~\bibnamefont {Peters}},\ }\href {\doibase
  10.1134/S2075108711030102} {\bibfield  {journal} {\bibinfo  {journal}
  {Gyroscopy and Navigation}\ }\textbf {\bibinfo {volume} {2}},\ \bibinfo
  {pages} {170} (\bibinfo {year} {2011})}\BibitemShut {NoStop}%
\bibitem [{\citenamefont {Bragg}(1912)}]{bragg}%
  \BibitemOpen
  \bibfield  {author} {\bibinfo {author} {\bibfnamefont {W.~L.}\ \bibnamefont
  {Bragg}},\ }\href@noop {} {\bibfield  {journal} {\bibinfo  {journal} {Proc.
  Cambridge Philos. Soc.}\ }\textbf {\bibinfo {volume} {17}},\ \bibinfo {pages}
  {43} (\bibinfo {year} {1912})}\BibitemShut {NoStop}%
\bibitem [{\citenamefont {Friedrich}\ \emph {et~al.}(1913)\citenamefont
  {Friedrich}, \citenamefont {Knipping},\ and\ \citenamefont {Laue}}]{laue}%
  \BibitemOpen
  \bibfield  {author} {\bibinfo {author} {\bibfnamefont {W.}~\bibnamefont
  {Friedrich}}, \bibinfo {author} {\bibfnamefont {P.}~\bibnamefont {Knipping}},
  \ and\ \bibinfo {author} {\bibfnamefont {M.}~\bibnamefont {Laue}},\ }\href
  {\doibase 10.1002/andp.19133461004} {\bibfield  {journal} {\bibinfo
  {journal} {Annalen der Physik}\ }\textbf {\bibinfo {volume} {346}},\ \bibinfo
  {pages} {971} (\bibinfo {year} {1913})}\BibitemShut {NoStop}%
\bibitem [{\citenamefont {Rauch}\ \emph {et~al.}(1974)\citenamefont {Rauch},
  \citenamefont {Treimer},\ and\ \citenamefont {Bonse}}]{Rauch74}%
  \BibitemOpen
  \bibfield  {author} {\bibinfo {author} {\bibfnamefont {H.}~\bibnamefont
  {Rauch}}, \bibinfo {author} {\bibfnamefont {W.}~\bibnamefont {Treimer}}, \
  and\ \bibinfo {author} {\bibfnamefont {U.}~\bibnamefont {Bonse}},\ }\href
  {\doibase http://dx.doi.org/10.1016/0375-9601(74)90132-7} {\bibfield
  {journal} {\bibinfo  {journal} {Physics Letters A}\ }\textbf {\bibinfo
  {volume} {47}},\ \bibinfo {pages} {369 } (\bibinfo {year}
  {1974})}\BibitemShut {NoStop}%
\bibitem [{\citenamefont {Rauch}\ and\ \citenamefont
  {Werner}(2000)}]{rauch+werner00}%
  \BibitemOpen
  \bibfield  {author} {\bibinfo {author} {\bibfnamefont {H.}~\bibnamefont
  {Rauch}}\ and\ \bibinfo {author} {\bibfnamefont {S.~A.}\ \bibnamefont
  {Werner}},\ }\href@noop {} {\emph {\bibinfo {title} {Neutron interferometry:
  Lessons in experimental quantum mechanics}}}\ (\bibinfo  {publisher} {Oxford
  University Press},\ \bibinfo {address} {Oxford},\ \bibinfo {year}
  {2000})\BibitemShut {NoStop}%
\bibitem [{\citenamefont {Cronin}\ \emph {et~al.}(2009)\citenamefont {Cronin},
  \citenamefont {Schmiedmayer},\ and\ \citenamefont {Pritchard}}]{cronin09}%
  \BibitemOpen
  \bibfield  {author} {\bibinfo {author} {\bibfnamefont {A.~D.}\ \bibnamefont
  {Cronin}}, \bibinfo {author} {\bibfnamefont {J.}~\bibnamefont
  {Schmiedmayer}}, \ and\ \bibinfo {author} {\bibfnamefont {D.~E.}\
  \bibnamefont {Pritchard}},\ }\href {\doibase 10.1103/RevModPhys.81.1051}
  {\bibfield  {journal} {\bibinfo  {journal} {Rev. Mod. Phys.}\ }\textbf
  {\bibinfo {volume} {81}},\ \bibinfo {pages} {1051} (\bibinfo {year}
  {2009})}\BibitemShut {NoStop}%
\bibitem [{\citenamefont {Gould}\ \emph {et~al.}(1986)\citenamefont {Gould},
  \citenamefont {Ruff},\ and\ \citenamefont {Pritchard}}]{gould86}%
  \BibitemOpen
  \bibfield  {author} {\bibinfo {author} {\bibfnamefont {P.~L.}\ \bibnamefont
  {Gould}}, \bibinfo {author} {\bibfnamefont {G.~A.}\ \bibnamefont {Ruff}}, \
  and\ \bibinfo {author} {\bibfnamefont {D.~E.}\ \bibnamefont {Pritchard}},\
  }\href {\doibase 10.1103/PhysRevLett.56.827} {\bibfield  {journal} {\bibinfo
  {journal} {Phys. Rev. Lett.}\ }\textbf {\bibinfo {volume} {56}},\ \bibinfo
  {pages} {827} (\bibinfo {year} {1986})}\BibitemShut {NoStop}%
\bibitem [{\citenamefont {Martin}\ \emph {et~al.}(1988)\citenamefont {Martin},
  \citenamefont {Oldaker}, \citenamefont {Miklich},\ and\ \citenamefont
  {Pritchard}}]{martin}%
  \BibitemOpen
  \bibfield  {author} {\bibinfo {author} {\bibfnamefont {P.~J.}\ \bibnamefont
  {Martin}}, \bibinfo {author} {\bibfnamefont {B.~G.}\ \bibnamefont {Oldaker}},
  \bibinfo {author} {\bibfnamefont {A.~H.}\ \bibnamefont {Miklich}}, \ and\
  \bibinfo {author} {\bibfnamefont {D.~E.}\ \bibnamefont {Pritchard}},\ }\href
  {\doibase 10.1103/PhysRevLett.60.515} {\bibfield  {journal} {\bibinfo
  {journal} {Phys. Rev. Lett.}\ }\textbf {\bibinfo {volume} {60}},\ \bibinfo
  {pages} {515} (\bibinfo {year} {1988})}\BibitemShut {NoStop}%
\bibitem [{\citenamefont {Rasel}\ \emph {et~al.}(1995)\citenamefont {Rasel},
  \citenamefont {Oberthaler}, \citenamefont {Batelaan}, \citenamefont
  {Schmiedmayer},\ and\ \citenamefont {Zeilinger}}]{rasel95}%
  \BibitemOpen
  \bibfield  {author} {\bibinfo {author} {\bibfnamefont {E.~M.}\ \bibnamefont
  {Rasel}}, \bibinfo {author} {\bibfnamefont {M.~K.}\ \bibnamefont
  {Oberthaler}}, \bibinfo {author} {\bibfnamefont {H.}~\bibnamefont
  {Batelaan}}, \bibinfo {author} {\bibfnamefont {J.}~\bibnamefont
  {Schmiedmayer}}, \ and\ \bibinfo {author} {\bibfnamefont {A.}~\bibnamefont
  {Zeilinger}},\ }\href {\doibase 10.1103/PhysRevLett.75.2633} {\bibfield
  {journal} {\bibinfo  {journal} {Phys. Rev. Lett.}\ }\textbf {\bibinfo
  {volume} {75}},\ \bibinfo {pages} {2633} (\bibinfo {year}
  {1995})}\BibitemShut {NoStop}%
\bibitem [{\citenamefont {Fray}\ \emph {et~al.}(2004)\citenamefont {Fray},
  \citenamefont {Diez}, \citenamefont {H\"ansch},\ and\ \citenamefont
  {Weitz}}]{fray04}%
  \BibitemOpen
  \bibfield  {author} {\bibinfo {author} {\bibfnamefont {S.}~\bibnamefont
  {Fray}}, \bibinfo {author} {\bibfnamefont {C.~A.}\ \bibnamefont {Diez}},
  \bibinfo {author} {\bibfnamefont {T.~W.}\ \bibnamefont {H\"ansch}}, \ and\
  \bibinfo {author} {\bibfnamefont {M.}~\bibnamefont {Weitz}},\ }\href
  {\doibase 10.1103/PhysRevLett.93.240404} {\bibfield  {journal} {\bibinfo
  {journal} {Phys. Rev. Lett.}\ }\textbf {\bibinfo {volume} {93}},\ \bibinfo
  {pages} {240404} (\bibinfo {year} {2004})}\BibitemShut {NoStop}%
\bibitem [{\citenamefont {Haslinger}\ \emph {et~al.}(2013)\citenamefont
  {Haslinger}, \citenamefont {D{\"o}rre}, \citenamefont {Geyer}, \citenamefont
  {Rodewald}, \citenamefont {Nimmrichter},\ and\ \citenamefont
  {Arndt}}]{haslinger13}%
  \BibitemOpen
  \bibfield  {author} {\bibinfo {author} {\bibfnamefont {P.}~\bibnamefont
  {Haslinger}}, \bibinfo {author} {\bibfnamefont {N.}~\bibnamefont
  {D{\"o}rre}}, \bibinfo {author} {\bibfnamefont {P.}~\bibnamefont {Geyer}},
  \bibinfo {author} {\bibfnamefont {J.}~\bibnamefont {Rodewald}}, \bibinfo
  {author} {\bibfnamefont {S.}~\bibnamefont {Nimmrichter}}, \ and\ \bibinfo
  {author} {\bibfnamefont {M.}~\bibnamefont {Arndt}},\ }\href
  {http://www.nature.com/nphys/journal/v9/n3/full/nphys2542.html} {\bibfield
  {journal} {\bibinfo  {journal} {Nature Phys.}\ }\textbf {\bibinfo {volume}
  {9}},\ \bibinfo {pages} {144} (\bibinfo {year} {2013})}\BibitemShut {NoStop}%
\bibitem [{\citenamefont {Giltner}\ \emph {et~al.}(1995)\citenamefont
  {Giltner}, \citenamefont {McGowan},\ and\ \citenamefont {Lee}}]{Giltner95}%
  \BibitemOpen
  \bibfield  {author} {\bibinfo {author} {\bibfnamefont {D.~M.}\ \bibnamefont
  {Giltner}}, \bibinfo {author} {\bibfnamefont {R.~W.}\ \bibnamefont
  {McGowan}}, \ and\ \bibinfo {author} {\bibfnamefont {S.~A.}\ \bibnamefont
  {Lee}},\ }\href {\doibase 10.1103/PhysRevA.52.3966} {\bibfield  {journal}
  {\bibinfo  {journal} {Phys. Rev. A}\ }\textbf {\bibinfo {volume} {52}},\
  \bibinfo {pages} {3966} (\bibinfo {year} {1995})}\BibitemShut {NoStop}%
\bibitem [{\citenamefont {M\"uller}\ \emph
  {et~al.}(2008{\natexlab{a}})\citenamefont {M\"uller}, \citenamefont {Chiow},\
  and\ \citenamefont {Chu}}]{quasi-Bragg}%
  \BibitemOpen
  \bibfield  {author} {\bibinfo {author} {\bibfnamefont {H.}~\bibnamefont
  {M\"uller}}, \bibinfo {author} {\bibfnamefont {S.-W.}\ \bibnamefont {Chiow}},
  \ and\ \bibinfo {author} {\bibfnamefont {S.}~\bibnamefont {Chu}},\ }\href
  {\doibase 10.1103/PhysRevA.77.023609} {\bibfield  {journal} {\bibinfo
  {journal} {Phys. Rev. A}\ }\textbf {\bibinfo {volume} {77}},\ \bibinfo
  {pages} {023609} (\bibinfo {year} {2008}{\natexlab{a}})}\BibitemShut
  {NoStop}%
\bibitem [{\citenamefont {Lett}\ \emph {et~al.}(1988)\citenamefont {Lett},
  \citenamefont {Watts}, \citenamefont {Westbrook}, \citenamefont {Phillips},
  \citenamefont {Gould},\ and\ \citenamefont {Metcalf}}]{Lett88}%
  \BibitemOpen
  \bibfield  {author} {\bibinfo {author} {\bibfnamefont {P.~D.}\ \bibnamefont
  {Lett}}, \bibinfo {author} {\bibfnamefont {R.~N.}\ \bibnamefont {Watts}},
  \bibinfo {author} {\bibfnamefont {C.~I.}\ \bibnamefont {Westbrook}}, \bibinfo
  {author} {\bibfnamefont {W.~D.}\ \bibnamefont {Phillips}}, \bibinfo {author}
  {\bibfnamefont {P.~L.}\ \bibnamefont {Gould}}, \ and\ \bibinfo {author}
  {\bibfnamefont {H.~J.}\ \bibnamefont {Metcalf}},\ }\href {\doibase
  10.1103/PhysRevLett.61.169} {\bibfield  {journal} {\bibinfo  {journal} {Phys.
  Rev. Lett.}\ }\textbf {\bibinfo {volume} {61}},\ \bibinfo {pages} {169}
  (\bibinfo {year} {1988})}\BibitemShut {NoStop}%
\bibitem [{\citenamefont {Dalibard}\ and\ \citenamefont
  {Cohen-Tannoudji}(1989)}]{Dalibard89}%
  \BibitemOpen
  \bibfield  {author} {\bibinfo {author} {\bibfnamefont {J.}~\bibnamefont
  {Dalibard}}\ and\ \bibinfo {author} {\bibfnamefont {C.}~\bibnamefont
  {Cohen-Tannoudji}},\ }\href {\doibase 10.1364/JOSAB.6.002023} {\bibfield
  {journal} {\bibinfo  {journal} {J. Opt. Soc. Am. B}\ }\textbf {\bibinfo
  {volume} {6}},\ \bibinfo {pages} {2023} (\bibinfo {year} {1989})}\BibitemShut
  {NoStop}%
\bibitem [{\citenamefont {Bordé}(1989)}]{borde89}%
  \BibitemOpen
  \bibfield  {author} {\bibinfo {author} {\bibfnamefont {C.}~\bibnamefont
  {Bordé}},\ }\href {\doibase http://dx.doi.org/10.1016/0375-9601(89)90537-9}
  {\bibfield  {journal} {\bibinfo  {journal} {Physics Letters A}\ }\textbf
  {\bibinfo {volume} {140}},\ \bibinfo {pages} {10 } (\bibinfo {year}
  {1989})}\BibitemShut {NoStop}%
\bibitem [{\citenamefont {Kasevich}\ and\ \citenamefont
  {Chu}(1991)}]{kasevich91}%
  \BibitemOpen
  \bibfield  {author} {\bibinfo {author} {\bibfnamefont {M.}~\bibnamefont
  {Kasevich}}\ and\ \bibinfo {author} {\bibfnamefont {S.}~\bibnamefont {Chu}},\
  }\href {\doibase 10.1103/PhysRevLett.67.181} {\bibfield  {journal} {\bibinfo
  {journal} {Phys. Rev. Lett.}\ }\textbf {\bibinfo {volume} {67}},\ \bibinfo
  {pages} {181} (\bibinfo {year} {1991})}\BibitemShut {NoStop}%
\bibitem [{\citenamefont {Peters}\ \emph {et~al.}(1999)\citenamefont {Peters},
  \citenamefont {Chung},\ and\ \citenamefont {Chu}}]{peters99}%
  \BibitemOpen
  \bibfield  {author} {\bibinfo {author} {\bibfnamefont {A.}~\bibnamefont
  {Peters}}, \bibinfo {author} {\bibfnamefont {K.~Y.}\ \bibnamefont {Chung}}, \
  and\ \bibinfo {author} {\bibfnamefont {S.}~\bibnamefont {Chu}},\ }\href@noop
  {} {\bibfield  {journal} {\bibinfo  {journal} {Nature (London)}\ }\textbf
  {\bibinfo {volume} {400}},\ \bibinfo {pages} {849} (\bibinfo {year}
  {1999})}\BibitemShut {NoStop}%
\bibitem [{\citenamefont {Moler}\ \emph {et~al.}(1992)\citenamefont {Moler},
  \citenamefont {Weiss}, \citenamefont {Kasevich},\ and\ \citenamefont
  {Chu}}]{moler92}%
  \BibitemOpen
  \bibfield  {author} {\bibinfo {author} {\bibfnamefont {K.}~\bibnamefont
  {Moler}}, \bibinfo {author} {\bibfnamefont {D.~S.}\ \bibnamefont {Weiss}},
  \bibinfo {author} {\bibfnamefont {M.}~\bibnamefont {Kasevich}}, \ and\
  \bibinfo {author} {\bibfnamefont {S.}~\bibnamefont {Chu}},\ }\href {\doibase
  10.1103/PhysRevA.45.342} {\bibfield  {journal} {\bibinfo  {journal} {Phys.
  Rev. A}\ }\textbf {\bibinfo {volume} {45}},\ \bibinfo {pages} {342} (\bibinfo
  {year} {1992})}\BibitemShut {NoStop}%
\bibitem [{\citenamefont {Kozuma}\ \emph {et~al.}(1999)\citenamefont {Kozuma},
  \citenamefont {Deng}, \citenamefont {Hagley}, \citenamefont {Wen},
  \citenamefont {Lutwak}, \citenamefont {Helmerson}, \citenamefont {Rolston},\
  and\ \citenamefont {Phillips}}]{Kozuma}%
  \BibitemOpen
  \bibfield  {author} {\bibinfo {author} {\bibfnamefont {M.}~\bibnamefont
  {Kozuma}}, \bibinfo {author} {\bibfnamefont {L.}~\bibnamefont {Deng}},
  \bibinfo {author} {\bibfnamefont {E.~W.}\ \bibnamefont {Hagley}}, \bibinfo
  {author} {\bibfnamefont {J.}~\bibnamefont {Wen}}, \bibinfo {author}
  {\bibfnamefont {R.}~\bibnamefont {Lutwak}}, \bibinfo {author} {\bibfnamefont
  {K.}~\bibnamefont {Helmerson}}, \bibinfo {author} {\bibfnamefont {S.~L.}\
  \bibnamefont {Rolston}}, \ and\ \bibinfo {author} {\bibfnamefont {W.~D.}\
  \bibnamefont {Phillips}},\ }\href {\doibase 10.1103/PhysRevLett.82.871}
  {\bibfield  {journal} {\bibinfo  {journal} {Phys. Rev. Lett.}\ }\textbf
  {\bibinfo {volume} {82}},\ \bibinfo {pages} {871} (\bibinfo {year}
  {1999})}\BibitemShut {NoStop}%
\bibitem [{\citenamefont {Torii}\ \emph {et~al.}(2000)\citenamefont {Torii},
  \citenamefont {Suzuki}, \citenamefont {Kozuma}, \citenamefont {Sugiura},
  \citenamefont {Kuga}, \citenamefont {Deng},\ and\ \citenamefont
  {Hagley}}]{torii00}%
  \BibitemOpen
  \bibfield  {author} {\bibinfo {author} {\bibfnamefont {Y.}~\bibnamefont
  {Torii}}, \bibinfo {author} {\bibfnamefont {Y.}~\bibnamefont {Suzuki}},
  \bibinfo {author} {\bibfnamefont {M.}~\bibnamefont {Kozuma}}, \bibinfo
  {author} {\bibfnamefont {T.}~\bibnamefont {Sugiura}}, \bibinfo {author}
  {\bibfnamefont {T.}~\bibnamefont {Kuga}}, \bibinfo {author} {\bibfnamefont
  {L.}~\bibnamefont {Deng}}, \ and\ \bibinfo {author} {\bibfnamefont {E.~W.}\
  \bibnamefont {Hagley}},\ }\href {\doibase 10.1103/PhysRevA.61.041602}
  {\bibfield  {journal} {\bibinfo  {journal} {Phys. Rev. A}\ }\textbf {\bibinfo
  {volume} {61}},\ \bibinfo {pages} {041602} (\bibinfo {year}
  {2000})}\BibitemShut {NoStop}%
\bibitem [{\citenamefont {Debs}\ \emph {et~al.}(2011)\citenamefont {Debs},
  \citenamefont {Altin}, \citenamefont {Barter}, \citenamefont {D\"oring},
  \citenamefont {Dennis}, \citenamefont {McDonald}, \citenamefont {Anderson},
  \citenamefont {Close},\ and\ \citenamefont {Robins}}]{Debs11}%
  \BibitemOpen
  \bibfield  {author} {\bibinfo {author} {\bibfnamefont {J.~E.}\ \bibnamefont
  {Debs}}, \bibinfo {author} {\bibfnamefont {P.~A.}\ \bibnamefont {Altin}},
  \bibinfo {author} {\bibfnamefont {T.~H.}\ \bibnamefont {Barter}}, \bibinfo
  {author} {\bibfnamefont {D.}~\bibnamefont {D\"oring}}, \bibinfo {author}
  {\bibfnamefont {G.~R.}\ \bibnamefont {Dennis}}, \bibinfo {author}
  {\bibfnamefont {G.}~\bibnamefont {McDonald}}, \bibinfo {author}
  {\bibfnamefont {R.~P.}\ \bibnamefont {Anderson}}, \bibinfo {author}
  {\bibfnamefont {J.~D.}\ \bibnamefont {Close}}, \ and\ \bibinfo {author}
  {\bibfnamefont {N.~P.}\ \bibnamefont {Robins}},\ }\href {\doibase
  10.1103/PhysRevA.84.033610} {\bibfield  {journal} {\bibinfo  {journal} {Phys.
  Rev. A}\ }\textbf {\bibinfo {volume} {84}},\ \bibinfo {pages} {033610}
  (\bibinfo {year} {2011})}\BibitemShut {NoStop}%
\bibitem [{\citenamefont {Szigeti}\ \emph {et~al.}(2012)\citenamefont
  {Szigeti}, \citenamefont {Debs}, \citenamefont {Hope}, \citenamefont
  {Robins},\ and\ \citenamefont {Close}}]{momentum-width}%
  \BibitemOpen
  \bibfield  {author} {\bibinfo {author} {\bibfnamefont {S.~S.}\ \bibnamefont
  {Szigeti}}, \bibinfo {author} {\bibfnamefont {J.~E.}\ \bibnamefont {Debs}},
  \bibinfo {author} {\bibfnamefont {J.~J.}\ \bibnamefont {Hope}}, \bibinfo
  {author} {\bibfnamefont {N.~P.}\ \bibnamefont {Robins}}, \ and\ \bibinfo
  {author} {\bibfnamefont {J.~D.}\ \bibnamefont {Close}},\ }\href
  {http://stacks.iop.org/1367-2630/14/i=2/a=023009} {\bibfield  {journal}
  {\bibinfo  {journal} {New Journal of Physics}\ }\textbf {\bibinfo {volume}
  {14}},\ \bibinfo {pages} {023009} (\bibinfo {year} {2012})}\BibitemShut
  {NoStop}%
\bibitem [{\citenamefont {M\"uller}\ \emph
  {et~al.}(2008{\natexlab{b}})\citenamefont {M\"uller}, \citenamefont {Chiow},
  \citenamefont {Long}, \citenamefont {Herrmann},\ and\ \citenamefont
  {Chu}}]{mueller08-24hbarK}%
  \BibitemOpen
  \bibfield  {author} {\bibinfo {author} {\bibfnamefont {H.}~\bibnamefont
  {M\"uller}}, \bibinfo {author} {\bibfnamefont {S.-w.}\ \bibnamefont {Chiow}},
  \bibinfo {author} {\bibfnamefont {Q.}~\bibnamefont {Long}}, \bibinfo {author}
  {\bibfnamefont {S.}~\bibnamefont {Herrmann}}, \ and\ \bibinfo {author}
  {\bibfnamefont {S.}~\bibnamefont {Chu}},\ }\href {\doibase
  10.1103/PhysRevLett.100.180405} {\bibfield  {journal} {\bibinfo  {journal}
  {Phys. Rev. Lett.}\ }\textbf {\bibinfo {volume} {100}},\ \bibinfo {pages}
  {180405} (\bibinfo {year} {2008}{\natexlab{b}})}\BibitemShut {NoStop}%
\bibitem [{\citenamefont {Peters}\ \emph {et~al.}(2001)\citenamefont {Peters},
  \citenamefont {Chung},\ and\ \citenamefont {Chu}}]{peters01}%
  \BibitemOpen
  \bibfield  {author} {\bibinfo {author} {\bibfnamefont {A.}~\bibnamefont
  {Peters}}, \bibinfo {author} {\bibfnamefont {K.~Y.}\ \bibnamefont {Chung}}, \
  and\ \bibinfo {author} {\bibfnamefont {S.}~\bibnamefont {Chu}},\ }\href
  {http://stacks.iop.org/0026-1394/38/i=1/a=4} {\bibfield  {journal} {\bibinfo
  {journal} {Metrologia}\ }\textbf {\bibinfo {volume} {38}},\ \bibinfo {pages}
  {25} (\bibinfo {year} {2001})}\BibitemShut {NoStop}%
\bibitem [{\citenamefont {Dickerson}\ \emph {et~al.}(2013)\citenamefont
  {Dickerson}, \citenamefont {Hogan}, \citenamefont {Sugarbaker}, \citenamefont
  {Johnson},\ and\ \citenamefont {Kasevich}}]{dickerson13}%
  \BibitemOpen
  \bibfield  {author} {\bibinfo {author} {\bibfnamefont {S.~M.}\ \bibnamefont
  {Dickerson}}, \bibinfo {author} {\bibfnamefont {J.~M.}\ \bibnamefont
  {Hogan}}, \bibinfo {author} {\bibfnamefont {A.}~\bibnamefont {Sugarbaker}},
  \bibinfo {author} {\bibfnamefont {D.~M.~S.}\ \bibnamefont {Johnson}}, \ and\
  \bibinfo {author} {\bibfnamefont {M.~A.}\ \bibnamefont {Kasevich}},\ }\href
  {\doibase 10.1103/PhysRevLett.111.083001} {\bibfield  {journal} {\bibinfo
  {journal} {Phys. Rev. Lett.}\ }\textbf {\bibinfo {volume} {111}},\ \bibinfo
  {pages} {083001} (\bibinfo {year} {2013})}\BibitemShut {NoStop}%
\bibitem [{\citenamefont {L\'ev\`eque}\ \emph {et~al.}(2009)\citenamefont
  {L\'ev\`eque}, \citenamefont {Gauguet}, \citenamefont {Michaud},
  \citenamefont {Pereira Dos~Santos},\ and\ \citenamefont
  {Landragin}}]{Leveque}%
  \BibitemOpen
  \bibfield  {author} {\bibinfo {author} {\bibfnamefont {T.}~\bibnamefont
  {L\'ev\`eque}}, \bibinfo {author} {\bibfnamefont {A.}~\bibnamefont
  {Gauguet}}, \bibinfo {author} {\bibfnamefont {F.}~\bibnamefont {Michaud}},
  \bibinfo {author} {\bibfnamefont {F.}~\bibnamefont {Pereira Dos~Santos}}, \
  and\ \bibinfo {author} {\bibfnamefont {A.}~\bibnamefont {Landragin}},\ }\href
  {\doibase 10.1103/PhysRevLett.103.080405} {\bibfield  {journal} {\bibinfo
  {journal} {Phys. Rev. Lett.}\ }\textbf {\bibinfo {volume} {103}},\ \bibinfo
  {pages} {080405} (\bibinfo {year} {2009})}\BibitemShut {NoStop}%
\bibitem [{\citenamefont {Dubetsky}\ and\ \citenamefont
  {Berman}(2002)}]{Dubetsky02}%
  \BibitemOpen
  \bibfield  {author} {\bibinfo {author} {\bibfnamefont {B.}~\bibnamefont
  {Dubetsky}}\ and\ \bibinfo {author} {\bibfnamefont {P.~R.}\ \bibnamefont
  {Berman}},\ }\href {\doibase 10.1103/PhysRevA.66.045402} {\bibfield
  {journal} {\bibinfo  {journal} {Phys. Rev. A}\ }\textbf {\bibinfo {volume}
  {66}},\ \bibinfo {pages} {045402} (\bibinfo {year} {2002})}\BibitemShut
  {NoStop}%
\bibitem [{\citenamefont {Malossi}\ \emph {et~al.}(2010)\citenamefont
  {Malossi}, \citenamefont {Bodart}, \citenamefont {Merlet}, \citenamefont
  {L\'ev\`eque}, \citenamefont {Landragin},\ and\ \citenamefont
  {Santos}}]{Malossi}%
  \BibitemOpen
  \bibfield  {author} {\bibinfo {author} {\bibfnamefont {N.}~\bibnamefont
  {Malossi}}, \bibinfo {author} {\bibfnamefont {Q.}~\bibnamefont {Bodart}},
  \bibinfo {author} {\bibfnamefont {S.}~\bibnamefont {Merlet}}, \bibinfo
  {author} {\bibfnamefont {T.}~\bibnamefont {L\'ev\`eque}}, \bibinfo {author}
  {\bibfnamefont {A.}~\bibnamefont {Landragin}}, \ and\ \bibinfo {author}
  {\bibfnamefont {F.~P.~D.}\ \bibnamefont {Santos}},\ }\href {\doibase
  10.1103/PhysRevA.81.013617} {\bibfield  {journal} {\bibinfo  {journal} {Phys.
  Rev. A}\ }\textbf {\bibinfo {volume} {81}},\ \bibinfo {pages} {013617}
  (\bibinfo {year} {2010})}\BibitemShut {NoStop}%
\bibitem [{\citenamefont {Le~Coq}\ \emph {et~al.}(2006)\citenamefont {Le~Coq},
  \citenamefont {Retter}, \citenamefont {Richard}, \citenamefont {Aspect},\
  and\ \citenamefont {Bouyer}}]{LeCoq}%
  \BibitemOpen
  \bibfield  {author} {\bibinfo {author} {\bibfnamefont {Y.}~\bibnamefont
  {Le~Coq}}, \bibinfo {author} {\bibfnamefont {J.}~\bibnamefont {Retter}},
  \bibinfo {author} {\bibfnamefont {S.}~\bibnamefont {Richard}}, \bibinfo
  {author} {\bibfnamefont {A.}~\bibnamefont {Aspect}}, \ and\ \bibinfo {author}
  {\bibfnamefont {P.}~\bibnamefont {Bouyer}},\ }\href {\doibase
  10.1007/s00340-006-2363-2} {\bibfield  {journal} {\bibinfo  {journal}
  {Applied Physics B}\ }\textbf {\bibinfo {volume} {84}},\ \bibinfo {pages}
  {627} (\bibinfo {year} {2006})}\BibitemShut {NoStop}%
\bibitem [{Note1()}]{Note1}%
  \BibitemOpen
  \bibinfo {note} {The injected polarizations also need to be orthogonal for
  the case of linear polarizations. In contrast, the linear polarizations must
  be parallel for Raman transitions. This difference is a consequence of the
  change of total angular momentum by $\Delta F = \pm 1$ in the latter case,
  whereas the internal state remains unaltered for Bragg
  diffraction.}\BibitemShut {Stop}%
\bibitem [{\citenamefont {Chiow}\ \emph {et~al.}(2011)\citenamefont {Chiow},
  \citenamefont {Kovachy}, \citenamefont {Chien},\ and\ \citenamefont
  {Kasevich}}]{chiow11}%
  \BibitemOpen
  \bibfield  {author} {\bibinfo {author} {\bibfnamefont {S.-w.}\ \bibnamefont
  {Chiow}}, \bibinfo {author} {\bibfnamefont {T.}~\bibnamefont {Kovachy}},
  \bibinfo {author} {\bibfnamefont {H.-C.}\ \bibnamefont {Chien}}, \ and\
  \bibinfo {author} {\bibfnamefont {M.~A.}\ \bibnamefont {Kasevich}},\ }\href
  {\doibase 10.1103/PhysRevLett.107.130403} {\bibfield  {journal} {\bibinfo
  {journal} {Phys. Rev. Lett.}\ }\textbf {\bibinfo {volume} {107}},\ \bibinfo
  {pages} {130403} (\bibinfo {year} {2011})}\BibitemShut {NoStop}%
\bibitem [{\citenamefont {Bernhardt}\ and\ \citenamefont
  {Shore}(1981)}]{bernhardt}%
  \BibitemOpen
  \bibfield  {author} {\bibinfo {author} {\bibfnamefont {A.~F.}\ \bibnamefont
  {Bernhardt}}\ and\ \bibinfo {author} {\bibfnamefont {B.~W.}\ \bibnamefont
  {Shore}},\ }\href {\doibase 10.1103/PhysRevA.23.1290} {\bibfield  {journal}
  {\bibinfo  {journal} {Phys. Rev. A}\ }\textbf {\bibinfo {volume} {23}},\
  \bibinfo {pages} {1290} (\bibinfo {year} {1981})}\BibitemShut {NoStop}%
\bibitem [{\citenamefont {Marte}\ and\ \citenamefont
  {Stenholm}(1992)}]{stenholm}%
  \BibitemOpen
  \bibfield  {author} {\bibinfo {author} {\bibfnamefont {M.}~\bibnamefont
  {Marte}}\ and\ \bibinfo {author} {\bibfnamefont {S.}~\bibnamefont
  {Stenholm}},\ }\href {\doibase 10.1007/BF00325391} {\bibfield  {journal}
  {\bibinfo  {journal} {Applied Physics B}\ }\textbf {\bibinfo {volume} {54}},\
  \bibinfo {pages} {443} (\bibinfo {year} {1992})}\BibitemShut {NoStop}%
\bibitem [{\citenamefont {Brion}\ \emph {et~al.}(2007)\citenamefont {Brion},
  \citenamefont {Pedersen},\ and\ \citenamefont {Mølmer}}]{Brion07}%
  \BibitemOpen
  \bibfield  {author} {\bibinfo {author} {\bibfnamefont {E.}~\bibnamefont
  {Brion}}, \bibinfo {author} {\bibfnamefont {L.~H.}\ \bibnamefont {Pedersen}},
  \ and\ \bibinfo {author} {\bibfnamefont {K.}~\bibnamefont {Mølmer}},\ }\href
  {http://stacks.iop.org/1751-8121/40/i=5/a=011} {\bibfield  {journal}
  {\bibinfo  {journal} {Journal of Physics A: Mathematical and Theoretical}\
  }\textbf {\bibinfo {volume} {40}},\ \bibinfo {pages} {1033} (\bibinfo {year}
  {2007})}\BibitemShut {NoStop}%
\bibitem [{\citenamefont {Bogoliubov}\ and\ \citenamefont
  {Mitropolsky}(1961)}]{bogoliubov}%
  \BibitemOpen
  \bibfield  {author} {\bibinfo {author} {\bibfnamefont {N.~N.}\ \bibnamefont
  {Bogoliubov}}\ and\ \bibinfo {author} {\bibfnamefont {Y.~A.}\ \bibnamefont
  {Mitropolsky}},\ }\href@noop {} {\emph {\bibinfo {title} {Asymptotic methods
  in the theory of non-linear oscillations}}}\ (\bibinfo  {publisher}
  {Hindustan Publishing Corpn.},\ \bibinfo {address} {Delhi},\ \bibinfo {year}
  {1961})\BibitemShut {NoStop}%
\bibitem [{\citenamefont {Schleich}(2001)}]{schleich}%
  \BibitemOpen
  \bibfield  {author} {\bibinfo {author} {\bibfnamefont {W.~P.}\ \bibnamefont
  {Schleich}},\ }\href@noop {} {\emph {\bibinfo {title} {{Quantum Optics in
  Phase Space}}}}\ (\bibinfo  {publisher} {Wiley-VCH},\ \bibinfo {address}
  {Weinheim},\ \bibinfo {year} {2001})\BibitemShut {NoStop}%
\bibitem [{\citenamefont {Meneghini}\ \emph {et~al.}(2000)\citenamefont
  {Meneghini}, \citenamefont {Jex}, \citenamefont {A.~H.~van Leeuwen},
  \citenamefont {R.~Kasimov}, \citenamefont {P.~Schleich},\ and\ \citenamefont
  {P.~Yakovlev}}]{Jex}%
  \BibitemOpen
  \bibfield  {author} {\bibinfo {author} {\bibfnamefont {S.}~\bibnamefont
  {Meneghini}}, \bibinfo {author} {\bibfnamefont {I.}~\bibnamefont {Jex}},
  \bibinfo {author} {\bibfnamefont {K.}~\bibnamefont {A.~H.~van Leeuwen}},
  \bibinfo {author} {\bibfnamefont {M.}~\bibnamefont {R.~Kasimov}}, \bibinfo
  {author} {\bibfnamefont {W.}~\bibnamefont {P.~Schleich}}, \ and\ \bibinfo
  {author} {\bibfnamefont {V.}~\bibnamefont {P.~Yakovlev}},\ }\href
  {http://www.lasphys.com} {\bibfield  {journal} {\bibinfo  {journal} {Laser
  Physics}\ }\textbf {\bibinfo {volume} {10}},\ \bibinfo {pages} {116}
  (\bibinfo {year} {2000})}\BibitemShut {NoStop}%
\bibitem [{\citenamefont {Chudesnikov}\ and\ \citenamefont
  {Yakovlev}(1991)}]{chudesnikovBragg}%
  \BibitemOpen
  \bibfield  {author} {\bibinfo {author} {\bibfnamefont {D.}~\bibnamefont
  {Chudesnikov}}\ and\ \bibinfo {author} {\bibfnamefont {V.}~\bibnamefont
  {Yakovlev}},\ }\href@noop {} {\bibfield  {journal} {\bibinfo  {journal}
  {Laser Phys}\ }\textbf {\bibinfo {volume} {1}},\ \bibinfo {pages} {110}
  (\bibinfo {year} {1991})}\BibitemShut {NoStop}%
\bibitem [{\citenamefont {Kazantsev}\ \emph {et~al.}(1989)\citenamefont
  {Kazantsev}, \citenamefont {Surdutovich},\ and\ \citenamefont
  {Yakovlev}}]{YakovlevBook}%
  \BibitemOpen
  \bibfield  {author} {\bibinfo {author} {\bibfnamefont {A.}~\bibnamefont
  {Kazantsev}}, \bibinfo {author} {\bibfnamefont {G.}~\bibnamefont
  {Surdutovich}}, \ and\ \bibinfo {author} {\bibfnamefont {V.}~\bibnamefont
  {Yakovlev}},\ }\href@noop {} {\emph {\bibinfo {title} {Mechanical Action of
  Light on Atoms}}}\ (\bibinfo  {publisher} {World Scientific Publishing
  Company Incorporated},\ \bibinfo {year} {1989})\BibitemShut {NoStop}%
\bibitem [{\citenamefont {James}\ and\ \citenamefont {Jerke}(2007)}]{James}%
  \BibitemOpen
  \bibfield  {author} {\bibinfo {author} {\bibfnamefont {D.~F.}\ \bibnamefont
  {James}}\ and\ \bibinfo {author} {\bibfnamefont {J.}~\bibnamefont {Jerke}},\
  }\href {\doibase 10.1139/p07-060} {\bibfield  {journal} {\bibinfo  {journal}
  {Canadian Journal of Physics}\ }\textbf {\bibinfo {volume} {85}},\ \bibinfo
  {pages} {625} (\bibinfo {year} {2007})}\BibitemShut {NoStop}%
\bibitem [{\citenamefont {Clad\'e}\ \emph {et~al.}(2006)\citenamefont
  {Clad\'e}, \citenamefont {de~Mirandes}, \citenamefont {Cadoret},
  \citenamefont {Guellati-Kh\'elifa}, \citenamefont {Schwob}, \citenamefont
  {Nez}, \citenamefont {Julien},\ and\ \citenamefont {Biraben}}]{clade}%
  \BibitemOpen
  \bibfield  {author} {\bibinfo {author} {\bibfnamefont {P.}~\bibnamefont
  {Clad\'e}}, \bibinfo {author} {\bibfnamefont {E.}~\bibnamefont
  {de~Mirandes}}, \bibinfo {author} {\bibfnamefont {M.}~\bibnamefont
  {Cadoret}}, \bibinfo {author} {\bibfnamefont {S.}~\bibnamefont
  {Guellati-Kh\'elifa}}, \bibinfo {author} {\bibfnamefont {C.}~\bibnamefont
  {Schwob}}, \bibinfo {author} {\bibfnamefont {F.}~\bibnamefont {Nez}},
  \bibinfo {author} {\bibfnamefont {L.}~\bibnamefont {Julien}}, \ and\ \bibinfo
  {author} {\bibfnamefont {F.}~\bibnamefont {Biraben}},\ }\href {\doibase
  10.1103/PhysRevA.74.052109} {\bibfield  {journal} {\bibinfo  {journal} {Phys.
  Rev. A}\ }\textbf {\bibinfo {volume} {74}},\ \bibinfo {pages} {052109}
  (\bibinfo {year} {2006})}\BibitemShut {NoStop}%
\bibitem [{\citenamefont {Gauguet}\ \emph {et~al.}(2008)\citenamefont
  {Gauguet}, \citenamefont {Mehlst\"aubler}, \citenamefont {L\'ev\`eque},
  \citenamefont {Le~Gou\"et}, \citenamefont {Chaibi}, \citenamefont {Canuel},
  \citenamefont {Clairon}, \citenamefont {Dos~Santos},\ and\ \citenamefont
  {Landragin}}]{gauguet}%
  \BibitemOpen
  \bibfield  {author} {\bibinfo {author} {\bibfnamefont {A.}~\bibnamefont
  {Gauguet}}, \bibinfo {author} {\bibfnamefont {T.~E.}\ \bibnamefont
  {Mehlst\"aubler}}, \bibinfo {author} {\bibfnamefont {T.}~\bibnamefont
  {L\'ev\`eque}}, \bibinfo {author} {\bibfnamefont {J.}~\bibnamefont
  {Le~Gou\"et}}, \bibinfo {author} {\bibfnamefont {W.}~\bibnamefont {Chaibi}},
  \bibinfo {author} {\bibfnamefont {B.}~\bibnamefont {Canuel}}, \bibinfo
  {author} {\bibfnamefont {A.}~\bibnamefont {Clairon}}, \bibinfo {author}
  {\bibfnamefont {F.~P.}\ \bibnamefont {Dos~Santos}}, \ and\ \bibinfo {author}
  {\bibfnamefont {A.}~\bibnamefont {Landragin}},\ }\href {\doibase
  10.1103/PhysRevA.78.043615} {\bibfield  {journal} {\bibinfo  {journal} {Phys.
  Rev. A}\ }\textbf {\bibinfo {volume} {78}},\ \bibinfo {pages} {043615}
  (\bibinfo {year} {2008})}\BibitemShut {NoStop}%
\bibitem [{\citenamefont {L\"ammerzahl}\ and\ \citenamefont
  {Bord{\'e}}(1995)}]{laemmerzahl95}%
  \BibitemOpen
  \bibfield  {author} {\bibinfo {author} {\bibfnamefont {C.}~\bibnamefont
  {L\"ammerzahl}}\ and\ \bibinfo {author} {\bibfnamefont {C.~J.}\ \bibnamefont
  {Bord{\'e}}},\ }\href {\doibase
  http://dx.doi.org/10.1016/0375-9601(95)00402-O} {\bibfield  {journal}
  {\bibinfo  {journal} {Physics Letters A}\ }\textbf {\bibinfo {volume}
  {203}},\ \bibinfo {pages} {59 } (\bibinfo {year} {1995})}\BibitemShut
  {NoStop}%
\end{thebibliography}%
\end{document}